\documentclass[12pt]{article}
\usepackage[linesnumbered,ruled]{algorithm2e}
\usepackage{titlesec}
\SetAlCapNameFnt{\small	}
\SetAlCapFnt{\small	}
\usepackage[title]{appendix}
\usepackage{soul}
\usepackage{amsmath,amssymb,amsbsy,amsthm,calc,cases,enumerate,url,framed,fancyhdr,lipsum,tikz,enumitem,natbib,etoolbox,bbm,booktabs,mathtools}
\usepackage{graphicx}
\usepackage[pageanchor,hypertexnames=false,hidelinks]{hyperref}
\usepackage{minitoc}

\usepackage[a4paper,margin=1in]{geometry}

\usepackage[font=small,skip=0pt]{caption}

\newcommand{\unparskip}{\vspace{-\parskip}}

\usepackage{setspace}

\newtheorem{theorem}{Theorem}%  meant for continuous numbers
\newtheorem{corollary}{Corollary}
\newtheorem{lemma}{Lemma}%
\usepackage{bbm}
\usepackage{stmaryrd}
\newtheorem{proposition}{Proposition}%
\newtheorem{assumption}{Assumption}%

\BeforeBeginEnvironment{proof}{\unparskip}
\BeforeBeginEnvironment{proof_outline}{\unparskip}
\BeforeBeginEnvironment{theorem}{\vspace{0.5\parskip}}
\BeforeBeginEnvironment{corollary}{\vspace{0.5\parskip}}
\BeforeBeginEnvironment{lemma}{\vspace{0.5\parskip}}
\BeforeBeginEnvironment{proposition}{\vspace{0.5\parskip}}
  
\BeforeBeginEnvironment{thm}{\vspace{0.5\parskip}}
\BeforeBeginEnvironment{cor}{\vspace{0.5\parskip}}
\BeforeBeginEnvironment{prop}{\vspace{0.5\parskip}}
\BeforeBeginEnvironment{lem}{\vspace{0.5\parskip}}

\DeclareMathOperator*{\argmax}{arg\,max}
\DeclareMathOperator*{\argmin}{arg\,min}

\newcommand{\dotcup}

\usepackage{xr}
\usepackage{tikz}
\usetikzlibrary{fit}
\newenvironment{proofoutline}{\proof}{\endproof}
\usepackage{mathrsfs}

\theoremstyle{definition}
\newtheorem{example}{Example}
\theoremstyle{plain}
\newtheorem{definition}{Definition}

\usepackage[normalem]{ulem}

\begin{document}

\title{Greedy Empirical Bayes for Density Estimation}

\author{Taole Sha$^{1}$, \ Shibo Yu$^{1}$ \ and Edwin Fong$^{1,}$\thanks{Corresponding author. Email: chefong@hku.hk}  \\ \\
{\small $^1$Department of Statistics and Actuarial Science, University of Hong Kong}\\
}
\date{}

\maketitle

\begin{abstract}
Bayesian nonparametric density estimation is challenging in high dimensions, where traditional approaches like the Dirichlet process (DP) mixture rely on computationally intensive Markov chain Monte Carlo or approximations. We introduce the \textit{partition posterior bootstrap} (PPB), where a flexible prior is constructed by smoothing a DP over a partition of the sample space. Conditioned on this partition, exact posterior sampling reduces to a simple Bayesian bootstrap-like algorithm. Importantly, we select the partition using a greedy empirical Bayes (GEB) algorithm which scales to high dimensions and large sample sizes. GEB greedily maximizes the closed-form marginal likelihood via local split decisions arising naturally from its factorization, yielding an adaptive binary tree partition. Our key theoretical result is that the GEB tree belongs to an admissible family with high probability, splitting all identifiable signals while avoiding over-splitting noise. As a result, the PPB posterior with the GEB tree attains adaptive contraction rates. Our theory holds under weak, verifiable assumptions, which we demonstrate in numerous examples. Empirically, PPB delivers accurate density estimates and uncertainty quantification at a fraction of the computational cost of other approaches, making it a practical off-the-shelf choice for high-dimensional Bayesian density estimation with theoretical guarantees.
\end{abstract}
\newpage

\section{Introduction}
Density estimation given independent and identically distributed (i.i.d.) observations from a distribution $P$ is of key importance in statistics, where downstream tasks such as prediction or functional estimation depend on having an accurate density estimate. As datasets grow in size, it is important to identify flexible methods which remain scalable in high dimensions or with large sample sizes, while providing uncertainty quantification.

Bayesian nonparametric methods are popular for density estimation due to their flexibility, regularization of the density estimates, and built-in uncertainty quantification.
A classic example is the Dirichlet process mixture model (DPMM) \citep{dpmm, dpmm2} which models the density as an infinite mixture model, and typically relies on Markov chain Monte Carlo (MCMC) for inference. Besides mixture modeling, another popular class of methods is based on P\'{o}lya trees (PT; \cite{PT_conju}) which recursively partition the sample space. While PTs are conjugate and computationally scalable, their performance relies on a fixed partition which limits adaptivity, leading to extensions such as Optional P\'{o}lya Trees (OPT; \cite{optional_PT}) which allow adaptive choice of the splitting variable and stopping.
 However, exact posterior inference for OPTs requires recursive marginalization over possible partition trees, making computation prohibitively expensive in higher dimensions. More recent extensions, such as hidden Markov P\'{o}lya trees (HMPT; \cite{LM_Hidden}), additionally allow adaptive splitting locations and employ sequential Monte Carlo (SMC) to approximate the posterior over partitions to improve scalability in high dimensions. 
 
While the more adaptive Bayesian nonparametric methods offer considerable flexibility, they face various computational challenges due to their reliance on MCMC, SMC or expensive recursive marginalization. In addition to their computational demands, existing Bayesian nonparametric methods often require careful specification of prior hyperparameters and algorithm settings, which can influence both estimation accuracy and computational efficiency. Outside the context of density estimation, but motivated by similar challenges in computation and hyperparameter tuning, recent work has revisited the Bayesian bootstrap \citep{BB, weighted_likelihood_bootstrap} as a means for default and scalable nonparametric inference on Euclidean functionals. \citet{fong2019scalable} elicit a simple Dirichlet process (DP) prior on the probability measure and leverage the conjugacy of the DP to obtain a highly scalable posterior bootstrap algorithm for inference on parameters which does not require MCMC. Another key advantage is that only a scalar prior count hyperparameter and base measure are required to be specified. The scalability and ease of the posterior bootstrap have led to applications in robust inference \citep{dellaporta2022robust}, high-dimensional regression \citep{newton2021weighted,nie2023bayesian}, and the use of generative models for prior construction \citep{o2025ai}. Connections to the P\'{o}lya urn  have also spurred recent advances in predictive Bayes \citep{fortini2020quasi,fong2023martingale}. However, due to the reliance on the DP, the posterior bootstrap cannot naturally handle the case where the target functional requires a density, or the density itself is of interest.

\subsection{Our contribution}\label{sec: intro: our}
In this work, we follow in the spirit of the posterior bootstrap, and consider a simple default prior on the density by smoothing a DP over a partition of the sample space, yielding a closed-form posterior we term the partition posterior bootstrap (PPB). The key to maintaining scalability is to treat the partition of the sample space as a nonparametric hyperparameter, whose efficient selection forms the focus of the paper. This avoids the need for sampling or integration over nonparametric latent structures, such as the mixture allocations in DPMMs or the partitions of the sample space in OPTs and HMPTs.
Our approach is a natural generalization of Bayesian histogram approaches as introduced in \citet{Hjort1996}, where Dirichlet conjugacy is also exploited. In particular, \citet{SIMENSEN2026108367} proposed a univariate Bayesian irregular histogram estimator which adaptively selects the number and locations of bins through approximate maximum a posteriori model selection, using a combination of search heuristics and dynamic programming. However, their optimization procedure exploits the natural ordering of intervals in one dimension and does not readily extend to multivariate settings. 

Our main methodological contribution is a scalable selection procedure for the partition: we exploit a natural recursive factorization of the marginal likelihood under the smoothed DP prior, yielding closed-form Bayes factors for local splitting decisions. This motivates a deterministic \textit{greedy empirical Bayes} (GEB) algorithm that recursively constructs a binary tree partition of the space. At each node, we evaluate the Bayes factors associated with potential splitting variables and locations, selecting the split with the largest Bayes factor and stopping if none exceeds one. Thus, both the partition structure and stopping decisions are automatically determined through local Bayesian model comparisons. Crucially, our GEB algorithm remains computationally scalable in higher dimensions and with large sample sizes, and is orders of magnitude faster than competing Bayesian methods while maintaining excellent performance in practice.

Our main theoretical contribution is to establish guarantees for the PPB posterior based on the partition actually selected by the GEB algorithm. This contrasts with existing results which study either globally optimal estimators (sieve MLE or MAP; \citealp{liu2023convergence,SIMENSEN2026108367}) or full posteriors over partitions \citep{Castillo,liu2023convergence}, both of which are computationally challenging or infeasible to obtain in higher dimensions. Our results can also be viewed as extending the asymptotic theory of empirical Bayes \citep{PETRONES.2014BaeB,RizzelliStefano2026EBiB} to the setting of data-adaptive recursive partitions. To handle the greedy nature of our algorithm, we introduce a family of \textit{admissible trees} and show that the GEB partition belongs to this family with high probability, capturing all identifiable signals while avoiding unnecessary splitting in regions dominated by noise.
We then establish contraction rates for the resulting PPB posterior, demonstrating adaptation to $\beta$-Hölder continuity for $0<\beta\leq 1$, anisotropic smoothness, axis-aligned discontinuities, and variable sparsity.

We conclude the introduction with a visualization of the partition obtained by our GEB algorithm in a simple simulation.
\begin{example}[GEB partition]
\label{ex:partition}
Consider the two-dimensional density formed as a product of independent marginals, each equal to the mixture $0.5 \,\textnormal{Beta}(2,2) + 0.5\,\text{Uniform}(0,1)$. Figure~\ref{fig:intro_partition} compares the true density and three histogram estimates based on $n = 100{,}000$ samples and different partitions. Panel (b) shows a standard equal-width histogram with a \(7 \times 7 = 49\) grid, which struggles to capture the density's geometry. Panel (c) depicts a partition guided by the true density's contour level sets, which acts as an ideal but infeasible benchmark. Panel (d) then illustrates the GEB partition, which chooses 49 rectangular bins and approximates the contours of the true density well. The estimation quality is verified in Table \ref{tab:intro_loglik}, where the average test log-likelihood on held-out data is noticeably higher for the GEB partition than for the equal-width partition and close to the contour partition. Importantly, the number of bins is chosen automatically, and the total run-time is only $0.1$ seconds.
\begin{figure}[!h]
    \centering
    \includegraphics[width=\textwidth]{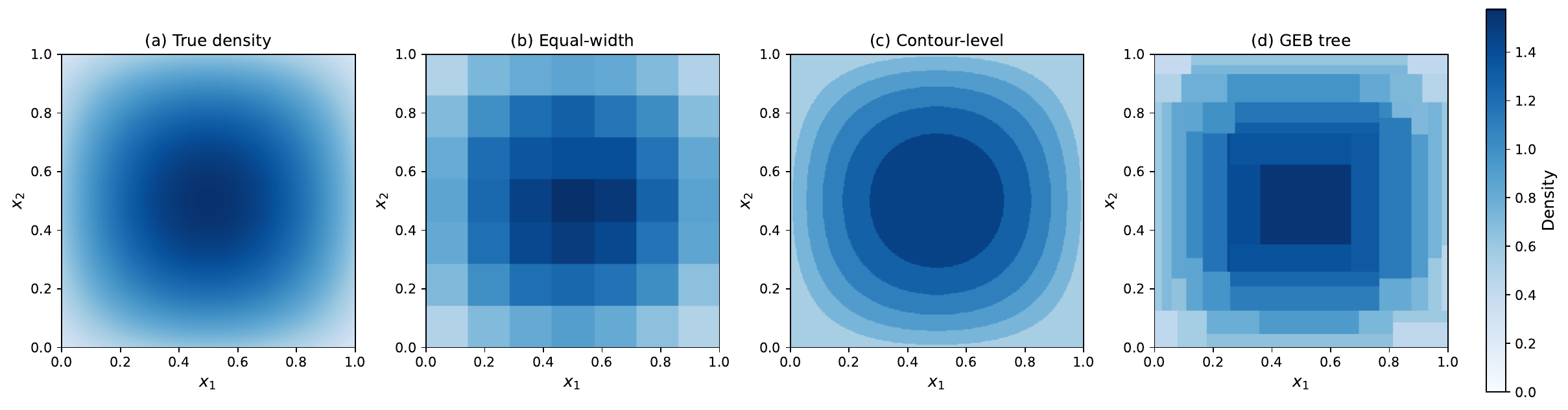}
    \caption{Density estimates for a two-dimensional product mixture.}
    \label{fig:intro_partition}
\end{figure}
\begin{table}[!h]
\centering
\small
\caption{Held-out average test log-likelihood.}
\begin{tabular}{lcccc}
\toprule
& True density & Equal-width & Contour-level & GEB tree \\
\midrule
Avg. test log-likelihood ($\times 10^{-2}$)& 5.30& 4.64& 5.03& 4.93\\
\bottomrule
\end{tabular}
\label{tab:intro_loglik}
\end{table}
\end{example}

\newpage
We now outline the structure of the paper. Section~\ref{sec:PPB} introduces the smoothed DP prior and the resulting PPB algorithm. Section~\ref{sec: tree empirical bayes} then addresses the central challenge of choosing the partition with our GEB algorithm, and Section~\ref{sec: Theory} establishes the theoretical guarantees of the approach. Finally, Section~\ref{sec: experiments} presents simulation and real-data experiments demonstrating the method's performance and substantial computational savings over existing Bayesian nonparametric alternatives. We conclude with a discussion in Section~\ref{sec: discussion}.

\section{Partition posterior bootstrap}\label{sec:PPB}
Let $P$ be the data-generating distribution with Lebesgue density $p$ from which we observe $n$ i.i.d. observations $x_{1:n} = (x_1,\ldots,x_n)$ where $x_i \in \mathcal{A} \subset \mathbb{R}^D$. For a probability measure $\mu$ and measurable sets $A, B$ with $\mu(A) > 0$, we adopt the shorthand $\mu_A := \mu(A)$ and $\mu_{B \mid A} := {\mu_{A\cap B}}/{\mu_A}$ throughout this paper. Let $\mathcal{P} = \{A_1, \ldots, A_M\}$ denote a measurable partition of the sample space $\mathcal{A}$, and let $N_A=\sum_{i=1}^n\mathbbm{1}(x_i\in A)$.

The posterior bootstrap begins by specifying the nonparametric prior $\mathrm{DP}(\alpha, H)$, where $\alpha$ is the concentration parameter and $H$ is a base measure representing our prior guess of $P$. It is well known that the posterior given $x_{1:n}$ is another Dirichlet process. However, a draw from a DP is almost surely discrete, making it unsuitable for modeling a continuous density $p$ directly. To address this, we consider augmenting the DP prior with a smoothing partition $\mathcal{P}$, which preserves tractability in posterior sampling and admits a natural marginal likelihood which we exploit in Section \ref{sec: tree empirical bayes}.

We now augment the DP prior of the posterior bootstrap with a partition $\mathcal{P}$, where we assume $H_A > 0$ for every $A \in \mathcal{P}$. Let $H$ have a Lebesgue density $h$, and assume that $P \ll H$. We then construct our smoothed DP prior on the density $p$ as 
\begin{align}
f(x) = \sum_{A \in \mathcal{P}} G_A\, h(x \mid A), \qquad G\sim \text{DP}\left(\alpha, H\right),
\end{align}
where $h(x \mid A)$ is the conditional density of the form
$$h(x \mid A) = \mathbbm{1}(x \in A) \, \frac{h(x)}{H_A}.$$  The partition $\mathcal{P}$ thus controls the resolution at which the density $f$ can depart from the base density, where the shape of $h$ is retained locally within each $A$.

As in the posterior bootstrap, the underlying random probability measure is specified through a single $\text{DP}(\alpha,H)$ prior. The additional ingredient is the partition $\mathcal{P}$ used to smooth this measure, which will be learned using a GEB algorithm in Section \ref{sec: tree empirical bayes}. Our prior still contains simple and interpretable elements, and another useful feature of our specific prior is that it preserves the base density as the prior mean for any choice of $\mathcal{P}$, that is
$$
\mathbb E[f(x)\mid\mathcal P]
=\sum_{A\in\mathcal P} H_A\,h(x\mid A)
=h(x).
$$
Our construction can be viewed as a natural extension of the Bayesian histogram formulations of \citet[Section 2.1]{Hjort1996} and \cite{SIMENSEN2026108367} by replacing the uniform density within each cell with the conditional base density $h(\cdot\mid A)$, giving the above prior mean property. When $H$ is uniform, the construction coincides with the usual Dirichlet histogram model, and more generally our prior belongs to the class of smoothed Dirichlet process priors discussed by \citet{Hjort1996}. The particular partition smoothing used here enables a simple posterior bootstrap algorithm for inference on the density or its functionals, which we outline in the next subsection. As we will see in Section 3, the above prior has a marginal likelihood structure that is particularly suitable for our GEB algorithm for the partition.

\subsection{Posterior computation} 
Since $(G_A: A \in\mathcal P)
\sim\mathrm{Dirichlet}\left(\alpha H_A: A \in \mathcal{P}\right)$, posterior inference conditioned on $\mathcal{P}$ follows from well-known properties of Dirichlet conjugacy which is usually leveraged for Bayesian histograms. A draw from the posterior distribution of $p$ given $x_{1:n}$ can be represented as
\begin{align*}
f(x) = \sum_{A \in \mathcal{P}} \widetilde{G}_A\, h(x \mid A), \qquad \widetilde{G} \sim \text{DP}\left(\alpha + n, \frac{\alpha H +\sum_{i=1}^n\delta_{x_i}}{\alpha + n} \right).
\end{align*}
We choose the above representation for the posterior as it retains the smoothed DP interpretation. Unlike the posterior bootstrap, our prior induces a posterior predictive density and marginal likelihood. The posterior predictive density given $\mathcal{P}$, which is the posterior mean of $f$ given $\mathcal{P}$, is simply a reweighted version of the base density $h$:
\begin{align}\label{eq: posterior predictive}
    \bar{f}(x) = \sum_{A \in \mathcal{P}} \bar{G}_A \, h(x \mid A), \qquad
    \bar{G}_A = \frac{\alpha H_A + N_A}{\alpha + n}. 
\end{align}
The marginal likelihood for $\mathcal{P}$, which will be the main focus of Section \ref{sec: tree empirical bayes}, is then
\begin{align}\label{eq: marginal likelihood}
    \mathcal{M}(\mathcal{P}) = \frac{\Gamma(\alpha)}{\Gamma(\alpha+n)} \prod_{A \in \mathcal{P}} \left( \frac{\Gamma(\alpha H_A + N_A)}{\Gamma(\alpha H_A)} \, \prod_{i: x_i \in A}h(x_i \mid A) \right).
\end{align}

Conditional on $\mathcal{P}$, posterior sampling involves a partition posterior bootstrap as outlined in Algorithm~\ref{alg: ppb}. 
Functionals with closed-form expressions can be evaluated directly from each density draw. For a risk minimizer $\theta(f)=\argmin_{\theta\in\Theta}\int\ell(x,\theta)\,f(x)\,dx$, the expected loss can instead be approximated by Monte Carlo through $S^{-1}\sum_{s=1}^S\ell(x_s,\theta)$, where $x_s$ is drawn by first selecting a cell $A$ with probability $G_A^{(b)}$ and then drawing a sample from $H(\cdot\mid A)$.
\begin{figure}[!ht]
  \small
  \centering
  \begin{minipage}{.7\linewidth}
    \begin{algorithm}[H]
      \caption{Partition posterior bootstrap}\label{alg: ppb}
      Compute cell counts $N_A = \sum_{i=1}^n \mathbbm{1}(x_i\in A)$ for all $A\in\mathcal P$\\
      \For{$b\gets 1$ \textbf{to} $B$\vspace{1mm}}{
        Draw $(G_A^{(b)}: A\in\mathcal P)\sim\mathrm{Dirichlet}\left(\alpha H_A+N_A : A\in\mathcal P\right)$\vspace{1.5mm} \\
        Set $f^{(b)}(x)=\sum_{A\in\mathcal P}G_A^{(b)}\,h(x\mid A)$\vspace{2mm} \\
          Compute $\theta^{(b)}=\theta(f^{(b)})$ \vspace{2mm}
      }
      \Return $\{f^{(1)},\ldots,f^{(B)}\}$ and $\{\theta^{(1)},\ldots,\theta^{(B)}\}$
    \end{algorithm}
  \end{minipage}
\end{figure}

  \section{Greedy empirical Bayes for partition selection}\label{sec: tree empirical bayes}

 In this section, we view $\mathcal{P}$ as a nonparametric structural hyperparameter to be selected using the marginal likelihood $\mathcal{M}(\mathcal{P})$, which is a natural criterion due to connections to prediction \citep{marginal_cv}. Since the space of partitions is enormous, global optimization is infeasible in practice.
We thus seek a greedy alternative, as the PPB marginal likelihood lends itself to a natural scalable algorithm based on recursive partitioning. 

\subsection{Split Bayes factors}
We begin by studying the key quantity related to the marginal likelihood defined in (\ref{eq: marginal likelihood}), which drives our greedy empirical Bayes algorithm. Consider a partition $\mathcal{P}$, and define the refined partition $\mathcal{P}'$ which splits a cell $A\in \mathcal{P}$ into disjoint cells $A_0,A_1$ so $A = A_0 \cup A_1$, that is $\mathcal{P}' = \mathcal{P} \setminus \{A\} \cup \{A_0, A_1\}$. We now define the ratio of marginal likelihoods
$\mathcal{B}_A(A_0,A_1) :=  {\mathcal{M}(\mathcal{P'})}/{\mathcal{M}(\mathcal{P})}$ as the \textit{split Bayes factor} at $A$, which measures the evidence in favor of splitting $A$ into $A_0,A_1$ relative to leaving $A$ unsplit. Crucially, the split Bayes factor satisfies the identity
\begin{align}\label{eq: node_bf}
    \mathcal{B}_A(A_0,A_1) = \frac{B\left(\alpha H_{A_0} + N_{A_0}, \alpha H_{A_1} + N_{A_1}\right)}{B\left(\alpha H_{A_0},\alpha H_{A_1} \right)} \, 
    \left(\frac{1}{H_{A_0 \mid A}}\right)^{N_{A_0}}\, \left(\frac{1}{H_{A_1 \mid A}}\right)^{N_{A_1}},
\end{align}
where $B(h_1,h_2)$ is the beta function. The above identity is the cornerstone of our greedy approach: $\mathcal{B}_A(A_0,A_1)$ depends solely on the counts and base probabilities of the proposed child cells $A_0,A_1$, and is unaffected by the observations or partition outside $A$. Furthermore, $\mathcal{B}_A$ has a natural interpretation as a binomial Bayes factor: conditional on $N_A$, we are comparing a binomial model for $N_{A_0}$ with a $\mathrm{Beta}(\alpha H_{A_0},\alpha H_{A_1})$ prior on the success probability against the null model where this probability is fixed at $H_{A_0\mid A}$. 

Before moving on to the GEB algorithm, we illustrate how the above interpretable form of the split Bayes factor arises from our prior, and more specifically from well-known properties of the Dirichlet distribution. We can express the marginal likelihood as
\begin{align*}
    \mathcal{M}(\mathcal{P}) = \left[\prod_{i = 1}^n h(x_i)\right] E\left[\prod_{A \in \mathcal{P}} \left(\frac{G_A}{H_A}\right)^{N_A}\right],
\end{align*}
where the expectation is over $(G_A: A \in\mathcal P)
\sim\mathrm{Dirichlet}\left(\alpha H_A: A \in \mathcal{P}\right)$. Now the split of $A$ into $A_0$ and $A_1$ would introduce the following term into the expectation:
\begin{align*}
    \prod_{j = 0}^1\left(\frac{G_{A_j}}{H_{A_j}}\right)^{N_{A_j}} = \left(\frac{G_A}{H_A}\right)^{N_A} \prod_{j = 0}^1 \left(\frac{G_{A_j \mid A}}{H_{A_j \mid A}}\right)^{N_{A_j}}.
\end{align*}
By the aggregation property, replacing $G_{A_0}$ and $G_{A_1}$ by their sum $G_A$ recovers the original Dirichlet distribution on $\mathcal P$, since $\alpha H_{A_0}+\alpha H_{A_1}=\alpha H_A$. By neutrality, the conditional probabilities $(G_{A_0\mid A},G_{A_1\mid A})$ are independent of the aggregated vector $(G_C:C\in\mathcal P)\sim\mathrm{Dirichlet}\left(\alpha H_C: C \in \mathcal{P}\right)$, with
$G_{A_0\mid A}\sim\mathrm{Beta}(\alpha H_{A_0},\alpha H_{A_1})$ and
$G_{A_1\mid A}=1-G_{A_0\mid A}$. 
Consequently, the factor $(G_A/H_A)^{N_A}$ restores the contribution of $A$ to $\mathcal M(\mathcal P)$, while the expectation of the remaining product factors out by independence, yielding the stated split Bayes factor.

Finally, we highlight some connections to related ideas in the PT literature. The leveraged independence property is closely connected to
tail-freeness \citep[Definition~3.11]{ghosal2017fundamentals}, and our prior can be regarded as a partially specified P\'olya tree \citep[Section~3.7.3]{ghosal2017fundamentals}. {Exploiting this tail-free structure to obtain a nodewise factorization is a common starting point for inference in PT models. For example, OPT uses local factors to recursively marginalize over stopping and splitting configurations \citep{optional_PT}, while \citet{ma2024partial} factorize a partial likelihood into nodewise terms. 
However, PT models typically use nodewise factors as computational steps for global inferential quantities, whereas GEB will directly use the local evidence for constructing $\mathcal{P}$.}

\subsection{Greedy empirical Bayes procedure}

We have just shown that once a partition is held fixed, any further refinement of the partition contributes to the marginal likelihood only locally through the split Bayes factor. The core of our GEB algorithm is thus to recursively partition the sample space by sequentially maximizing the split Bayes factor until we reach a partition where it is no longer advantageous to refine further. To cement the idea, let $T$ be a binary tree whose root node is $\mathcal{A}$ and leaves form the partition $\mathcal{P}$. We will subsequently refer to $A$ as a node of the tree $T$, and restrict splits of $A$ to be axis-aligned, so that its child nodes must take the form
$$
A_0 = \{x \in A : x^{(d)} \leq q\}, \quad A_1 = \{x \in A : x^{(d)} > q\},
$$
where $q \in \mathbb{R}$ is the split location, $d \in \{1,\ldots,D\}$ is the split dimension, and $x^{(d)}$ denotes the $d$-th element of the vector $x$. In addition, for regularity, we restrict attention to
eligible splits, meaning those producing two nonempty
children with $N_{A_0}\geq1$ and $N_{A_1}\geq1$.
We call a node $A$ eligible if it admits at least one
eligible split. As a result, we will henceforth write the split Bayes factor as $\mathcal{B}_A(d,q)$ to highlight the potential splits. Finally, for a constant $c \in \mathbb{R}$, let the choice $q = c_{-}$ denote a split of the form $A_0 = \{x \in A: x^{(d)} < c\}$ and $A_1 = \{x \in A: x^{(d)} \geq c\}$ with $\mathcal{B}_A(d,c_{-}) = \lim_{q \uparrow c} \mathcal{B}_A(d,q)$. 

Our GEB algorithm is then illustrated in Algorithm \ref{alg:geb}. At each eligible node $A$, we maximize the split Bayes factor, which in turn greedily maximizes the marginal likelihood $\mathcal{M}(\mathcal{P})$, hence the GEB name. Importantly, the split Bayes factor provides an interpretable stopping criterion based on the binomial Bayes factor, as observing $\mathcal{B}_A(\widehat{d}, \widehat{q}) \leq 1$ indicates that splitting cannot further increase the marginal likelihood. A desirable property of the GEB algorithm is that both the split and stopping decisions are fully governed by the marginal likelihood, without the need for heuristic stopping rules as is common in tree algorithms.
\begin{figure}[!h]
  \small
  \centering
  \begin{minipage}{.7\linewidth}
    \renewcommand{\thempfootnote}{\arabic{mpfootnote}}
    \begin{algorithm}[H]
      \caption{Greedy empirical Bayes (GEB)}\label{alg:geb}
      Initialize $T$ with root node $\mathcal{A}$ so $\mathcal{P} = \{\mathcal{A}\}$\\
      \While{$\mathcal P$ \textnormal{contains any unvisited eligible leaves}}{
      Visit an unvisited eligible leaf $A$\\
        Compute $\mathcal{B}_A(d,q)$ for all eligible splits $(d,q)$ \\
        Set $(\widehat{d}, \widehat{q}) = \argmax_{d,q} \mathcal{B}_A(d,q)$\footnote{We also require a deterministic tie-breaking rule, but the choice is not important; see Appendix \ref{app: split and stop sufficient}.} \\
        \If{$\mathcal{B}_A(\widehat{d}, \widehat{q}) > 1$}{
          Split $A$ into $A_0, A_1$ at $(\widehat{d}, \widehat{q})$, and update $T$ and $\mathcal{P}$
        }
      }
      \Return $T$ and $\mathcal{P}$\vspace{0.5mm}
    \end{algorithm}
  \end{minipage}
  \vspace{-5mm}
\end{figure}

A useful sanity check for the GEB algorithm is that under axis-aligned splits that produce non-empty child nodes, the maximum of the split Bayes factor can be computed using finitely many evaluations, and the GEB tree will always have finite depth. 
\begin{theorem}[Split Bayes factor]
\label{theorem: maximum data point}
 Consider axis-aligned splits producing non-empty children. 
If $N_A \leq 2$, no eligible split has $\mathcal{B}_A(d,q) > 1$. Otherwise, the maximum is almost surely of the form $\mathcal{B}_A(d,x_i^{(d)})$ or $\mathcal{B}_A(d,(x_{i}^{(d)})_{-})$ for some $d$ and observation $x_i \in A$, excluding the sample extrema in coordinate \(d\) within \(A\). Thus, at most $2D(N_A-2)$ evaluations suffice.
\end{theorem}
The detailed proof is given in Appendix~\ref{app: split selection}. In practice however, it is convenient both computationally and theoretically to  restrict candidate splits to a deterministic $m$-grid along each dimension, that is
\begin{align}\label{eq:m_grid}
    q_r = l^{(d)} + r  \, \frac{w^{(d)}}{m} \qquad \text{for } r = 1, \ldots, m-1,
\end{align}
where $m \geq 2$ is a chosen integer, and $l^{(d)}$ and $w^{(d)}$ are the lower bound and width of $A$ along dimension $d$. Note that if $\mathcal{A}$ is a rectangle, every node $A$ is also a rectangle. Again, splits that leave either child empty are not eligible, and a node is eligible if at least one candidate is eligible. In particular, the computation is reduced to at most $D(m-1)$ eligible choices at each node, and our theory in Section~\ref{sec: Theory} requires a fixed and finite $m$.

Intuitively, the split Bayes factor balances evidence of local departures from \(h(\cdot\mid A)\) against the penalty for further refinement. By choosing both the split dimension and location, GEB can place split boundaries near potential discontinuities, vary node sizes across regions,  and avoid unnecessary splits along uninformative coordinates or in regions where a coarse approximation already suffices. This suggests adaptation to the density’s discontinuities, smoothness, and effective dimension, and Section \ref{sec: Theory} formalizes this intuition under structural assumptions and establishes adaptive posterior contraction rates.

\subsection{Practical considerations}\label{sec:practical}

In this section, we discuss some practical choices for the hyperparameters in the PPB model. We begin with a brief discussion on the choice of the base measure. In many cases, it is convenient to map data to the unit cube $[0,1]^D$ using coordinate-wise differentiable transformations with strictly positive derivative, and take $H$ to be uniform on $\mathcal A=[0,1]^D$. The density on the original scale can be computed using the Jacobian, and the choice of $H$ is therefore replaced by the choice of the transform.
 Section~\ref{sec: experiments} describes the specific transforms used in our experiments.

In addition to $H$, GEB requires only two additional hyperparameters: the concentration parameter
$\alpha >0$ and the candidate-grid size $m$, where we recommend $\alpha=1$ and $m=32$ as robust, computationally efficient defaults.
The parameter $\alpha$ adds pseudo-observations from $H$, where a larger $\alpha$ supports finer partitions while increasing shrinkage towards $h$. As a result, increasing $\alpha$ balances two opposing effects: performance may improve as finer partitions capture
structure, but may deteriorate when shrinkage and sampling variability due to fine partitions become excessive. Small values such as $\alpha=1$ mitigate this deterioration and provide a robust default.
Increasing $m$ permits more precise split locations at greater computational cost, with diminishing gains once the grid resolves the relevant structure of the density. For example, densities with sharp local features may prefer larger $m$. Following \citet{LM_Hidden}, we use $m=32$ to balance resolution and
computation. Note that Theorem~\ref{theorem: maximum data point} allows for finite computation even if we take $m \to \infty$. Finally, while our defaults work well in practice,  one can run GEB for different values of \((\alpha,m)\) and select the pair yielding the highest marginal likelihood; see Appendix~\ref{app:geb-sensitivity} for a sensitivity study.

\section{Theory}\label{sec: Theory}
In this section, we study the asymptotic properties of the GEB algorithm, and in particular establish adaptive posterior contraction rates for the induced PPB around the true density. As discussed earlier, unlike analyses based on globally optimal trees, our theory applies directly to the tree selected by GEB which can actually be obtained in practice. As a result, we require a proof technique based on local optimality and a set of greedy-specific assumptions, the exposition of which forms the focus of this section. For convenience, we assume for the theoretical results that $\mathcal{A} = [0,1]^D$, $H$ is the uniform distribution, and splitting is based on the $m$-grid as outlined in (\ref{eq:m_grid}).
We now consider $X_i \overset{\textnormal{i.i.d.}}{\sim} P$ for $i = 1,\ldots,n$, where statements in probability are with respect to the product measure.

\subsection{Signal and bias-variance tradeoff}
The core of our GEB algorithm is the split Bayes factor, which drives both splitting and stopping decisions. To understand the nature of the algorithm, we will need to introduce the notion of \textit{signal} at a given node $A$. Let $K(p_1; \,p_2)$ denote the Kullback-Leibler (KL) divergence between densities $p_1$ and $p_2$. We will define the \textit{signal} of a node $A$ as
\begin{align}\label{eq:true_signal}
    S_A := P_A\,  K\left\{p(\cdot \mid A); \, h(\cdot \mid A)\right\}.
\end{align}
The signal is thus larger if $h(\cdot \mid A)$ is a poor local approximation of $p(\cdot \mid A)$, but this is down-weighted by the probability mass of $A$ for smaller nodes. To see why $S_A$ is a useful notion of signal, for a given partition $\mathcal{P}$, consider the noiseless partition density $p^{\dagger}(x) = \sum_{A \in \mathcal{P}} P_A \, h(x \mid A)$. The KL divergence between $p$ and $p^{\dagger}$ can be shown to satisfy 
{
\begin{align}\label{eq:KL_signal}
K(p; \, p^{\dagger}) = \sum_{A \in \mathcal{P}} S_A, 
\end{align}
}using the fact $p(x) =  P_A\, p(x \mid A)$ for $x \in A$ for a measure $P$ with density $p$. Each node $A \in \mathcal{P}$ thus contributes $S_A$ to the overall KL divergence, balancing local fit with node size.

A counterpart to the signal is the \textit{split gain} at a node $A$ for a split $(d,q)$, defined as
\begin{align}\label{eq:split_gain}
    s_A(d,q) := P_A \left[P_{A_0 \mid A} \log \frac{P_{A_0 \mid A}}{H_{A_0 \mid A}} + \left(1-P_{A_0 \mid A} \right)\log  \left(\frac{1- P_{A_0 \mid A} }{1 - H_{A_0 \mid A} }\right)\right].
\end{align}
Analogous to the signal, the term in the square brackets is the KL divergence between the Bernoulli distributions with success probabilities $P_{A_0 \mid A}$ and $H_{A_0 \mid A}$. Crucially, the split gain and signal are related by the lemma below.
\begin{lemma}[Split gain]\label{lem:split_gain}
Consider splitting a node $A$ at $(d,q)$ into children $A_0$ and $A_1$. The split gain then satisfies 
    \begin{align*}
        S_{A_0}+S_{A_1} =  S_A - s_A(d,q).
    \end{align*}
\end{lemma}
Splitting a node $A$ at $(d,q)$ thus reduces the KL divergence in (\ref{eq:KL_signal}) exactly by $s_A(d,q)$; further details are provided in Appendix \ref{app: KL_monotonicity}. A key assumption, which we outline in Section~\ref{sec: assumptions}, is that the maximum split gain is sufficiently large relative to $S_A$, so that greedy splits can substantially reduce the node's contribution to the KL divergence.

Finally, we connect the split gain to the split Bayes factor.
For sufficiently large $n$ and for nodes with $N_A > 0$, we show in Appendix \ref{app: technical results} that the split Bayes factor satisfies 
\begin{align}\label{eq:splitBF_decomp}
        P_A\left[\frac{\log \mathcal{B}_A(d,q)}{N_A}\right] &= s_A(d,q)-\frac{1}{2n}\log\frac{n}{H_A} + \varepsilon_{A, n}(d,q),
\end{align}
where the second negative term is the Bayesian penalty, and the remainder $\varepsilon_{A, n}(d,q)$ accounts for sampling and approximation errors. The above decomposition reveals a competition between the split gain and the complexity penalty for a given $n$. For strong signals, the split gain $s_A(d,q)$ generally dominates (\ref{eq:splitBF_decomp}), leading to $\mathcal{B}_A(d,q) > 1$ which drives splitting. On the other hand, for smaller nodes and/or weak signals, the negative penalty will tend to drive $\mathcal{B}_A(d,q) < 1$ which favors stopping.
A key technical step (Appendix \ref{app: decomposition}) is to control the remainder terms under both the strong and weak signal regimes, so that the split gain or penalty dominates respectively with high probability.

Importantly, strong and weak signals must be understood relative to the sample size $n$. Although the population signals are defined using the true node probabilities $P_A$, these are estimated in practice (e.g. with $N_A/n$), and the data must support both splitting and the estimation of the resulting density. Further splitting reduces the approximation error in (\ref{eq:KL_signal}), but leaves smaller nodes with fewer observations in each child for estimating node probabilities. The split Bayes factor thus reflects this bias–variance tradeoff: for a fixed node and split, larger $n$ reduces both the penalty and the remainder with high probability, allowing weaker signals to justify further splitting. This motivates the $n$-dependent signal threshold introduced in the next section.

\subsection{Near-optimal and admissible trees}
This section introduces the key oracle machinery behind our proof, leveraging the intuition from the split Bayes factor decomposition in (\ref{eq:splitBF_decomp}). In particular, in order to handle the greedy nature of our GEB algorithm, we will need to define an oracle class of trees based on local criteria without assuming any kind of global optimality.

We begin by defining a useful class of trees which captures a local version of near-optimality. Let $\delta \in (0,1)$ be a slack parameter which governs the near-optimality of our oracle class. In most cases, the value of $\delta$ is unimportant and we can take $\delta = 1/2$ for convenience. One exception where $\delta$ must be selected appropriately is if $p$ is discontinuous as in Example \ref{ex:step}, but we will tend to omit the dependence on $\delta$ for convenience. We now define the class of near-optimal trees denoted as $\mathcal{T}^*$, which is similar to the tree growing rules defined in \citet[Condition 5]{chi2022asymptotic}.
\begin{definition}[Near-optimal trees]\label{df: near-optimal}
  Fix $\delta \in (0,1)$. Let $\mathcal{T}^*$ consist of all finite trees rooted at $\mathcal{A}$ such that, at every
internal node $A$, the chosen split $(\hat{d},\hat{q})$ satisfies
$$
    s_A(\hat{d},\hat{q})
    \geq (1-\delta)\,\max_{d,q} s_A(d,q),
$$
where the maximum is taken over the candidate $m$-grid splits of $A$.
\end{definition}
The class $\mathcal{T}^*$ is defined entirely at the population level, depending only on $p$, $m$, and $\delta$, which we will use to construct our assumptions in Section \ref{sec: assumptions}. While $\mathcal{T}^*$ constrains the choice of splits, it allows for arbitrarily deep trees, which does not take into account the aforementioned bias-variance tradeoff. In order to handle this, we will define a threshold to indicate that the signal $S_A$ of a node is large enough relative to $n$. 

Specifically, a node $A$ is deemed \textit{identifiable} if its signal satisfies
\begin{align}\label{eq:ident}
    S_A \geq \frac{C_0\log n}{n},
\end{align}
where $C_0 > 0$ is a sufficiently large constant that only depends on $p$, $m$ and $\delta$ as defined in Appendix \ref{app: technical results}. To see why this is a suitable threshold, we refer back to the decomposition of the split Bayes factor in (\ref{eq:splitBF_decomp}). Since $s_A(d,q) \leq S_A$ by Lemma \ref{lem:split_gain}, no split gain can overcome the negative penalty unless $S_A$ is at least of order $(\log n)/n$. Furthermore, since the split gain is nonnegative, each child has signal no greater than its parent, so once a node falls below the identifiability threshold, all its descendants remain non-identifiable. 

We are now ready to introduce the class of \textit{admissible trees} $\mathcal{T}_n$ based on the above threshold.
\begin{definition}[Admissible trees]\label{def: admissible trees}
    Let $k_{\max}$ be the fixed positive integer depending only on $p$, $m$ and $\delta$ as defined in Appendix \ref{app: technical results}.  Let $\mathcal T_n$ consist of all trees constructed as follows:
\begin{enumerate}
    \item Start with a near-optimal tree $T\in\mathcal T^*$ whose internal nodes are identifiable and whose leaves are non-identifiable at $n$.
    \item Refine each leaf of $T$ using fewer than $k_{\max}$ additional $m$-grid splits.
\end{enumerate}
\end{definition}

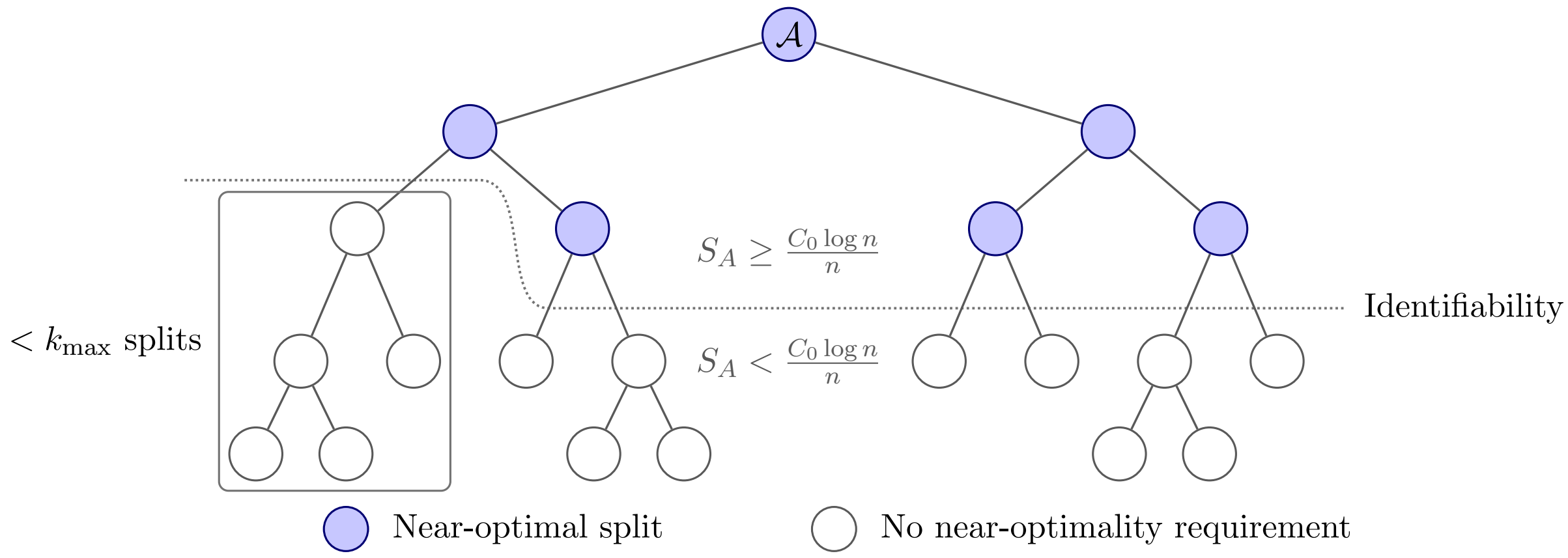
\begin{figure}
  \centering
  \vspace{5mm}
  \resizebox{\linewidth}{!}{%
\begin{tikzpicture}[
  x=0.85cm, y=1cm,
  nodepoint/.style={circle,draw=black!65,line width=0.65pt,
    fill=white,minimum size=6mm,inner sep=0pt},
  nearoptimal/.style={nodepoint,fill=blue!22!white,draw=blue!45!black},
  branch/.style={draw=black!65,line width=0.7pt},
  every node/.style={font=\small}
]
  % Shaded nodes are identifiable and split near-optimally.
  \node[nearoptimal] (r)  at (0,0) {$\mathcal A$};
  \node[nearoptimal] (a)  at (-4.25,-1.1) {};
  \node[nearoptimal] (b)  at ( 4.25,-1.1) {};
  \node[nodepoint] (a0) at (-5.75,-2.2) {};
  \node[nearoptimal] (a1) at (-2.75,-2.2) {};
  \node[nearoptimal] (b0) at ( 2.75,-2.2) {};
  \node[nearoptimal] (b1) at ( 5.75,-2.2) {};
  % The first non-identifiable nodes are a0 and l3,...,l8.
  % Nodes l1 and l2 are additional descendants of a0.
  \node[nodepoint] (l1) at (-6.5,-3.7) {};
  \node[nodepoint] (l2) at (-5.0,-3.7) {};
  \node[nodepoint] (l3) at (-3.5,-3.7) {};
  \node[nodepoint] (l4) at (-2.0,-3.7) {};
  \node[nodepoint] (l5) at ( 2.0,-3.7) {};
  \node[nodepoint] (l6) at ( 3.5,-3.7) {};
  \node[nodepoint] (l7) at ( 5.0,-3.7) {};
  \node[nodepoint] (l8) at ( 6.5,-3.7) {};
  % The dotted line separates strong signals from weak signals.
  \draw[black!55,line width=0.8pt,densely dotted]
    (-8.05,-1.65) -- (-4.12,-1.65)
    .. controls (-3.43,-1.65) and (-3.85,-3.1) .. (-3.20,-3.1)
    -- (7.4,-3.1);
  \node[anchor=west,inner sep=0pt]
    at (7.65,-3.1) {Identifiability};
  \node[text=black!65] at (0,-2.45) {$S_A\geq \frac{C_0\log n}{n}$};
  \node[text=black!65] at (0,-3.7) {$S_A<\frac{C_0\log n}{n}$};
  \foreach \parent/\child in {
    r/a,r/b,a/a0,a/a1,b/b0,b/b1,
    a0/l1,a0/l2,a1/l3,a1/l4,b0/l5,b0/l6,b1/l7,b1/l8}
    \draw[branch] (\parent) -- (\child);
  % A few additional m-grid splits, with no near-optimality constraint.
  % One split below each of l1, l4, and l7.
  \node[nodepoint] (l10)  at (-7.1,-4.75) {};
  \node[nodepoint] (l11)  at (-5.9,-4.75) {};
  \node[nodepoint] (l40)  at (-2.6,-4.75) {};
  \node[nodepoint] (l41)  at (-1.4,-4.75) {};
  \node[nodepoint] (l70)  at ( 4.4,-4.75) {};
  \node[nodepoint] (l71)  at ( 5.6,-4.75) {};
  \foreach \parent/\child in {
    l1/l10,l1/l11,l4/l40,l4/l41,l7/l70,l7/l71}
    \draw[branch] (\parent) -- (\child);
  % Outline the full subtree rooted at the first non-identifiable node a0.
  \node[draw=black!55, rounded corners=3pt, line width=0.65pt,
        inner sep=3pt, fit=(a0)(l1)(l2)(l10)(l11)]
        (refinement) {};
  \node[anchor=east, inner sep=0pt]
    at ([xshift=-5pt]refinement.west) {$<k_{\max}$ splits};
  % Legend.
  \node[nearoptimal,minimum size=5mm] at (-5.9,-5.6) {};
  \node[anchor=west] at (-5.45,-5.6) {Near-optimal split};
  \node[nodepoint,minimum size=5mm] at (0.6,-5.6) {};
  \node[anchor=west] at (1.05,-5.6) {No near-optimality requirement};
\end{tikzpicture}
}\vspace{3mm}
  \caption{An illustration of an admissible tree, where the boxed subtree contains 2 splits.}
  \label{fig:admissible_tree}
\end{figure}

Figure \ref{fig:admissible_tree} illustrates an admissible tree: all identifiable nodes are split near-optimally, while subsequent non-identifiable subtrees have limited size. This reflects the desired bias-variance tradeoff, where nodes with strong signals are split to reduce approximation bias, and limited splitting on nodes with weak signals controls the estimation variance. In Section~\ref{sec: main_results}, we show that under our
assumptions, GEB selects an admissible tree with high probability and
the resulting PPB posterior thus attains adaptive contraction rates.

\subsection{Assumptions}\label{sec: assumptions}
We now outline the assumptions required for our main results. Throughout, we assume that $p$, $m$ and $\delta$ are fixed.
\begin{assumption}[Bounded density]
\label{as:bounded density}
There exist constants $0<C_l\leq C_u<\infty$ such that $C_l\leq p(x)\leq C_u$ for all $x\in[0,1]^D$.
\end{assumption}
\begin{assumption}[Greedy identifiability]
\label{as:greedy identifiable}
There exists a constant $\kappa\in(0,1]$ such that,
for every tree $T\in\mathcal T^*$ and every node $A\in T$
with $S_A>0$, 
\begin{align*}
    \max_{d,q} s_A(d,q) \geq \kappa S_A, 
\end{align*}
where the maximum is over the candidate $m$-grid splits of $A$.
\end{assumption}
\begin{assumption}[Signal decay]
\label{as:decay}
There exist constants $C_1>0$ and $\gamma\in(1,\infty]$
such that, for every tree $T\in\mathcal T^*$ and every
node $A\in T$ with $S_A>0$,
\begin{align*}
    S_A \leq \left(C_1P_A\right)^\gamma. 
\end{align*}
For $\gamma=\infty$, we interpret the condition as
$P_A\geq C_1^{-1}$ for every such node.
\end{assumption}

Assumption \ref{as:bounded density} ensures node probabilities are comparable to volumes, i.e. $C_l H_A\leq P_A\leq C_uH_A$. As Assumption \ref{as:bounded density} is standard, we will not discuss it further. Assumption \ref{as:greedy identifiable} addresses the greedy
nature of the algorithm, while Assumption \ref{as:decay}
controls approximation difficulty.

Assumption \ref{as:greedy identifiable} requires that a
single candidate split can capture a fixed fraction of
the signal at each node under consideration. A similar `sufficient impurity decrease' assumption is made in the random forest literature \citep{chi2022asymptotic,mazumder2023convergence}. We see from Lemma \ref{lem:split_gain} that if Assumption \ref{as:greedy identifiable} holds, a near-optimal split would give 
\begin{align*}
    S_{A_0} + S_{A_1} \leq \left[1-(1-\delta)\kappa\right] S_A, 
\end{align*}
so repeated near-optimal splitting of nodes with $S_A > 0$ can drive the signals, and thus the KL divergence from (\ref{eq:KL_signal}), to 0. To illustrate how Assumption \ref{as:greedy identifiable} can fail, consider a non-uniform univariate density symmetric
about $x=0.5$. When $m=2$, the only candidate split of $\mathcal A$ is at $q=0.5$, for which $P_{A_0\mid\mathcal A}=H_{A_0\mid\mathcal A}=1/2$. Consequently, $s_{\mathcal A}(1,0.5)=0$ despite
$S_{\mathcal A}>0$. Fortunately, under additional regularity conditions, a sufficiently fine grid can ensure greedy identifiability, as we illustrate shortly in the examples.

 For $1 < \gamma < \infty$, Assumption \ref{as:decay} bounds the signal by $(P_A)^\gamma$ up to a multiplicative
constant. For $0<P_A<1$, a larger exponent $\gamma>1$ gives
an upper bound on $S_A$ that decays faster as $P_A$
approaches zero. Since the signals sum to
the KL approximation error in (\ref{eq:KL_signal}), a larger $\gamma$ thus intuitively corresponds to faster decay of approximation
bias as $\mathcal{P}$ is refined, indicating an easier
approximation problem.
As we will see shortly, for H\"older continuous densities
with exponent $\beta\in(0,1]$, we can take $\gamma=1+2\beta$,
directly linking smoothness to signal decay. The endpoint
$\gamma=\infty$ includes suitably aligned piecewise-constant
densities for which a finite partition achieves zero
approximation error. Assumption \ref{as:decay}
is similar in flavor to the global approximation decay
rate of \citet[Equation 6]{liu2023convergence}.

Next, we illustrate examples of verifying Assumptions \ref{as:greedy identifiable} and \ref{as:decay} in the  $D=1$ case, with derivations in Appendix \ref{app:univ}. We defer the more nuanced multivariate case to Section~\ref{sec:check_assumptions}. 

\subsubsection{H\"{o}lder continuous with $0 < \beta \leq 1$}\label{sec:holder}
In the univariate case, we have the useful sufficient condition for Assumption \ref{as:decay}.
\begin{proposition}\label{prop:holder_1d}
Suppose $D=1$, Assumption~\ref{as:bounded density} holds and $H = \mathrm{Unif}[0,1]$.
If $p$ is $\beta$-H\"older continuous for some
$\beta\in(0,1]$, then Assumption~\ref{as:decay} holds
with $\gamma=1+2\beta$.
\end{proposition}
In the univariate case in particular, we can verify the signal decay condition over all intervals $A \subseteq [0,1]$, so we need not consider the class of near-optimal trees $\mathcal{T}^*$. However, in the multivariate case, the near-optimal tree plays a role, as we will see in Section \ref{sec:check_assumptions}.

While Assumption \ref{as:decay} essentially holds for free in the univariate H\"{o}lder continuous case (given Assumption \ref{as:bounded density}), Assumption \ref{as:greedy identifiable} does not. Below, we provide a sufficient condition which can be checked more easily than Assumption \ref{as:greedy identifiable}, based on the \textit{locally reverse Poincar\'{e}} inequality (LRP) condition introduced by \citet{mazumder2023convergence} in the regression context. 
\begin{proposition}\label{prop:var_1d}
Let $D=1$, $H=\mathrm{Unif}[0,1]$, and $p$
be absolutely continuous on $[0,1]$. Suppose there exists
$C_2>0$ such that, for every interval $A\subseteq[0,1]$
with $H_A>0$,
\begin{align*}
\left(\int_A |p^{\prime}(x)|\,dx\right)^2
\le
\frac{C_2}{H_A}
\int_A
\left(p(x)-\frac{P_A}{H_A}\right)^2dx.
\end{align*}
Then Assumption~\ref{as:greedy identifiable} holds
if $m\geq\max\{2,\lceil C_2\rceil\}$. 
\end{proposition}
In the above, $P_A/H_A$ is the mean of the density $p(X)$ under $X \sim \mathrm{Unif}(A)$, so the right-hand side is a scaled variance term. 
The LRP condition thus controls the total variation of $p$ relative to its variance under the uniform distribution, thereby limiting excessive oscillations or highly localized changes.
While the above is still technical, we can verify the conditions of Proposition~\ref{prop:var_1d} for a few examples that are H\"{o}lder continuous for $\beta \in (0,1]$.  

\begin{example}[Monotonic]
Let $p$ be a monotone, absolutely continuous density satisfying $0<c\le |p^{\prime}(x)|\le C<\infty$ almost everywhere. Then the LRP condition of Proposition~\ref{prop:var_1d} holds. If Assumption~\ref{as:bounded density} holds, then Assumption~\ref{as:decay} holds with $\gamma=3$ since $p$ is Lipschitz. The LRP condition for this class was also verified in \citet[Example 3.1]{mazumder2023convergence}.
\end{example}

\begin{example}[Densities with a cusp]
Consider $p(x) \propto 1 + c|x -x_0|^{\beta}$ for $c \in  (-1,\infty)\setminus\{0\}$, $x_0 \in [0,1]$ and $\beta \in (0,1]$. Clearly $p$ is bounded above and away from 0, is absolutely continuous and has H\"{o}lder exponent $\beta$. It thus satisfies Assumptions \ref{as:bounded density} and \ref{as:decay} with $\gamma = 1 + 2\beta$, and one can further show that $p$ also satisfies the LRP condition of Proposition~\ref{prop:var_1d}.
\end{example}

\subsubsection{{$C^{k}[0,1]$} with $k \geq 1$}\label{ex:ck}
We now consider the smoother case $p\in C^k[0,1]$ with $k\ge1$, which is always Lipschitz.
Under Assumption~\ref{as:bounded density}, applying Proposition~\ref{prop:holder_1d} gives Assumption~\ref{as:decay} with decay exponent $\gamma=3$. Fortunately in this setting, we can also establish Assumption \ref{as:greedy identifiable} using an interpretable non-vanishing derivative condition which implies the above LRP condition.
\begin{proposition}\label{prop:ck}
Let $D=1$, $H=\mathrm{Unif}[0,1]$, and $p\in C^k[0,1]$ for some integer $k\geq 1$. If additionally for every $x \in[0,1]$, we have 
\begin{align*}
    \max_{1\leq j\leq k}
    \left|p^{(j)}(x)\right|>0, 
\end{align*}
then Assumption~\ref{as:greedy identifiable} holds
for sufficiently large $m$, where the threshold depends on $p$ and $k$. 
\end{proposition}
For $k=1$, the condition implies that $p$ is strictly monotone, with derivative either strictly positive or negative. For larger $k$, stationary points are permitted provided they are not flat through order $k$, where all
derivatives of orders $1,\ldots,k$ vanish. It nevertheless covers a broad class of densities, including $C^2$ strongly convex densities and multimodal densities. 
\begin{example}[Polynomial]
Let $p(x)$ be a polynomial density of degree $r \geq 1$ on $[0,1]$. Since $p^{(r)}(x) = r!  \,a_r \neq 0$, where $a_r$ is the leading coefficient, Proposition~\ref{prop:ck} holds with $k = r$.  Polynomials were also shown to satisfy LRP in \citet[Example 3.3]{mazumder2023convergence}.
\end{example}
\begin{example}[Sinusoidal]
Let $p(x) = 1 + a \sin(2\pi x)$ for $0 < |a| < 1$.  We can then apply Proposition~\ref{prop:ck} with $k = 2$  since $\max\{|\cos(2\pi x)|, |\sin(2\pi x)|\} > 0$ for all $x \in[0,1]$. 
\end{example}
More generally, every non-constant density that is real analytic on a neighborhood of $[0,1]$ satisfies the condition of Proposition~\ref{prop:ck} for some finite $k$; this covers many examples including (truncated) Gaussian mixtures and exponentials of non-constant polynomials. Finally, while Assumption \ref{as:greedy identifiable} is much simpler to verify in the $C^k[0,1]$ setting, higher smoothness alone does not improve the $\gamma=3$ exponent, as we are ultimately still limited by the piecewise-constant approximation (e.g. see \citet{Castillo}).

\subsubsection{Step density}
Finally, a step density example illustrates the role of choosing $\delta$ appropriately and restricting the assumptions to near-optimal trees, which allows for $\gamma = \infty$. 
\begin{example}[Step density]\label{ex:step}
Let $m \geq 2$, and consider the univariate step density given by
$p(x)=p_1\,\mathbbm{1}\left(x\in\left[0,\frac{1}{m}\right]\right)
+p_2\,\mathbbm{1}\left(x\in\left(\frac{1}{m},1\right]\right)$,
where $p_1, p_2 > 0$, $p_1\ne p_2$, and $p_1 + p_2(m-1) = m$ so $p$ is a density. In this case, one can show that choosing $\delta=1/m$ ensures that any split
of the root in a near-optimal tree must occur at $q=1/m$.
All descendants then lie in constant-density regions,
so the root is the only node with $S_A>0$.
Consequently, Assumptions~\ref{as:greedy identifiable}
and~\ref{as:decay} hold with $\gamma=\infty$. Unlike in the univariate smooth examples,
Assumption~\ref{as:decay} cannot hold uniformly over all intervals here, illustrating the role of the near-optimal tree restriction.
\end{example}

\subsection{Main results}\label{sec: main_results}
We are now ready to state our main result, which guarantees that the GEB algorithm selects a member of the admissible tree family $\mathcal{T}_n$ with high probability.
\begin{theorem}[Admissibility of the GEB tree]\label{theorem:geb_tree_bound}
Fix $m\geq2$ and $\alpha>0$, and suppose Assumptions~\ref{as:bounded density},
\ref{as:greedy identifiable}, and~\ref{as:decay} hold. Let $T_n$ denote the GEB tree obtained from Algorithm \ref{alg:geb} with $m$-grid splitting as in (\ref{eq:m_grid}). Then we have $\mathbb P(T_n\in\mathcal T_n)\longrightarrow1$ as $n \to \infty$.
\end{theorem}
\begin{proofoutline}
We control the error in the log split Bayes factor expansion
\eqref{eq:splitBF_decomp} and bound the complexity of the two parts
of an admissible tree shown in Figure~\ref{fig:admissible_tree}.
Locally, we bound the probabilities of failing to split an identifiable
node near-optimally and of making at least $k_{\max}$ splits below a first non-identifiable node. A union bound over the possible nodes then turns these local bounds into the global admissibility guarantee
above.
\end{proofoutline}

We highlight that the Bayesian penalty in the split Bayes factor is what discourages excess splitting once the remaining signal is weak, so the resulting tree remains within the admissible tree class with high probability. A technical point is that it may be possible to relax Assumption \ref{as:decay} for the above, although it greatly simplifies the proof and is necessary for the next result. We will now see that the admissible tree class allows for adaptive posterior contraction rates to be achieved by the PPB given the GEB partition. 
\begin{theorem}[Posterior contraction]
\label{th:posterior_contraction}
Fix $m\geq2$ and $\alpha>0$, and suppose Assumptions~\ref{as:bounded density},
\ref{as:greedy identifiable}, and~\ref{as:decay} hold. Let $f$ be a density drawn from the PPB posterior $\Pi(\cdot\mid X_{1:n})$ based on the leaf partition of the GEB tree $T_n$ from Theorem~\ref{theorem:geb_tree_bound}. For a sufficiently large constant $M > 0$, 
$$
 \Pi\left\{
 f:\|p-f\|_{L_1}\geq
 M\left(\frac{\log n}{n}\right)^{(1-1/\gamma)/2}
 \mid X_{1:n}
 \right\}\to0 \qquad \text{in } \mathbb P\text{-probability}.
$$
\end{theorem}
\begin{proofoutline}
We leverage the two requirements of admissibility to control the bias--variance tradeoff. The identifiability threshold limits the signal left in each leaf and hence the approximation bias, while $k_{\max}$ limits the extra splitting and thus the number of additional leaf probabilities to estimate, helping control the sampling error.  Together these bounds give the stated rate for the posterior mean in
\eqref{eq: posterior predictive}. Finally, we use the Dirichlet posterior variances of the leaf probabilities
to show that the posterior concentrates around the mean.
\end{proofoutline}
We see the bias-variance tradeoff of the admissible tree class gives an adaptive posterior contraction rate of $n^{-(\gamma-1)/(2\gamma)}$ up to logarithmic factors, where a larger exponent $\gamma > 1$ improves the rate. For example, with $\gamma = 1 + 2\beta$ in the H\"{o}lder case in Section \ref{sec:holder}, we have a posterior contraction rate of $n^{-\beta/(1+2\beta)}$, which agrees with the minimax rate for H\"{o}lder-smooth densities with $\beta \in (0,1]$. We note that this may not be the minimax rate for the class of H\"{o}lder-smooth densities which satisfy Assumption \ref{as:greedy identifiable}, as this restricted class is smaller.
Likewise, the step function case in Example \ref{ex:step} gives a parametric rate of $n^{-1/2}$. In the next section, we demonstrate multivariate examples of adaptation.

\subsection{Verifying assumptions for $D > 1$}\label{sec:check_assumptions}
As alluded to earlier, Assumptions \ref{as:greedy identifiable} and \ref{as:decay} are less straightforward to verify in the multivariate case due to pathologies that can occur even when $D = 2$. For Assumption \ref{as:greedy identifiable}, interactions between the coordinates can cancel out marginal trends, preventing detection via axis-aligned splits. As an example, if $p$ is a bivariate copula density, which has uniform marginals, and $H = \mathrm{Unif}[0,1]^2$, the split gain $s_\mathcal{A}(d,q)$ is equal to 0 for all root splits even if $p \neq h$.  A similar counterexample for random forests in regression is considered in \citet[Remark 3]{chi2022asymptotic}. For Assumption~\ref{as:decay}, the challenge is that an elongated rectangle can have small $P_A$ while retaining large variation along an informative coordinate. Thus, unlike the univariate case, smoothness alone does not ensure the required signal decay over all rectangles, and restricting to near-optimal nodes when verifying Assumption \ref{as:decay} is needed. 

Despite the above challenges, we are able to provide general, verifiable sufficient conditions for Assumptions \ref{as:greedy identifiable} and \ref{as:decay}, albeit with slightly less generality than in the $D = 1$ case. In particular, we can rely on a multivariate version of the LRP condition (from Proposition \ref{prop:var_1d}), and an additional \textit{marginal visibility} condition unique to the multivariate case. The latter condition ensures that averaging over other coordinates does not hide trends, which combats examples like the bivariate copula density. 
In the interest of space, we defer a detailed discussion of the general assumptions to Appendix \ref{app:mv},  
and instead demonstrate the wide range of examples covered by our conditions below.

In all the below, we assume $D \geq 2$, $H = \mathrm{Unif}[0,1]^D$, and Assumption \ref{as:bounded density} holds. Some of the below examples, such as the additive, monotonic and convex examples, are similar to those considered in \citet{chi2022asymptotic,mazumder2023convergence} for regression.

\begin{example}[Product and additive densities]\label{ex:add}
Suppose the density has either an additive structure, $p(x) \propto c + \sum_{d = 1}^D g_d(x^{(d)})$, or a product structure, $p(x) \propto \prod_{d = 1}^D g_d(x^{(d)})$. If each component $g_d$ is absolutely continuous and satisfies the LRP condition in Proposition~\ref{prop:var_1d}, then Assumption~\ref{as:greedy identifiable} holds for sufficiently large $m$. Furthermore, if each $g_d$ is $\beta_d\,$-H\"{o}lder continuous for $\beta_d \in (0,1]$, then Assumption \ref{as:decay} holds with $\gamma = 1 + 2\beta_0/D$ where $\beta_0 = \left(D^{-1}\sum_{d = 1}^D \beta_d^{-1}\right)^{-1}$. Note that $\beta_0$ governs the minimax rate for the anisotropic H\"older class \citep{shen2013adaptive}. 
\end{example}

\begin{example}[Coordinate-wise monotonicity]
Suppose $p\in C^1([0,1]^D)$ and $|{\partial p(x)}/{\partial x^{(d)}}| > 0$
for every $x\in[0,1]^D$ and every $d=1,\ldots,D$. By continuity, each partial derivative has a fixed sign, so $p$ is strictly monotone in each coordinate with the others fixed.
Then, for sufficiently large $m$, Assumptions~\ref{as:greedy identifiable} and~\ref{as:decay}
hold with $\gamma=1+2/D$.
\end{example}

\begin{example}[Coordinate-wise convexity]\label{ex:convex}
Suppose $p\in C^2([0,1]^D)$ and
$\left|\partial^2 p(x)/\partial(x^{(d)})^2\right|>0$
for every $x\in[0,1]^D$ and every $d=1,\ldots,D$. Then, for sufficiently large $m$, Assumptions~\ref{as:greedy identifiable} and~\ref{as:decay}
hold with $\gamma=1+2/D$. Note that this allows for non-additive examples, and contains strong convexity as a special case. 
\end{example}

\begin{example}[Variable sparsity]\label{ex:sparse}
Suppose $p(x)=g(x^{(1:D_0)})$ for some $1\leq D_0<D$, so the remaining coordinates are independent uniform noise. If $g$ satisfies Assumptions~\ref{as:bounded density},
\ref{as:greedy identifiable}, and~\ref{as:decay} in $D_0$ dimensions, then $p$ satisfies the assumptions with the same $m$, $\kappa$ and $\gamma$. 
\end{example}

In Appendix \ref{app:copula}, we include an illustrative $D=2$ quadratic example showing how interactions affect marginal visibility, and how the copula pathology can be alleviated with a coordinate transformation.
Importantly, Examples~\ref{ex:add} and~\ref{ex:sparse} illustrate multivariate adaptation of the PPB posterior based on the GEB partition to anisotropic smoothness and variable sparsity. These features need not be specified to the algorithm, provided the stated assumptions hold for the chosen $m$.
In fact, the anisotropic decay exponent extends beyond additive and product densities under appropriate assumptions, as we show in Corollary~\ref{corr:mv_decay_aniso} in the Appendix.
In the variable sparsity case, if $g$ is Lipschitz and satisfies additional assumptions, the decay exponent improves from $\gamma=1+2/D$ to $\gamma=1+2/D_0$, yielding a faster contraction rate. We thus see that the GEB partition enables adaptation even in challenging multivariate settings.

\section{Experiments}
\label{sec: experiments}

Our experiments consider three separate structural regimes in the spirit of the examples in Sections~\ref{sec: assumptions} and \ref{sec:check_assumptions}, namely densities with axis-aligned discontinuities, smooth densities, and low-dimensional densities embedded in noise. The simulations are intentionally designed to exhibit these features, while the real datasets provide corresponding examples of the respective features in practice. For the density estimation experiments, we compare our approach to three baselines over 20 replicates, reporting the average test log score,
$n_{\mathrm{test}}^{-1}\sum_{i=1}^{n_{\mathrm{test}}}\log\hat p(x_i^{\mathrm{test}})$,
where larger values indicate better density estimation. A final Laplace distribution simulation illustrates uncertainty quantification for the integrated squared density as the functional of interest with the PPB. All experiments are run on the CPU of a Windows 11 machine with an AMD Ryzen 7 H 260 processor and 32GB of RAM.

\subsection{Candidate models}

We compare the following estimators: (1) \textbf{GEB} (ours), with shrinkage
\(\alpha=1\) and grid size \(m=32\); (2) \textbf{HMPT}
\citep{LM_Hidden}, a Bayesian nonparametric tree model that accommodates both flexible partitioning and adaptive local smoothness, using SMC for inference; (3) \textbf{PT}, a conventional P\'olya tree with midpoint splits; and (4)
\textbf{DPMM}, a Dirichlet process mixture of Gaussian location--scale
kernels. For the DPMM, we use diagonal covariance matrices and truncated variational inference for the density estimation experiments, and MCMC for more accurate uncertainty quantification in the later posterior uncertainty simulation. Full hyperparameter settings for all methods are provided in
Appendix~\ref{app:model_hyperparameters}. For all examples, we use the posterior predictive density as the estimate. 

Since the tree methods are defined on \([0,1]^D\), we map each coordinate to
the unit interval using coordinate-wise transforms before fitting
GEB, HMPT, and PT. We consider two transforms, the first of which is
the Gaussian transform
\(T_j^{\mathrm{G}}(x^{(j)})
    = \Phi\bigl((x^{(j)}-\mu^{(j)})/\sigma^{(j)}\bigr)\), where $\Phi$ is the standard Gaussian cumulative distribution function (CDF), and $\mu^{(j)}$ and $\sigma^{(j)}$ are the training sample mean and standard deviation of coordinate $j$. The second transform we consider is
 the mixture transform \(T_j^{\mathrm{M}}(x^{(j)})
    = (1-\epsilon)\,F_{\mathrm{Unif}}\bigl(x^{(j)};  a^{(j)},  b^{(j)}\bigr)
      +\epsilon\,\Phi\bigl((x^{(j)}-\mu^{(j)})/\sigma^{(j)}\bigr)\),
where $ a^{(j)}, b^{(j)}$ are the training sample minimum and maximum of coordinate $j$, $F_{\mathrm{Unif}}(x;a,b)$ denotes the CDF of
$\mathrm{Unif}(a,b)$, and $\epsilon \in (0,1)$ is a small number. We set $\epsilon = 0.001$, and the small Gaussian component in $T_j^{\mathrm M}$ allows the fitted density to assign positive density to future observations outside the training sample range. We use the suffixes -G and -M to indicate the Gaussian and the mixture transforms respectively in the simulation results. For our results, we map the fitted densities back to the untransformed data scale using the Jacobian of the coordinate-wise transformation. 

\subsection{Simulation study}
For our simulation datasets, we let $n = n_{\mathrm test} = 5{,}000$, and consider three data-generating distributions. The first setting is
\(0.9\,\mathrm{Unif}([0.2,0.8]^D) + 0.1\,\mathrm{Unif}([0,1]^D)\) with $D = 5$, which contains axis-aligned discontinuities along the boundary of a high-density central rectangle. The second setting is a smooth $D = 2$ Gaussian mixture with density  \(0.5\,\mathcal N_2(\mu_0,\Sigma_0) + 0.5\,\mathcal N_2(\mu_1,\Sigma_1)\), where \(\mu_0=(-2,-1)^\top\), \(\mu_1=(2,1)^\top\), and
$\Sigma_0=
\begin{psmallmatrix}0.70&0.35\\0.35&0.80\end{psmallmatrix}$ and  $\Sigma_1=
\begin{psmallmatrix}\hphantom{-}0.60&-0.25\\-0.25& \hphantom{-}0.90\end{psmallmatrix}$. Finally, we consider a sparse setting with $D=50$, where the first $D_0=2$ coordinates follow the same Gaussian mixture and the remaining 48 coordinates are independent standard Gaussian noise, independent of the first two coordinates.

Table~\ref{tab:sim_logscore} reports the average test log scores and run-times for the three simulation settings. In the axis-aligned setting, GEB-M achieves the highest score, followed by HMPT-M. The mixture transform largely preserves the sharp density changes at the boundary of the central rectangle, yielding a setting similar to Example~\ref{ex:step}, which can be captured by a small number of informative splits. For the two-dimensional Gaussian mixture, DPMM achieves the highest score, while the tree methods remain competitive. The Gaussian transform provides modest improvements for GEB and HMPT relative to the mixture transform. In the sparse 50-dimensional setting, the Gaussian transform makes the independent noise coordinates approximately uniform, yielding a setting similar to Example~\ref{ex:sparse}. GEB-G achieves the highest score, closely followed by HMPT-G. All three tree methods perform substantially better with the Gaussian transform than with the mixture transform, consistent with the benefit of adapting to active coordinates.

Across the three settings, GEB combines competitive or superior predictive accuracy with substantially shorter run-times than both HMPT and DPMM, and all GEB configurations take less than one second on average to run. While the PT is faster still, it generally achieves lower test log scores. As an aside, DPMM’s shorter run-time in the sparse setting reflects faster variational convergence to a simpler mixture when the noise coordinates are included.

\begin{table}[!htb]
\centering
\small
\setlength{\tabcolsep}{3pt}
\captionsetup{font=small}
\caption{Simulation average test log scores and wall-clock run-times.
Standard errors are given in
parentheses. Best results are in bold. \vspace{2mm}}
\label{tab:sim_logscore}
\label{tab:sim_time}
\begin{tabular}{@{}l@{\quad} r@{\ }l r@{\ }l r@{\ }l@{\hspace{1.5em}} c c c@{}}
\toprule
& \multicolumn{6}{c@{\hspace{1.5em}}}{Test log score}
& \multicolumn{3}{c}{Run-time (seconds)} \\
\cmidrule(lr){2-7}\cmidrule{8-10}
Method
& \multicolumn{2}{c}{Axis-aligned}
& \multicolumn{2}{c}{Smooth}
& \multicolumn{2}{c@{\hspace{1.5em}}}{Sparse}
& Axis-aligned & Smooth & Sparse \\
\midrule
GEB-M
& \textbf{1.96} & \textbf{(0.005)}
& -3.21 & (0.004)
& -92.4 & (0.04)
& 0.06 & 0.05 & 0.83 \\
HMPT-M
& 1.91 & (0.005)
& -3.20 & (0.004)
& -96.5 & (0.07)
& 67 & 189 & 257 \\
PT-M
& 0.90 & (0.005)
& -3.21 & (0.004)
& -100.2 & (0.08)
& \textbf{0.02} & \textbf{0.02} & \textbf{0.03} \\
\midrule
GEB-G
& 1.59 & (0.007)
& -3.19 & (0.003)
& \textbf{-71.3} & \textbf{(0.02)}
& 0.11 & 0.05 & 0.12 \\
HMPT-G
& 1.40 & (0.007)
& -3.18 & (0.003)
& -71.4 & (0.02)
& 274 & 324 & 795 \\
PT-G
& 1.21 & (0.005)
& -3.21 & (0.003)
& -72.0 & (0.02)
& \textbf{0.02} & \textbf{0.02} & \textbf{0.03} \\
\midrule
DPMM
& 1.50 & (0.006)
& \textbf{-3.14} & \textbf{(0.003)}
& -71.6 & (0.02)
& 46 & 34 & 1.57 \\
\bottomrule
\end{tabular}
\end{table}

\subsection{Real-data experiments}

We next apply the methods to the POWER, Dry Bean, and ISOLET datasets from the UCI Machine Learning Repository \citep{uci_repository}, chosen to reflect the three structural regimes considered above. After preprocessing, POWER contains \(2{,}049{,}280\) total observations with \(D=6\). In each replication, we subsample \(100{,}000\) observations without replacement and split them equally into training and test sets. For Dry Bean ($D = 16$) and ISOLET ($D = 100$), we use all \(13{,}611\) and \(7{,}797\) observations respectively, dividing each dataset into approximately equal training and test sets.

\begin{figure}[!th]
  \centering 
  \includegraphics[width=\textwidth]{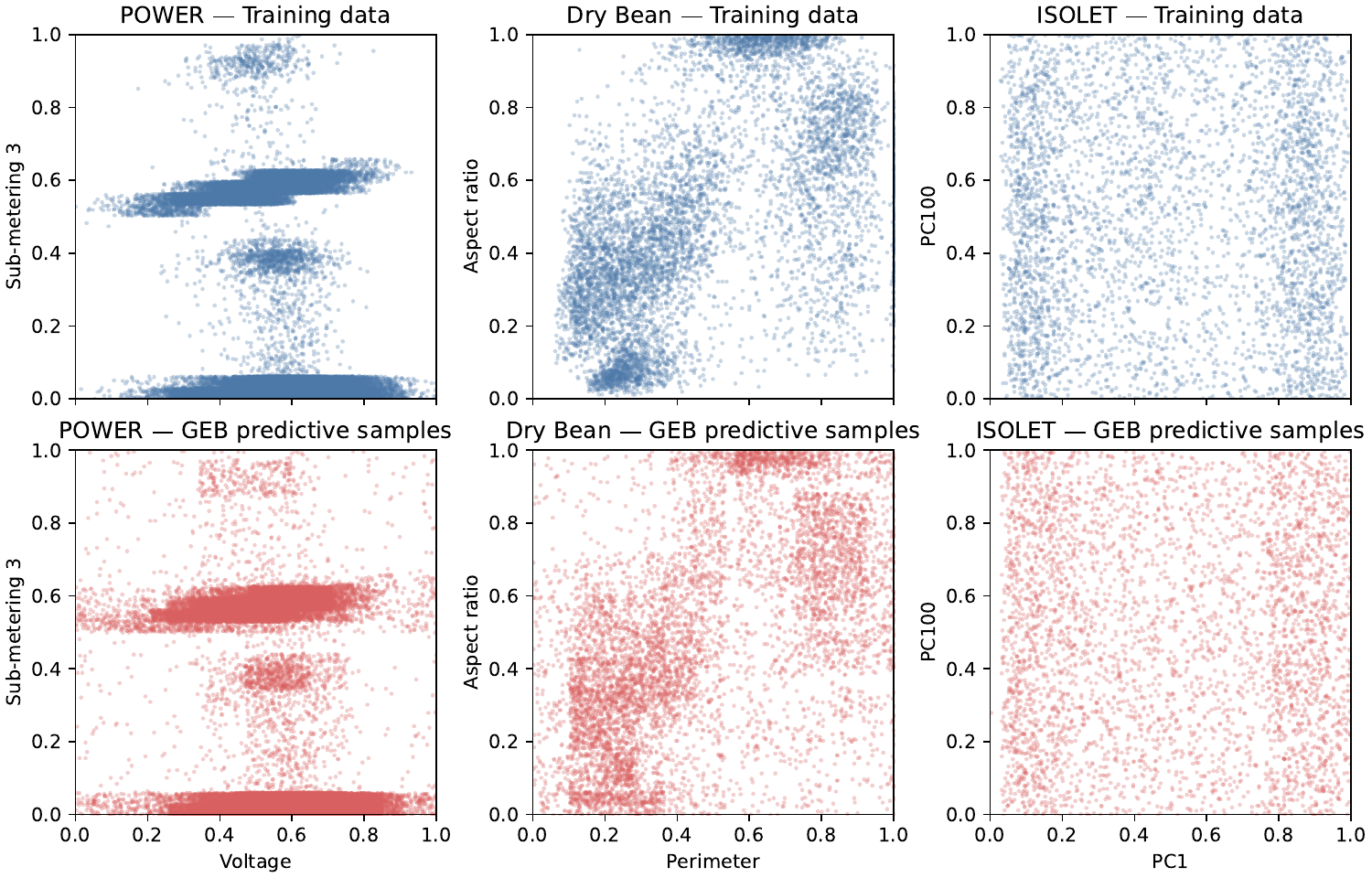}
\caption{Bivariate scatterplots of two representative coordinates
from POWER, Dry Bean, and ISOLET. The top row shows the training
data, and the bottom row shows samples of the same size from
the PPB posterior predictive density based on the GEB
partition. POWER is shown on the mixture-transformed scale,
while Dry Bean and ISOLET are shown on the Gaussian-transformed scale.}
  \label{fig:real-data-structure}
\end{figure}

Figure~\ref{fig:real-data-structure} (top) illustrates parallels between the real datasets and the three simulation settings using two representative coordinates from each dataset. In POWER, ``Voltage'' and ``Sub-metering~3'' form pronounced horizontal bands, which likely can be captured efficiently by axis-aligned splits. In Dry Bean, ``Perimeter'' and ``Aspect ratio'' show several overlapping clusters, consistent with the mixture structure suggested by the seven bean varieties. In ISOLET, the Gaussian-transformed observations show pronounced variation in density along PC1 but are spread approximately uniformly along PC100. This contrast is reminiscent of the sparse simulation setting, where some coordinates contain little structure. 

For the tree-based methods, we use the mixture transform for POWER, which preserves its banded structure, and the Gaussian transform for Dry Bean and ISOLET. In particular, the Gaussian transform maps approximately Gaussian coordinates to approximately uniform ones, allowing GEB to leave these coordinates largely unsplit. Before applying these transforms, POWER is preprocessed following  \citet{NIPS2017_6c1da886}, while ISOLET is reduced to 100 dimensions using PCA fitted to the standardized training data. Further preprocessing and evaluation details are provided in Appendix~\ref{app:preprocessing}.

\begin{table}[!htbp]
\centering
\small
\setlength{\tabcolsep}{3pt}
\captionsetup{font=small}
\caption{Real-data average test log scores and wall-clock run-times.
Standard errors are given in
parentheses. Best results are in bold. \vspace{2mm}}
\label{tab:real_logscore}
\begin{tabular}{@{}l@{\quad} r@{\ }l r@{\ }l r@{\ }l@{\hspace{1.5em}} c c c@{}}
\toprule
& \multicolumn{6}{c@{\hspace{1.5em}}}{Test log score}
& \multicolumn{3}{c}{Run-time (seconds)} \\
\cmidrule(lr){2-7}\cmidrule{8-10}
Method
& \multicolumn{2}{c}{POWER}
& \multicolumn{2}{c}{Dry Bean}
& \multicolumn{2}{c@{\hspace{1.5em}}}{ISOLET}
& POWER & Dry Bean & ISOLET \\
\midrule
GEB
& \textbf{0.051} & \textbf{(0.004)}
& -6.30 & (0.05)
& \textbf{-180.78} & \textbf{(0.11)}
& 0.65 & 0.62 & 0.64 \\
HMPT
& -0.256 & (0.016)
& -14.18 & (0.06)
& -181.25 & (0.11)
& 805 & 91 & 877 \\
PT
& -1.660 & (0.011)
& -13.50 & (0.02)
& -181.95 & (0.11)
& \textbf{0.31} & \textbf{0.04} & \textbf{0.05} \\
DPMM
& -0.307 & (0.003)
& \textbf{-2.71} & \textbf{(0.02)}
& -182.60 & (0.11)
& 200 & 22 & 1.41 \\
\bottomrule
\end{tabular}
\end{table}
Figure~\ref{fig:real-data-structure} (bottom) shows that samples from the PPB posterior predictive density closely match the main visual patterns in the training data, and the rectangular structure induced by GEB's axis-aligned partition is also visible in the generated samples. This visual agreement is accompanied by competitive performance on held-out data, as shown by the average test log scores in Table~\ref{tab:real_logscore}. On POWER, GEB achieves the highest score, consistent with its strong performance in the axis-aligned simulation setting. On Dry Bean, DPMM achieves the highest score, while GEB is the strongest tree-based method, substantially outperforming both HMPT and PT. On the 100-dimensional ISOLET dataset, GEB again achieves the highest score, followed by HMPT, extending its favorable predictive performance to the high-dimensional setting. Across all three datasets, GEB combines this predictive accuracy with run-times of less than one second on average, making it substantially faster than HMPT and DPMM. PT is once again the fastest method, but achieves noticeably lower test log scores than GEB on every dataset. As in the sparse simulation, fast variational convergence to a simple mixture explains DPMM’s short run-time on ISOLET. 

\subsection{Posterior uncertainty simulation}
After obtaining a partition with the GEB procedure, we can carry out posterior inference using the PPB. To assess the quality of its uncertainty quantification, we carry out a final simulation experiment under a $D = 1$ standard Laplace density $p(x)=\frac{1}{2}\exp(-|x|)$, with the functional of interest being the integrated squared density $\theta(f)=\int_{\mathbb{R}}f(x)^2\,dx$.
The true parameter value can be shown to be $\theta(p)=1/4$.

We run \(1{,}000\) repeats, each with \(n=5{,}000\)
training observations and \(1{,}000\) retained posterior draws.
The PPB samples of \(\theta\) are obtained using
Algorithm~\ref{alg: ppb}, whereas posterior density draws for the DPMM are generated
using the MCMC implementation in the \texttt{BNPmix} package. For both methods, we construct \(95\%\) central credible intervals for \(\theta\) from the posterior samples, and we apply the Gaussian transform to map to $[0,1]$ for PPB.

\begin{figure}[htb]
    \centering
    \includegraphics[width=0.55\linewidth]{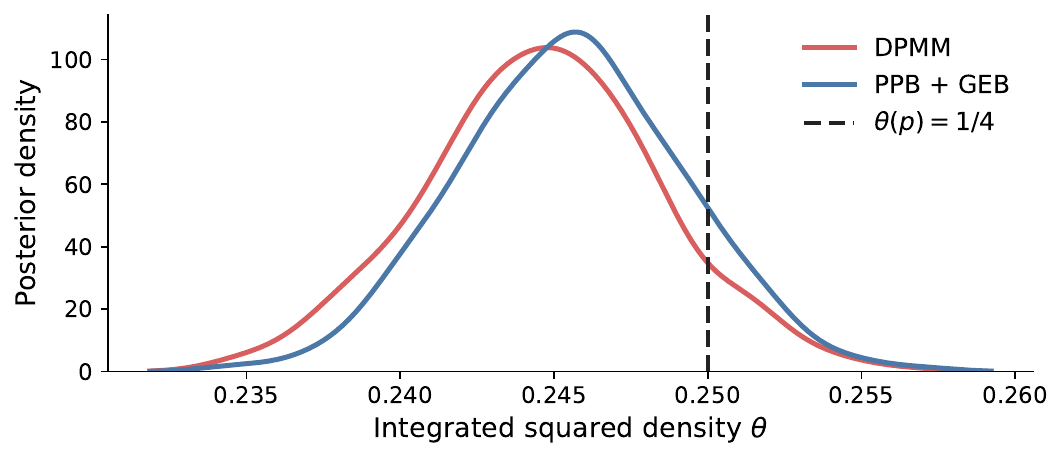}
    \caption{Posterior distribution of the integrated squared density under PPB + GEB and
    the DPMM for one simulated dataset.
    The dashed line indicates the true \(\theta(p)=1/4\).}  
    \label{fig:posterior-isd}
\end{figure}

Table~\ref{tab:posterior_uncertainty} shows that PPB achieves similar coverage of $\theta$ with similar interval lengths, and noticeably smaller bias of the posterior mean estimate for $\theta$.  The downward bias observed for both models is consistent with the discussion in \citet{semiparametric_correction}. The reported PPB run-time, which includes GEB partition selection, is substantially lower than that of DPMM, reflecting the efficiency of posterior sampling without MCMC. Finally, Figure~\ref{fig:posterior-isd} visually compares the two posterior distributions for one dataset, illustrating their similar spread.

\begin{table}[!htb]
\centering
\small
\setlength{\tabcolsep}{6pt}
\captionsetup{font=small}
\caption{Coverage of \(95\%\) credible intervals,
mean credible interval length, bias of the posterior mean, and run-times
over \(1{,}000\) replications. Standard errors are given in
parentheses. \vspace{2mm} }
\label{tab:posterior_uncertainty}
\begin{tabular}{@{}lrrrc@{}}
\toprule
Method & Coverage (\%) & Length ($\times10^{-2}$)
& Bias ($\times10^{-3}$) & Run-time (seconds) \\
\midrule
PPB + GEB
& 92.6 (0.8)
& 1.556 (0.002)
& $-0.071$ (0.136)
& 0.011 \\
DPMM
& 93.1 (0.8)
& 1.557 (0.002)
& $-1.056$ (0.129)
& 6.789 \\
\bottomrule
\end{tabular}
\end{table}

\section{Discussion}
\label{sec: discussion}
In this paper, we have introduced the partition posterior bootstrap (PPB) based on a smoothed Dirichlet process, which avoids the need for expensive MCMC and SMC for posterior computation. We introduce a greedy empirical Bayes (GEB) algorithm for selecting the partition, which scales very well to large samples and high-dimensional data. Empirically, GEB is orders of magnitude faster than existing Bayesian nonparametric alternatives, and remains competitive in prediction accuracy across challenging data regimes with minimal tuning. We thus view the PPB with the GEB partition as a practical off-the-shelf choice in settings where scalability and automatic tuning are paramount. Importantly, we are able to provide theoretical guarantees for the actual GEB algorithm and its associated PPB, demonstrating adaptive posterior contraction rates in challenging multivariate settings like anisotropic smoothness and variable sparsity, which may help explain its excellent practical performance.

A natural direction for future work is to extend the associated greedy empirical Bayes principle beyond density estimation. The interpretable local Bayes factors exploited here arise from the Dirichlet aggregation property, which may not be specific to density trees. Other models admitting a tractable tail-free factorization could benefit from the same structure, yielding scalable GEB procedures. Using GEB trees as part of a larger algorithm, such as boosting, may also form an interesting direction of future research.

\newpage
\section*{Acknowledgments}
Edwin Fong receives funding from the Research Grants Council of Hong Kong through the
Early Career Scheme (Grant 27304424) and the General Research Fund (Grant 17306925). During the preparation of this manuscript, ChatGPT Astra was used to assist with coding, proofreading, and some of the technical arguments for verifying the assumptions in the examples. The authors have reviewed, verified and take full responsibility for all content. All datasets used in this paper are publicly available. Code for reproducing the results can be found at \url{https://github.com/stlllll/GEB_density}.

\begingroup
\setstretch{1.0}
\bibliographystyle{abbrvnat}
\bibliography{paper-ref}
\endgroup

\newpage
\begin{appendices}
\counterwithin{assumption}{section}
\counterwithin{lemma}{section}
\counterwithin{theorem}{section}
\counterwithin{proposition}{section}
\counterwithin{corollary}{section}
\counterwithin{definition}{section}
\counterwithin{example}{section}
\setcounter{equation}{0}
\numberwithin{equation}{section}
\numberwithin{figure}{section}
\numberwithin{table}{section}

\section{Preliminaries and notation}
To handle the many constants involved throughout the Appendix, we write \(x \lesssim y\) to mean that there exists a constant \(C\), independent of node choice $A$ and sample size $n$, such that \(x \leq C y\). Likewise, we use \(x \gtrsim y\) for the reverse direction. When both directions hold, for possibly different constants, we write \(x \asymp y\). Unless otherwise stated, $A$ will be a positive-volume rectangle in $[0,1]^D$. With the exception of Section \ref{app:max_proof}, $H$ will be assumed to be $\mathrm{Unif}[0,1]^D$.

We also introduce some useful shorthand for our later proofs. 
For node $A$, we will write
\begin{align}\label{eq:pbar_sigma}
    \bar{p}_A=\frac{P_A}{H_A}, \qquad  \sigma_A^2 = \frac{1}{H_A}\int_A \left[p(x) - \bar{p}_A\right]^2\, dx.
\end{align}
Note that the above correspond to mean and variance of $p(X)$ under $H(\cdot \mid A)$. A useful property that we use repeatedly is the following: under Assumption \ref{as:bounded density}, we have $$P_A \asymp H_A, \qquad \bar{p}_A \asymp 1.$$

\section{Proof of Theorem \ref{theorem: maximum data point}}\label{app:max_proof}
\label{app: split selection}
To prove Theorem~\ref{theorem: maximum data point}, we first show that nodes containing at most two observations stop automatically. For larger nodes, we study the split Bayes factor between consecutive observed coordinates and show that its supremum is attained at, or immediately to the left of, an observed coordinate.

\subsection{Nodes with $N_A \leq 2$}
We first verify that the stopping rule is automatic for nodes
with $N_A \leq 2$. If $N_A\leq1$, then node $A$ is not eligible. Now suppose $N_A=2$. Every eligible split has
$N_{A_0}=N_{A_1}=1$. Using the beta function form of the split Bayes factor,
\begin{align*}
\mathcal B_A(d,q)
&=
\frac{
 \operatorname{B}(\alpha H_{A_0}+1,\alpha H_{A_1}+1)
}{
 \operatorname{B}(\alpha H_{A_0},\alpha H_{A_1})
}
\frac{H_A^2}{H_{A_0}H_{A_1}}\\
&=
\frac{\alpha^2H_{A_0}H_{A_1}}
     {\alpha H_A(\alpha H_A+1)}
\frac{H_A^2}{H_{A_0}H_{A_1}}=\frac{\alpha H_A}{\alpha H_A+1}<1.
\end{align*}
Thus every eligible split has $\mathcal{B}_A(d,q) < 1$, and the stopping rule therefore
retains $A$ as a terminal node. 

\subsection{Derivatives}
Decomposing the beta functions into gamma functions and using
$\Gamma(z+1)=z\Gamma(z)$ for $z>0$, we obtain
\begin{align*}
    \log \mathcal{B}_A(d,q)
    ={}&
    \sum_{j=0}^{N_{A_0}-1}\log(\alpha H_{A_0}+j)
    +
    \sum_{j=0}^{N_{A_1}-1}\log(\alpha H_{A_1}+j)
    \notag\\
    &-
    N_{A_0}\log H_{A_0}
    -
    N_{A_1}\log H_{A_1}
    +
    \log\left(
        \frac{\Gamma(\alpha H_A)}
             {\Gamma(\alpha H_A+N_A)}
        H_A^{N_A}
    \right).
\end{align*}
The last term is independent of $H_{A_0}$. The function above may be discontinuous at observed data coordinates, since crossing such a value changes $N_{A_0}$ and $N_{A_1}$ by one. For each $d$, moving $q$ between consecutive
observed coordinates leaves the child counts unchanged,
while the base probabilities may vary. On each such
interval, we may regard $\log \mathcal{B}_A(d,q)$ as a function of
$H_{A_0}$, keeping the child counts and $H_A$ fixed and
setting $H_{A_1}=H_A-H_{A_0}$. Since $H$ has a density, $H_{A_0}$ is continuous and
nondecreasing in $q$. Differentiating with respect to $H_{A_0}$ gives
\begin{align}\label{eq: log B derivative}
    \frac{\partial \log \mathcal{B}_A}{\partial H_{A_0}}
    =
    \sum_{j=1}^{N_{A_0}-1}
    \frac{-j}{(\alpha H_{A_0}+j)H_{A_0}}
    +
    \sum_{j=1}^{N_{A_1}-1}
    \frac{j}{(\alpha H_{A_1}+j)H_{A_1}}.
\end{align}
Differentiating again yields
\begin{align}\label{eq: log B second derivative}
    \frac{\partial^2 \log \mathcal{B}_A}{\partial H_{A_0}^2}
    ={}&
    \sum_{j=1}^{N_{A_0}-1}
    \frac{j(2\alpha H_{A_0}+j)}
         {(\alpha H_{A_0}+j)^2H_{A_0}^2}
    +
    \sum_{j=1}^{N_{A_1}-1}
    \frac{j(2\alpha H_{A_1}+j)}
         {(\alpha H_{A_1}+j)^2H_{A_1}^2}. 
\end{align}

\subsection{Boundary intervals}
We first handle the boundary cases, where one of the child nodes contains exactly one datum.
\begin{lemma}\label{lemma: boundary monotone}
    Let $N_A\geq3$ and $a<b$, and suppose the child counts remain constant for $q\in[a,b)$.
    Then we have $\sup_{q \in [a, b)} \log \mathcal{B}_A(d,q) = \log \mathcal{B}_A(d, b_{-})$ if $N_{A_0} = 1$,  and $\sup_{q \in [a, b)}\log \mathcal{B}_A(d,q) = \log \mathcal{B}_A(d, a)$ if $N_{A_1} = 1$.
\end{lemma}

\begin{proof}
    For the case $N_{A_0} = 1$, we have $N_{A_1}=N_A-1$. Substituting into
    \eqref{eq: log B derivative}, the first sum is empty and hence
    \[
        \frac{\partial \log \mathcal{B}_A}{\partial H_{A_0}}
        =
        \sum_{j=1}^{N_A-2}
        \frac{j}{(\alpha H_{A_1}+j)H_{A_1}}
        > 0.
    \]
Consequently, $\log\mathcal B_A(d,q)$ is nondecreasing in $q$ on $[a,b)$, and
\[
\sup_{q\in[a,b)}\log\mathcal B_A(d,q)
=
\log\mathcal B_A(d,b_-).
\]
This value is attained by the allowed boundary split
$q=b_-$. If instead $N_{A_1}=1$, then $N_{A_0}=N_A-1$. By symmetry, $\partial\log\mathcal B_A/\partial H_{A_0}<0$, so $\log\mathcal B_A(d,q)$ is nonincreasing in $q$
and its maximum on $[a,b)$ is attained at $a$.  
\end{proof}

\subsection{Interior intervals}
We next consider intervals where both children contain
at least two observations. The result then follows from
convexity in $H_{A_0}$.
\begin{lemma}\label{lemma: local minimum}
Let $N_A\geq4$ and $a<b$, and suppose the child counts remain constant for $q\in[a,b)$. Additionally assume $\min\{N_{A_0}, N_{A_1}\} \geq 2$. We then have 
    \begin{align*}
        \sup_{q \in [a, b)}\log \mathcal{B}_A(d,q) =
        \max\{\log \mathcal{B}_A(d, a),\log \mathcal{B}_A(d,b_{-})\}.
    \end{align*}
\end{lemma}

\begin{proof}
On the interval $[a,b)$, the child counts remain constant.
By \eqref{eq: log B second derivative},
$\log\mathcal B_A$ is strictly convex as a function of
$H_{A_0}$. Since $H_{A_0}$ is continuous and nondecreasing
in $q$, the supremum over $[a,b)$ equals the larger of
the endpoint values
$\log\mathcal B_A(d,a)$ and $\log\mathcal B_A(d,b_-)$.
This proves the claim.
\end{proof}

\subsection{Main proof}
For $N_A\geq3$, the observations almost surely have distinct
values in each coordinate, since $P$ has a Lebesgue density.
For each dimension $d$, consider the intervals between
consecutive observed coordinates within $A$.
Lemma~\ref{lemma: boundary monotone} shows that the sample
minimum and the left-hand value at the sample maximum
can be discarded from the search.
Lemma~\ref{lemma: local minimum} reduces each remaining
interval to its endpoint values, including the left-hand
value at its right endpoint.
Thus, the maximum is attained at an observed coordinate excluding the sample extrema, using one of the two boundary assignments. There are at most $2(N_A-2)$ such choices
per dimension, giving $2D(N_A-2)$ evaluations in total.

\section{Properties of the KL divergence}\label{app: KL_variance}
We present two useful results about the KL divergence which serve as tools for the later main proofs. 
\subsection{Proof of Lemma~\ref{lem:split_gain}}\label{app: KL_monotonicity}
We take $0\log 0=0$ and set $S_A=s_A(d,q)=0$ whenever $P_A=0$. For any node $A$, we have
    \begin{align*}
        S_A &=  \int_A p(x) \log p(x) \, dx - P_{A} \log \bar{p}_A.
    \end{align*}
If $P_A=0$, both children also have zero mass, so the claim follows. For $P_A>0$, since $A_0$ and $A_1$ partition $A$, additivity of the integral and the  definition of the split gain give
\begin{align*}
S_A-s_A(d,q)
&=\int_A p(x)\log p(x)\,dx
  -\sum_{i=0}^1 P_{A_i}\log\bar p_{A_i}\\
  &=\sum_{i=0}^1\left[
    \int_{A_i} p(x)\log p(x)\,dx
    -P_{A_i}\log\bar p_{A_i}
  \right]\\
&=S_{A_0}+S_{A_1}.
\end{align*}

\subsection{KL and variance}\label{app: variance bound KL}
Our next result relates the KL divergence, and thus the signal, to the variance of the density under the base measure. This will be particularly useful when verifying Assumptions \ref{as:greedy identifiable} and \ref{as:decay}. 

\begin{lemma}\label{lemma: variance KL}
Under Assumption \ref{as:bounded density}, we have
$$K\left\{p(\cdot \mid A); \, h(\cdot \mid A)\right\} \asymp \sigma_A^2.$$
\end{lemma}
\begin{proof}
The standard inequalities between the KL divergence, the Hellinger metric and the $\chi^2$-divergence give
\begin{align*}
    \int\left[\sqrt{f(x)} - \sqrt{g(x)}\right]^2\,dx \leq K(f; \, g) \leq \int \frac{[f(x) - g(x)]^2}{g(x)}\, dx.
\end{align*}
Plugging in the conditional densities, the upper bound becomes
\begin{align}\label{eq:chi2_bound}
    \int_A \frac{\left[p(x \mid A) - 1/H_A\right]^2}{1/H_A}\, dx = \frac{\sigma_A^2}{(\bar{p}_A)^2} \lesssim \sigma_A^2.
\end{align}
The lower bound becomes
\begin{align*}
    \int_A \left[\sqrt{p(x \mid A)} - \sqrt{1/H_A}\right]^2\, dx = \frac{1}{P_A} \int_A\left[\sqrt{p(x)} - \sqrt{\bar p_A}\right]^2 \, dx.
\end{align*}
For any $x \in A$, we have
\begin{align*}
    \left[\sqrt{p(x)} - \sqrt{\bar p_A}\right]^2 = \left[\frac{p(x) - \bar{p}_A}{\sqrt{p(x)} + \sqrt{\bar p_A}}\right]^2 \gtrsim [p(x) - \bar{p}_A]^2.
\end{align*}
We thus have
\begin{align*}
    \int_A \left[\sqrt{p(x \mid A)} - \sqrt{1/H_A}\right]^2\, dx \gtrsim \frac{1}{\bar{p}_A} \sigma_A^2 \gtrsim \sigma_A^2.
\end{align*}
This proves the result. 
\end{proof}

\section{Structural assumptions}
In this section, we provide detailed proofs and further examples for the results surrounding Assumptions \ref{as:greedy identifiable} and \ref{as:decay}. As in the main paper, we begin with the univariate case, before generalizing to the multivariate.
\subsection{Univariate case}\label{app:univ}
\subsubsection{Proof of Proposition \ref{prop:holder_1d}}
Let $A\subseteq [0,1]$ be any interval with $H_A > 0$. Using Lemma \ref{lemma: variance KL}, we have
\begin{align*}
    S_A &= P_AK\left\{p(\cdot \mid A);\,  h(\cdot \mid A)  \right\} \lesssim P_A \sigma_A^2.
\end{align*}
Next, for $x \in A$,
\begin{align*}
   \left| p(x) - \bar{p}_A\right| &= \frac{1}{H_A}\left|\int_A\left[p(x) - p(y)\right]\, dy\right|\\
   &\lesssim \frac{1}{H_A} \int_A \left|x - y\right|^\beta dy \notag
\end{align*}
where the second line follows from H\"{o}lder continuity. Since $|x - y| \leq H_A$, the above gives $\left| p(x) - \bar{p}_A\right|  \lesssim H_A^\beta$, which then implies
\begin{align*}
    \sigma_A^2 \lesssim \frac{H_A^{2\beta}}{H_A} \int_A dx = H_A^{2\beta}.
\end{align*}
Under Assumption \ref{as:bounded density}, we note that $H_A^{2\beta}\lesssim (P_A)^{2\beta}$, so
\begin{align*}
    S_A \lesssim (P_A)^{1 + 2\beta},
\end{align*}
which gives Assumption \ref{as:decay} with $\gamma = 1 + 2\beta$.

\subsubsection{Proof of Proposition \ref{prop:var_1d}}
Our proof uses an integration argument analogous to
Lemma B.2 of \citet{mazumder2023convergence},
and additional steps to handle the KL-based split gain and the $m$-grid. As $p$ is absolutely continuous and $D = 1$, it is almost everywhere (a.e.) differentiable. Let $A = [a,b]\subseteq [0,1]$ where $b > a$ and $S_A > 0$. 

We begin by applying Pinsker's inequality to the KL divergence in the split gain:
\begin{align}\label{eq:s1_pinsker}
    s_A(1,q) \geq 2 P_A \left(P_{A_0 \mid A} - H_{A_0 \mid A}\right)^2 = \frac{2}{P_A} \left(P_{A_0} - H_{A_0} \bar{p}_A\right)^2
\end{align}
for $q \in (a,b)$, where we can write
\begin{align*}
   F_A(q) :=  P_{A_0} - H_{A_0} \bar{p}_A = \int_a^q \left[p(x) - \bar{p}_A\right]dx.
\end{align*}
Note that $F_A(a) = F_A(b) = 0$ from the definition of $\bar{p}_A$. We thus want to lower bound $|F_A({q})|$ for an appropriately chosen $q$ in order to lower bound $s_A(1,q)$ for Assumption \ref{as:greedy identifiable}.

Since $p$ is continuous, $F_A$ is continuously differentiable, with $F_A'(x)=p(x)-\bar p_A$, so we have
$$
H_A\sigma_A^2 =\int_a^b [F_A'(x)]^2\,dx.
$$
 As $p$ is absolutely continuous by assumption, so is $F_A'$ with
$F_A''(x)=p^{\prime}(x)$ a.e., and this term arises in the left-hand side of the LRP inequality.
We now use integration by parts to recover both the $F_A$ and $p^{\prime}$ terms from the above equality:
\begin{align*}
\int_a^b [F_A'(x)]^2\,dx&=\left[F_A(x)F_A'(x)\right]_a^b
  -\int_a^b F_A(x)F_A''(x)\,dx\\
&=-\int_a^b F_A(x)\,p^{\prime}(x)\,dx,
\end{align*}
where the boundary term vanishes. Together, we have the key identity
\begin{align}\label{eq:HaSigma}
    H_A \sigma_A^2 = -\int_A F_A(x)\,p^{\prime}(x)\,dx.
\end{align}
Similar arguments connecting variances to derivatives through
integration by parts are used in the study of Stein-type
covariance identities, e.g. \citet[Lemma 2.18]{ernst2020}.

Next, since $F_A(x)$ is continuous and $A$ is compact, there must exist some $q$ such that $|F_A(q)|= \max_{x\in A}|F_A(x)|$ by the extreme value theorem.
 From (\ref{eq:HaSigma}), we then have
\begin{align*}
    H_A \sigma_A^2 \leq \left|F_A(q) \right| \int_A |p^{\prime}(x)|\, dx.
\end{align*}
Rearranging and applying the assumed LRP inequality then gives
\begin{align}\label{eq:FA_ineq}
    |F_A(q)| \geq \frac{H_A \sigma_A}{\sqrt{C_2}}.
\end{align}

Next, we will round $q$ to the nearest grid point, and use Lipschitz continuity to bound the change in $F_A$. Since $p$ is continuous on $A$, the mean value theorem for integrals ensures there exists $c \in A$
 such that $\bar{p}_A = p(c)$. For any $x \in A$, we then have
\begin{align}\label{eq:dev_bound}
    |p(x) - \bar{p}_A| &= \left|\int_c^x p^{\prime}(y)\, dy\right| \leq \int_A |p^{\prime}(y)| \,dy\leq \sqrt{C_2}\, \sigma_A
\end{align}
where the last inequality follows again from LRP. 
As a result, $|F_A'(x)| \leq \sqrt{C_2} \, \sigma_A$, so $F_A$ is Lipschitz with constant $\sqrt{C_2}\, \sigma_A$. Let $\hat q$ denote the nearest grid point to $q$, including the endpoints $a,b$ for now, so that $|\hat q-q|\leq H_A/(2m)$. The Lipschitz constant gives
\begin{align*}
    |F_A(\hat{q})-F_A({q})| \leq \frac{H_A\sqrt{C_2} \, \sigma_A}{2m} \leq \frac{C_2}{2m}\,|F_A(q)|
\end{align*}
where the last inequality follows from (\ref{eq:FA_ineq}). Since $m \geq C_2$ by assumption, we have
\begin{align*}
    |F_A(\hat{q})| \geq \frac{1}{2}\frac{H_A \sigma_A}{\sqrt{C_2}}.
\end{align*}
Note that the weaker condition $m > C_2 / 2$ also suffices for greedy identifiability, but our condition is convenient. Since $F_A(a) = F_A(b) = 0$ but $|F_A(\hat{q})|  > 0$, we must have $\hat{q} \in (a,b)$ as a valid candidate split. 

Finally, we connect the above to the split gain. Applying the above inequality to (\ref{eq:s1_pinsker}) gives
\begin{align*}
    s_A(1,\hat{q}) \geq \frac{H_A}{2\bar{p}_A}\frac{\sigma_A^2}{C_2} = \frac{P_A \sigma_A^2}{2(\bar{p}_A)^2 C_2}.
\end{align*}
Using the $\chi^2$-divergence upper bound on the KL and the equality statement in (\ref{eq:chi2_bound}), which does not require Assumption \ref{as:bounded density}, we have
${P_A\sigma_A^2}/{(\bar p_A)^2} \geq S_A$, so
\begin{align*}
    s_A(1,\hat{q}) \geq \frac{1}{2C_2}S_A
\end{align*}
which gives Assumption \ref{as:greedy identifiable} with $\kappa = \min\{1,1/(2C_2)\}$.

\subsubsection{Proof of Proposition \ref{prop:ck}}
To verify LRP, we will use the lower bound on the derivatives to ensure $p$ still has a trend at each $x$ in order to lower bound the variance.  
 By continuity of the derivatives, compactness of \([0,1]\), and the nonvanishing-derivative assumption, we have
 \begin{align*}
   t := \min_{x \in [0,1]} \max_{1 \leq j \leq k}\left| p^{(j)}(x)\right| > 0.
 \end{align*}
For each $x$, we have at least one $j_x$ such that $\left| p^{(j_x)}(x)\right| \geq t$, where $j_x$ may change with $x$. By continuity, we can choose
a relatively open interval \(I_x\subseteq[0,1]\) containing $x$ such that
\begin{align*}
    \left| p^{(j_x)}(y)\right| \geq \frac{t}{2}\quad \text{for } y \in I_x.
\end{align*}
The intervals $\{I_x:x\in[0,1]\}$ cover $[0,1]$, and by compactness, they admit a finite subcover $\{I_{x_l}\}_{l=1}^L$ depending only on $p$ and $k$.

For each interval $I_{x_l}$, we now leverage a sublevel set estimate given a lower bound on a derivative, which is used in Fourier analysis. Let $g:[a,b] \to \mathbb{R}$ be $j$-times differentiable on $(a,b)$ for $j \geq 1$, satisfying $|g^{(j)}(x)| \geq t/2 > 0$.  We then have from \citet[Lemma 1]{rogers2005sharp} that for any $u >0$,
\begin{align*}
    \left|\left\{y \in [a,b]: |g(y)| \leq u \right\} \right| \leq 2j \left(\frac{2u}{t}\right)^{1/j}.
\end{align*}
The above states that if a function has some derivative uniformly bounded away from 0 by assumption, it cannot remain close to 0 for too much of the interval. Since subtracting a constant from $p$ does not change its derivatives, for any interval $A \subseteq [0,1]$ with $H_A > 0$, we can substitute $g = p - \bar p_A$ and $j = j_{x_l}$, giving
\begin{align*}
    \left|\{x \in [0,1]: | p(x) - \bar{p}_A| \leq u \}\right| &\leq  \sum_{l = 1}^L    \left| \{x \in I_{x_l}: | p(x) - \bar{p}_A| \leq u \} \right| \\
    &\leq  \sum_{l = 1}^L 2j_{x_l} \left(\frac{2u}{t}\right)^{1/j_{x_l}}.
\end{align*}
Since $j_{x_l}\leq k$, choosing $0 < u \leq t/2$ gives
\begin{align*}
    \left|\{x \in [0,1]: | p(x) - \bar{p}_A| \leq u \}\right| &\leq 2kL\left(\frac{2u}{t}\right)^{1/k}.
    \end{align*}
    Note that the above bound holds uniformly over $A$. Next, we choose 
    \begin{align*}
        u = \frac{t}{2}\left(\frac{H_A}{4kL}\right)^{k}
    \end{align*}
    which satisfies $u \leq t/2$ since $H_A \leq 1$, thus giving 
    \begin{align*}
        \left|\{x \in [0,1]: | p(x) - \bar{p}_A| \leq u \}\right| \leq \frac{H_A}{2}.
    \end{align*}
    This then gives
    \begin{align*}
        \int_A \left[p(x) - \bar{p}_A\right]^2 \, dx \geq \frac{H_A u^2}{2},
    \end{align*}
    which then gives the key bound
    \begin{align}\label{eq:CK_lb}
        \sigma_A^2 \geq \frac{u^2}{2} \gtrsim H_A^{2k}.
    \end{align}

We now use a Sobolev interpolation inequality to bound the left-hand side of the LRP condition as a function of $H_A^k$ and $\sigma_A$. As $p - \bar{p}_A \in C^k[0,1]$, it lies in the Sobolev space $W^{k,1}(A)$, so we can apply \citet[Theorem 7.40, Equation (7.23)]{leoni2017first} for $k \geq 2$ giving
    \begin{align*}
        \int_A|p^{\prime}(x)|\, dx&\lesssim \left(\frac{H_A}{2}\right)^{-1} \int_A \left|p(x) - \bar{p}_A \right| dx + \left(\frac{H_A}{2}\right)^{k - 1} \int_A \left|p^{(k)}(x) \right|\, dx
    \end{align*} 
    where we have chosen the length as $H_A/2$ (written as $\ell$ in their notation). Since $\left|p^{(k)}(x) \right|$ is continuous, it is bounded on $[0,1]$. Cauchy-Schwarz applied to the first term then gives
    \begin{align*}
        \int_A|p^{\prime}(x)|\, dx&\lesssim \sigma_A + H_A^k.
    \end{align*}
For $k = 1$, continuity and compactness again give bounded $p^{\prime}$, so $\int_A |p^{\prime}(x)| dx \lesssim H_A$. Finally, \eqref{eq:CK_lb} gives $\int_A |p^{\prime}(x)|\,dx\lesssim\sigma_A$ in both cases. Squaring establishes the LRP condition, so Proposition~\ref{prop:var_1d} completes the proof for $k \geq 1$.

\subsubsection{H\"{o}lder examples}\label{app:holder_ex}
In this section, we verify the LRP condition for the two examples in the main paper. Again, let $A = [a,b]$ with $b > a$ throughout. 
\paragraph*{Monotonic} We begin with the monotonic example which is slightly easier. By the upper bound on the absolute value of the derivative, we immediately have
\begin{align}\label{eq:hold_ineq1}
    \left(\int_A |p^{\prime}(x)|\, dx\right)^2 \leq C^2 H_A^2. 
\end{align}
Monotonicity, absolute continuity and the bounds on the absolute value of the derivative give $c|x-y|\leq |p(x) - p(y) | \leq C|x-y|$. Next, we have the useful variance identity, $\operatorname{Var}[X] = \frac{1}{2} \mathbb E\left[(X - Y)^2\right]$, when $Y$ is an independent copy of $X$. Applying this to $\sigma_A^2$ with $X$ and $Y$ drawn independently from $\mathrm{Unif}[a,b]$, we have
\begin{align*}
    \sigma_A^2 = \frac{1}{2}\mathbb E\left[\left\{p(X) - p(Y)\right\}^2\right]&\geq \frac{c^2}{2} \mathbb E\left[(X - Y)^2\right] = \frac{c^2 H_A^2}{12}.
\end{align*}
Combining with (\ref{eq:hold_ineq1}) gives 
\begin{align*}
     \left(\int_A |p^{\prime}(x)|\, dx\right)^2 \leq 12\left(\frac{C}{c}\right)^2\sigma_A^2.
\end{align*}
The same constant was obtained in \citet[Example 3.1]{mazumder2023convergence}.

\paragraph*{Cusp} For the cusp example, first consider an interval
$I=[u,v]$ lying to the left of $x_0$. The density gives,
for almost every $x\in I$,
$$
\frac{|p^{\prime}(x)|}{|p(v)-p(u)|}
=
\frac{\beta(x_0-x)^{\beta-1}}
{(x_0-u)^\beta-(x_0-v)^\beta}
\ge \frac{\beta}{H_I},
$$
as the denominator satisfies
$$(x_0 - u)^\beta\left[1 - \left(\frac{x_0 - v}{x_0 - u}\right)^\beta\right]
\le (x_0-u)^{\beta-1}H_I \leq (x_0 - x)^{\beta-1}H_I,$$
where the first inequality uses $r^\beta\ge r$ for
$r\in[0,1]$, and the second uses $x\ge u$ and $\beta-1\le0$. Since $p$ is monotone on $I$, the same variance argument
as above yields
$$
\sigma_I^2
\ge \frac{\beta^2}{12}|p(v)-p(u)|^2.
$$
Moreover, $\int_I|p^{\prime}(x)|\,dx=|p(v)-p(u)|$, so
\begin{align}\label{eq:cusp_one_side}
\left(\int_I|p^{\prime}(x)|\,dx\right)^2
\le \frac{12}{\beta^2}\sigma_I^2.
\end{align}
The argument for $I$ on the right of $x_0$ is identical, so the above bound holds there too.

Now suppose $a<x_0<b$, and let $I$ be the longer of
$[a,x_0]$ and $[x_0,b]$. Since we have $\int_a^{x_0} \beta(x_0 -x)^{\beta-1}dx = (x_0 - a)^\beta$ and similarly for $[x_0,b]$, it follows that
\[
\int_A|p^{\prime}(x)|\,dx
\le 2\int_I|p^{\prime}(x)|\,dx.
\]
Conditioning a uniform draw from $A$ on whether it lies in $I$,
the law of total variance gives
\[
\sigma_A^2
\ge \frac{H_I}{H_A}\sigma_I^2 + \frac{H_{A \setminus I}}{H_A}\,\sigma_{A \setminus I}^2 \geq  \frac{H_I}{H_A}\sigma_I^2
\ge \frac12\sigma_I^2,
\]
where we used  $H_I\ge H_A/2$ in the last inequality. Applying (\ref{eq:cusp_one_side}) therefore yields
\[
\left(\int_A|p^{\prime}(x)|\,dx\right)^2
\le 4\left(\int_I|p^{\prime}(x)|\,dx\right)^2
\le \frac{48}{\beta^2}\sigma_I^2
\le \frac{96}{\beta^2}\sigma_A^2.
\]
This establishes the LRP condition with $C_2=96/\beta^2$.

\subsubsection{$C^k[0,1]$ examples}
For any non-constant density $p$ that is real analytic on an
open interval containing $[0,1]$, at each $x\in[0,1]$ some
derivative of positive order must be nonzero. By the same continuity and finite-subcover argument used in the proof of Proposition~\ref{prop:ck}, these
derivative orders can be chosen uniformly bounded by some
finite $k$. Thus the condition of Proposition~\ref{prop:ck} holds. 

\subsubsection{Step density}
For the step density in Example \ref{ex:step}, one can verify that the split gain at $q \geq 1/m$ is
\begin{align*}
    s_\mathcal{A}(1,q) = \left[1 - (1-q) p_2\right]\log \frac{1 - (1-q)p_2}{q} + (1-q) p_2\log p_2
\end{align*}
which follows as $P_{A_1} = (1-q)p_2$. One can verify that
$$
\frac{d^2}{dq^2}s_{\mathcal A}(1,q)
=
\frac{(1-p_2)^2}
{q^2[1-(1-q)p_2]}
\geq0,
$$
so $s_\mathcal{A}(1,q)$ is convex in $q$. At the optimal split $q = 1/m$, we have $s_\mathcal{A}(1,1/m) = S_{\mathcal{A}}$, and a continuous extension gives $s_\mathcal{A}(1,1) = 0$. Since $p_1 \neq p_2$, we must have $S_\mathcal{A} > 0$. By convexity, for $q = t/m + (1-t)$ for $0 \leq t \leq 1$, we have
\begin{align*}
    s_\mathcal{A}(1,q) \leq t S_{\mathcal{A}}.
\end{align*}
Since $q \geq 2/m$ corresponds to $t \leq (m-2)/(m-1)$, we thus have
\begin{align*}
     s_\mathcal{A}(1,q) \leq \frac{m-2}{m-1}S_{\mathcal{A}} <  \left(1- \frac{1}{m}\right)S_{\mathcal{A}} \quad \text{for } q \geq 2/m.
\end{align*}
Thus, $\delta=1/m$ ensures that $q=1/m$ is the only
near-optimal root split among the candidate grid points.

\subsection{Multivariate case}\label{app:mv}
In the multivariate setting, we require two new ingredients. The first is to define the multivariate version of the LRP condition, which is relatively natural and plays the same role as the univariate case. The second new ingredient is a notion of \textit{marginal visibility}, where differences between $p$ and the base density $h$ are visible to axis-aligned splits in order to combat the copula counterexample. We assume $H =\mathrm{Unif}[0,1]^D$ throughout.

Fortunately, we can borrow much of the univariate machinery.
To that end, let $A=\prod_{d=1}^D I_d$ be a rectangle with
$H_A>0$. For each dimension $d$, we define the coordinate average 
\begin{align*}
    \bar{p}_{A,d}(t)
    = \frac{1}{H_{A_{-d}}}
      \int_{A_{-d}} p(t,x_{-d})\,dx_{-d},
    \qquad t\in I_d,
\end{align*}
where $x_{-d}$ denotes $x$ with its $d$-th element removed,
and $A_{-d}=\prod_{j\neq d}I_j$. Since $p$ is integrable, Fubini's theorem ensures that $\bar p_{A,d}(t)$ is well-defined a.e. Note that the above is again the conditional mean of $p(t, X_{-d})$ under $H(\cdot \mid X_{-d} \in A_{-d})$. Thus, $\bar{p}_{A,d}(t)$ describes how the density varies
along coordinate $d$, after averaging uniformly over the
remaining coordinates within $A_{-d}$.

 A nice property satisfied by the above coordinate average is
\begin{align}\label{eq:coord_pA}
    \bar{p}_A =  \frac{1}{H_{I_d}} \int_{I_d}\,\bar{p}_{A,d}(t)\, dt,
\end{align}
so we can naturally define the coordinate-averaged variance of $\bar{p}_{A,d}(X^{(d)})$ under $H(\cdot \mid A)$:
\begin{align*}
    \sigma_{A,d}^2 = \frac{1}{H_{I_d}} \int_{I_d}\left[\bar{p}_{A,d}(t) -     \bar{p}_A\right]^2 \,dt.
\end{align*}
We can then simply define the LRP condition using $\bar{p}^{\prime}_{A,d}(t)$ and $\sigma_{A,d}^2$ as below. 
\begin{assumption}[Coordinate-averaged LRP]\label{as:coord_lrp}
    Let $H = \mathrm{Unif}[0,1]^D$ and $p \in W^{1,1}((0,1)^D)$. There exists
$C_2>0$ such that, for every rectangle $A\subseteq[0,1]^D$
with $H_A>0$ and each $d \in \{1,\ldots,D\}$,
\begin{align*}
\left(\int_{I_d} |\bar{p}_{A,d}^{\prime}(t)|\,dt\right)^2
\le
C_2\, \sigma_{A,d}^2.
\end{align*}
\end{assumption}
The Sobolev condition ensures that $p$ has a representative that is absolutely continuous on lines, with integrable partial derivatives \citep[Theorem~11.45]{leoni2017first}. These properties together with Fubini's theorem ensure that each coordinate average $\bar{p}_{A,d}$ admits an absolutely continuous representative using standard arguments. 

Our new marginal visibility condition then requires that the coordinate-averaged variance uniformly captures a fraction of the total variance $\sigma_A^2$.
\begin{assumption}[Marginal visibility]\label{as:marg_vis}
There exists $\eta \in (0,1]$ such that for every rectangle $A\subseteq[0,1]^D$
with $H_A>0$, we have
\begin{align*}
    \max_d \sigma_{A,d}^2 \geq \eta \sigma_A^2.
\end{align*}
\end{assumption}
In the case where $p$ is a non-constant bivariate copula density, taking $A = \mathcal{A}$ violates the above, as $\sigma_{\mathcal{A},1}^2 = \sigma_{\mathcal{A},2}^2 = 0$, while $\sigma_{\mathcal{A}}^2 > 0$. Note that when $D=1$, marginal visibility holds automatically with $\eta=1$, since $\sigma_{A,1}^2=\sigma_A^2$.

\subsubsection{Main results}
We then have our main results below, from which we derive our examples. 
\begin{proposition}\label{prop:mv_greedy}
Suppose $p$ satisfies Assumptions \ref{as:coord_lrp} and~\ref{as:marg_vis}, and let
$m\geq\max\{2,\lceil C_2\rceil\}$. Then Assumption \ref{as:greedy identifiable} holds. 
\end{proposition}
\begin{proof}
  Let $A$ be a rectangle with $S_A > 0$. For this $A$, choose $d$ to be the coordinate satisfying the marginal visibility inequality, and let $I_d = [a,b]$. We can then apply almost the exact same proof as Proposition \ref{prop:var_1d}. Starting from (\ref{eq:s1_pinsker}) with $s_A(d,q)$ and $q \in (a,b)$ instead, we additionally require the identities
  \begin{align*}
      P_{A_0} = \int_{a}^q H_{A_{-d}}\, \bar{p}_{A,d}(t)\, dt, \qquad H_{A_0} = H_{A_{-d}}(q-a).
  \end{align*}
which together give
\begin{align*}
    F_{A,d}(q) := P_{A_0} - H_{A_0}\bar{p}_A =  H_{A_{-d}}\int_a^q [\bar{p}_{A,d}(t) - \bar{p}_A]\, dt. 
\end{align*}
Again $F_{A,d}(a) = F_{A,d}(b) = 0$ from (\ref{eq:coord_pA}). Since $\bar p_{A,d}$ admits an absolutely continuous representative and satisfies Assumption~\ref{as:coord_lrp}, we can apply the exact same argument from Proposition~\ref{prop:var_1d} to $F_{A,d}/H_{A_{-d}}$.
Together with Pinsker's inequality, this gives the existence of a candidate grid point $\hat{q}\in(a,b)$ such that
\begin{align*}
    s_A(d,\hat{q}) \geq \frac{H_{I_d}^2 \, H_{A_{-d}}^2 \,  \sigma_{A,d}^2}{2P_A \, C_2} = \frac{H_A \, \sigma_{A,d}^2}{2\bar{p}_A \, C_2}\geq \frac{\eta H_A\, \sigma_A^2}{2\bar{p}_A C_2}
\end{align*}
where the final inequality is from marginal visibility. The same final step from the equality in (\ref{eq:chi2_bound}) gives
\begin{align*}
    s_A(d,\hat{q}) \geq \frac{\eta}{2C_2}S_A
\end{align*}
so Assumption \ref{as:greedy identifiable} holds with $\kappa = \min\{1,\eta/(2C_2)\}$.
  \end{proof}
The above proposition highlights that marginal visibility is the extra ingredient that allows us to extend the univariate LRP argument to establish greedy identifiability as in Assumption \ref{as:greedy identifiable} when $D>1$. We will demonstrate verification of the marginal visibility condition shortly.

We now proceed to show sufficient conditions for signal decay as in Assumption \ref{as:decay}. To begin, we need a technical lemma which controls the variance under nested rectangles, which will help combat the issue of elongated rectangles.
\newpage
\begin{lemma}\label{lem:nested_variance}
Let $p$ be continuous, and suppose Assumption \ref{as:coord_lrp} holds. For any nested rectangles $B \subseteq A$ with $H_A, H_B > 0$, we have
\begin{align*}
    \sigma_B^2 \lesssim \sigma_A^2.
\end{align*}
\end{lemma}
\begin{proof}
 First, we note that
\begin{align}\label{eq:var_B}
    \sigma_B^2 = \frac{1}{H_B}\int_B\left[p(x) - \bar{p}_B\right]^2\, dx \leq \frac{1}{H_B}\int_B\left[p(x) - \bar{p}_A\right]^2\, dx,
\end{align}
which follows as $\bar p_B$ minimizes $\int_B[p(x)-c]^2\,dx$ over constants $c\in\mathbb R$. We will now show that $\sup_{x \in A}\left[p(x) - \bar{p}_A\right]^2 \lesssim \sigma_A^2$ under Assumption \ref{as:coord_lrp}, which is sufficient for $\sigma_B^2 \lesssim \sigma_A^2$.

Write $A=\prod_{d=1}^D I_d$, and fix a coordinate $d$. Under Assumption \ref{as:coord_lrp}, we can apply the proof of Proposition~\ref{prop:var_1d} to $\bar p_{A,d}$ again as we did in Proposition \ref{prop:mv_greedy}. The bound  (\ref{eq:dev_bound}) thus gives
\begin{align*}
    |\bar p_{A,d}(t)-\bar p_A|
    \leq \sqrt{C_2}\,\sigma_{A,d}
\end{align*}
for any $t \in I_d$. For any constant $c\in\mathbb{R}$, the triangle inequality and AM-GM then give
\begin{align}
    |\bar p_{A,d}(t)-c|^2 &\leq \left[\sqrt{C_2}\, \sigma_{A,d} + |\bar{p}_A - c|\right]^2 \notag \\
    &\lesssim \sigma_{A,d}^2 + [\bar{p}_A - c]^2
    = \frac{1}{H_{I_d}}
    \int_{I_d}
    [\bar p_{A,d}(u)-c]^2\,du,\label{eq:coav_ineq}
\end{align}
where the final equality follows from $\mathbb E[(Y-c)^2] = \operatorname{Var}[Y] + (\mathbb E[Y] - c)^2$ and \eqref{eq:coord_pA}.

We now convert the above coordinate-averaged bound to a section-wise bound. Fix a coordinate $d$, a point $x_{-d}\in A_{-d}$, and a
constant $c\in\mathbb R$. For $\varepsilon>0$, define
$$
Q_\varepsilon
=
A_{-d}\cap
\prod_{j\neq d}
[x^{(j)}-\varepsilon,x^{(j)}+\varepsilon],
\qquad
A_\varepsilon=I_d\times Q_\varepsilon.
$$
The rectangles $A_{\varepsilon}$ have positive volume and remain inside $A$,
and $Q_\varepsilon$ shrinks to $x_{-d}$ as $\varepsilon\to 0$. Since $p$ is continuous, for each fixed $t\in I_d$,
$$
\lim_{\varepsilon \to 0}\bar p_{A_\varepsilon,d}(t)
= p(t,x_{-d}).
$$
Moreover, continuity and compact support imply that $p$
is bounded, so $\bar{p}_{A_{\varepsilon},d}$ is uniformly bounded.
We now apply \eqref{eq:coav_ineq} to $A_\varepsilon$ at a fixed $t$
and take $\varepsilon \to 0$, where dominated convergence for the right-hand side gives
\begin{align}\label{eq:section_wise}
|p(t,x_{-d})-c|^2
\lesssim
\frac{1}{H_{I_d}}
\int_{I_d}|p(u,x_{-d})-c|^2\,du
\end{align}
for every $t\in I_d$, $x_{-d}\in A_{-d}$ and
$c\in\mathbb R$, with an implicit constant depending only
on $C_2$.

We now apply \eqref{eq:section_wise} successively for each $d$. For any $x\in A$,
applying it to $x^{(1)}$ gives
$$
|p(x)-c|^2
\lesssim
\frac{1}{H_{I_1}}
\int_{I_1}
|p(u_1,x^{(2:D)})-c|^2\,du_1.
$$
For each $u_1$, applying \eqref{eq:section_wise} to $x^{(2)}$ in the integrand gives
$$
|p(x)-c|^2
\lesssim
\frac{1}{H_{I_1}H_{I_2}}
\int_{I_1}\int_{I_2}
|p(u_1,u_2,x^{(3:D)})-c|^2
\,du_2\,du_1.
$$
Repeating this argument through all $D$ coordinates gives
$$
|p(x)-c|^2
\lesssim
\frac{1}{H_A}\int_A|p(u)-c|^2\,du,
$$
with an implicit constant depending only on $C_2$ and $D$. Setting $c=\bar p_A$, we have
$$
\sup_{x\in A}[p(x)-\bar p_A]^2
\lesssim \sigma_A^2.
$$
As discussed earlier, the above is then sufficient for $\sigma_B^2 \lesssim \sigma_A^2$ from (\ref{eq:var_B}).
\end{proof}

We are now equipped to show the multivariate version of Proposition \ref{prop:holder_1d}.
\begin{proposition}\label{prop:mv_decay}
Suppose Assumption \ref{as:bounded density} and the conditions of Proposition \ref{prop:mv_greedy} hold. If $p$ is additionally $\beta$-H\"older continuous for some $\beta\in(0,1]$, then Assumption~\ref{as:decay} holds with $\gamma=1+2\beta/D$.
\end{proposition}
\begin{proof}
We first illustrate why the proof of Proposition \ref{prop:holder_1d} faces difficulties when $D > 1$. Starting with $S_A \lesssim P_A \sigma_A^2$,
a direct extension of the H\"older argument would require $\|x-y\|\lesssim H_A^{1/D}$ for all $x,y\in A$. However, this bound fails for general rectangles, as an elongated rectangle can have arbitrarily small volume while $\|x-y\|$ remains large. We must thus bound the variance another way.

We begin with some general properties under axis-aligned splitting. Let $A$ be an arbitrary rectangle with $S_A > 0$, where again we write $A = \prod_{d = 1}^D I_d$.  The $\chi^2$-divergence upper bound on the KL applied to the split gain gives
\begin{align*}
s_A(d,q)
&\leq P_A\sum_{i=0}^1
\frac{(P_{A_i\mid A}-H_{A_i\mid A})^2}
{H_{A_i\mid A}}=\frac{1}{\bar p_A}
\sum_{i=0}^1 H_{A_i}
(\bar p_{A_i}-\bar p_A)^2, 
\end{align*}
where we used
$P_{A_i\mid A}
=(H_{A_i}\bar p_{A_i})/(H_A\bar p_A)$. Since the split only changes coordinate $d$, let \(J_0,J_1\) be the subintervals of \(I_d\) induced by the split, so \(A_i=J_i\times A_{-d}\). We then have
$$
\bar p_{A_i}
=
\frac{1}{H_{J_i}H_{A_{-d}}}
\int_{J_i}\int_{A_{-d}}
p(t,x_{-d})\,dx_{-d}\,dt
=
\frac{1}{H_{J_i}}
\int_{J_i}\bar p_{A,d}(t)\,dt.
$$
Consequently, we have 
$$
\bar p_{A_i}-\bar p_A
=
\frac{1}{H_{J_i}}
\int_{J_i}
[\bar p_{A,d}(t)-\bar p_A]\,dt.
$$
Next, for any $d$ and $t \in I_d$, the H\"{o}lder assumption gives
\begin{align}\label{eq:mv_holder_ineq}
    \left|\bar{p}_{A,d}(t) - \bar{p}_{A} \right| &= \frac{1}{H_A}\left|\int_{A_{-d}} \int_{I_d}\left[{p}(t, x_{-d}) - {p}(u, x_{-d})\right] \, du\,dx_{-d} \right|\lesssim |I_d|^{\beta},
\end{align}
which follows from $|t - u| \leq |I_d|$. Together, these identities and the H\"older bound give
$|\bar p_{A_i}-\bar p_A|\lesssim |I_d|^\beta$, which then gives
\begin{align*}
s_A(d,q)
&\lesssim \frac{1}{\bar p_A}
\sum_{i=0}^1 H_{A_i}|I_d|^{2\beta}=\frac{H_A}{\bar p_A}|I_d|^{2\beta}\lesssim P_A|I_d|^{2\beta},
\end{align*}
where the final inequality uses Assumption \ref{as:bounded density}. Suppose now that $(\hat{d},\hat{q})$ is a near-optimal split of $A$. By Proposition \ref{prop:mv_greedy}, Assumption~\ref{as:greedy identifiable} holds, so the above inequality gives
\begin{align}\label{eq:split_gain_ineq}
S_A
\lesssim {s_A(\hat d,\hat q)}
\lesssim P_A|I_{\hat d}|^{2\beta}.
\end{align}
The above inequality illustrates that when the split coordinate is narrow, H\"older continuity limits
the split gain, and the remaining signal must be small.

We now leverage the above property of near-optimal splits to show Assumption \ref{as:decay}. Fix a near-optimal tree $T\in\mathcal T^*$ and a node $B\in T$, where
$B=\prod_{d=1}^D J_d$, $S_B>0$, and $B\neq\mathcal A$.
Let $\hat d$ be a coordinate of minimum width, so that $|J_{\hat{d}}| = \min_{d} |J_d|$.  Naturally, we then have
\begin{align}\label{eq:H_B_ineq}
     H_B \geq |J_{\hat{d}}|^D.
\end{align}
Along the path from the root to $B$ in the tree $T$, let $A = \prod_{d = 1}^D I_d$ be the last ancestor whose near-optimal split was along coordinate $\hat d$. Such an ancestor exists because $B\neq\mathcal A$ implies $|J_{\hat d}|<1$, so this coordinate must have been split along this path. While $A$ need not be the parent of $B$, the interval $J_{\hat d}$ coincides with the subinterval of $I_{\hat d}$ corresponding to the child of $A$ that contains $B$. By the $m$-grid restriction,
$$
|J_{\hat d}|\geq \frac{|I_{\hat d}|}{m}.
$$
Since the near-optimal split of $A$ is along coordinate $\hat{d}$, (\ref{eq:split_gain_ineq}) gives
\begin{align*}
    |J_{\hat d}| \gtrsim \left(\frac{S_A}{P_A}\right)^{1/(2\beta)}
\end{align*}
and Lemma \ref{lemma: variance KL} and the nested variance result from Lemma \ref{lem:nested_variance} (since $B \subseteq A$) then give
\begin{align*}
    \frac{S_A}{P_A} \gtrsim  \sigma_A^2 \gtrsim \sigma_B^2 \gtrsim \frac{S_B}{P_B}.
\end{align*}
Combining the above with (\ref{eq:H_B_ineq}) and $P_B \gtrsim H_B$ from Assumption \ref{as:bounded density} gives
\begin{align*}
    S_B \lesssim (P_B)^{1 + 2\beta/D}.
\end{align*}
Finally, we handle the edge cases. The case $S_B=0$ is immediate, and the root also satisfies the
desired bound, since $P_{\mathcal A}=1$ and
$S_{\mathcal A}<\infty$. 
\end{proof}

Finally, we consider a minor extension to the anisotropic H\"{o}lder setting where each coordinate may have different smoothness \citep{shen2013adaptive}. Following the coordinatewise formulation of
\citet{barron1999risk} and \citet{bhattacharya2014anisotropic},
we assume that there exist a constant $L\in(0,\infty)$ and exponents $(\beta_1,\ldots,\beta_D)\in(0,1]^D$ such that
\begin{align}\label{eq:aniso}
   \left|p(t,x_{-d})-p(u,x_{-d})\right|
   \leq L|t-u|^{\beta_d}
\end{align}
for every $d\in\{1,\ldots,D\}$, $t,u\in[0,1]$, and
$x_{-d}\in[0,1]^{D-1}$.
Fortunately, the above condition fits naturally into our proof technique, and we recover the familiar harmonic mean exponent. 
\begin{corollary}\label{corr:mv_decay_aniso}
Suppose Assumption \ref{as:bounded density} and the conditions of Proposition~\ref{prop:mv_greedy}
hold. If $p$ additionally satisfies the anisotropic
H\"older condition \eqref{eq:aniso} with exponents
$(\beta_1,\ldots,\beta_D)\in(0,1]^D$, then
Assumption~\ref{as:decay} holds with
$\gamma=1+2\beta_0/D$, where $\beta_0 = \left(D^{-1}\sum_{d = 1}^D \beta_d^{-1}\right)^{-1}.$
\end{corollary}
\begin{proof}
    The proof is identical to that of Proposition \ref{prop:mv_decay}, except for a few minor changes. First, following the section-wise definition of anisotropic H\"{o}lder smoothness, the argument in (\ref{eq:mv_holder_ineq}) can be tightened to have $|I_d|^{\beta_d}$ on the right hand side, and similarly for (\ref{eq:split_gain_ineq}). 
    
    The other difference is the final step when choosing the coordinate $\hat{d}$ for the ancestor argument. Instead of selecting $\hat{d}$ such that we have minimum width, we instead choose $\hat{d}$ to minimize $|J_{{d}}|^{\beta_{{d}}}$.  This allows for coordinates with larger exponents to have wider intervals. Finally, the volume bound in (\ref{eq:H_B_ineq}) is replaced with a tighter argument. Since $\hat d$ minimizes $|J_d|^{\beta_d}$,
we have $|J_d|\geq
(|J_{\hat d}|^{\beta_{\hat d}})^{1/\beta_d}$ for every $d$.
Multiplying over coordinates gives
$$
H_B=\prod_{d=1}^D|J_d|
\geq
\left(|J_{\hat d}|^{\beta_{\hat d}}\right)^{\sum_d1/\beta_d},
$$
which then gives the tighter bound
$$
\min_d|J_d|^{\beta_d}
=|J_{\hat d}|^{\beta_{\hat d}}
\leq H_B^{\beta_0/D}.
$$
The remainder of the proof is then straightforward with $\beta_0$ in place
of $\beta$.
\end{proof}

\subsubsection{Product and additive densities}
\paragraph*{Additive case} We consider the additive form first with $p(x) = Z^{-1}[c + \sum_{d = 1}^D g_d(x^{(d)})]$, where $Z$ is the normalizing constant. We use an additive variance decomposition similar to that in \citet[Lemma B.3]{mazumder2023convergence}. Since $\sigma_A^2$ is the variance under $H(\cdot \mid A)$, and the coordinates are independent under $H(\cdot \mid A)$, additivity of variance gives
\begin{align*}
    \sigma_A^2 = \frac{1}{Z^2}\sum_{d = 1}^D \operatorname{Var}_H[g_d(X^{(d)}) \mid A]
\end{align*}
where the above notation indicates  $X\sim H(\cdot \mid A)$. Next, we can see that 
$$\bar{p}_{A,d}(t) = Z^{-1}[c_{A,d} + g_d(t)],$$ 
where $c_{A,d}$ is a constant depending on $A$ and $d$, so we have
\begin{align*}
    \sigma_{A,d}^2 = \operatorname{Var}_H[\bar{p}_{A,d}(X^{(d)}) \mid A] = \frac{1}{Z^2}\operatorname{Var}_H[g_d(X^{(d)}) \mid A],
\end{align*}
as adding a constant to a random variable does not change its variance. We thus have $\sigma_A^2 = \sum_{d = 1}^D \sigma_{A,d}^2$. The above then gives $\max_d \sigma_{A,d}^2 \geq \sigma_A^2/D$, so Assumption \ref{as:marg_vis} is satisfied. 

Next, we verify Assumption \ref{as:coord_lrp}.
Since each $g_d$ is absolutely continuous, the additive form immediately gives
$p\in W^{1,1}((0,1)^D)$. Each coordinate average is also
absolutely continuous, as it is an affine transformation of $g_d$. Next, the LRP condition is invariant under rescaling and addition of constants, so it carries over from \(g_d\) to \(\bar p_{A,d}\) with the same constant. 

Finally, for anisotropic H\"older condition, we note that 
$$|p(t,x_{-d})-p(u,x_{-d})| =Z^{-1}|g_d(t)-g_d(u)|,$$
so Corollary \ref{corr:mv_decay_aniso} holds with exponents $(\beta_1,\ldots, \beta_D)$.

\paragraph*{Product case} For the product case, it is simpler to assume $g_d(t)$ are densities so $p(x) = \prod_{d = 1}^D g_d(x^{(d)})$, and naturally Assumption \ref{as:bounded density} holds for each component. Here, it is more convenient to work with the KL divergence then use Lemma \ref{lemma: variance KL} to connect it to $\sigma_A^2$. Since both $p(\cdot \mid A)$ and $h(\cdot \mid A)$ are product densities, KL additivity gives
\begin{align*}
    \sigma_A^2 &\asymp  \sum_{d = 1}^D K\left\{g_d(\cdot \mid I_d); \mathrm{Unif}(I_d)\right\} \asymp \sum_{d = 1}^D \operatorname{Var}_H[g_d(X^{(d)}) \mid A]
\end{align*}
which follows from applications of Lemma \ref{lemma: variance KL}. We can also verify that
\begin{align*}
    \bar{p}_{A,d}(t) = c_{A,d}\,  g_d(t), 
\end{align*}
where $c_{A,d}$ is bounded uniformly above and away from 0 from Assumption \ref{as:bounded density}, since it is an average of $\prod_{j \neq d}^D g_j(x^{(j)})$. We thus have
\begin{align*}
    \sigma_{A,d}^2 = \operatorname{Var}_H[\bar{p}_{A,d}(X^{(d)}) \mid A] \asymp \operatorname{Var}_H[g_d(X^{(d)}) \mid A].
\end{align*}
We thus have $\sigma_A^2 \asymp \sum_{d = 1}^D \sigma_{A,d}^2$, so again $\max_d \sigma_{A,d}^2 \gtrsim \sigma_A^2/D$ and Assumption \ref{as:marg_vis} holds.

Since each $g_d$ is absolutely continuous, the product form gives
$$
\frac{\partial p(x)}{\partial x^{(d)}}
=g_d'(x^{(d)})\prod_{j\ne d}g_j(x^{(j)})
\quad\text{a.e.}
$$
The above is integrable, so
$p\in W^{1,1}((0,1)^D)$, and each coordinate average is
absolutely continuous and inherits the LRP condition from
$g_d$ again by scale invariance. 
Assumption~\ref{as:coord_lrp} thus holds.

Finally, boundedness of the components gives
$$
|p(t,x_{-d})-p(u,x_{-d})|
=|g_d(t)-g_d(u)|\prod_{j\ne d}g_j(x^{(j)})
\lesssim |t-u|^{\beta_d},
$$
where we again use Assumption \ref{as:bounded density} to handle $\prod_{j\ne d}g_j(x^{(j)})$. The anisotropic H\"{o}lder condition \eqref{eq:aniso} thus holds with the component exponents.

\subsubsection{Coordinate-wise monotonicity}
From continuity of the partial derivative and the compact support, there exist constants such that $0 < c \leq \left|\partial p(x) / \partial x^{(d)}\right| \leq C< \infty$ for all $x$ and $d$. 
We first show that the coordinate averages inherit these
derivative bounds. Since $p\in C^1([0,1]^D)$, Leibniz's rule gives
\begin{align*}
    \bar p_{A,d}'(t)
    = \frac{1}{H_{A_{-d}}}
      \int_{A_{-d}}
      \frac{\partial p(t,x_{-d})}{\partial t}\,dx_{-d}
\end{align*}
for every $t$ in the interior of $I_d$.
Since the partial derivative has a fixed sign, averaging over $x_{-d}$
preserves the bounds $c\leq |\bar p_{A,d}'(t)|\leq C$.
The coordinate average is therefore monotone and continuously
differentiable, so the argument in
Section~\ref{app:holder_ex} gives
$$\sigma_{A,d}^2\gtrsim |I_d|^2.$$
Next, since $p\in C^1([0,1]^D)$ has bounded partial derivatives, the mean value theorem gives
$|p(x)-p(y)|\lesssim\|x-y\|$.
Thus, following the argument of
Proposition~\ref{prop:holder_1d},
\begin{align*}
|p(x)-\bar p_A|
&=\frac{1}{H_A}\left|\int_A[p(x)-p(y)]\,dy\right|\lesssim \frac{1}{H_A}\int_A\|x-y\|\,dy\leq \left(\sum_{d=1}^D|I_d|^2\right)^{1/2}.
\end{align*}
The above then implies $$\sigma_{A}^2 \lesssim \sum_{d = 1}^D |I_d|^2 \lesssim \sum_{d = 1}^D \sigma_{A,d}^2,$$ 
so we again have $\max_{d}\sigma_{A,d}^2 \gtrsim \sigma_A^2/D$, giving Assumption \ref{as:marg_vis}.

The LRP condition also holds for $\bar{p}_{A,d}$ following Section \ref{app:holder_ex} with a constant uniform over $A$ and $d$, so we have Assumption \ref{as:coord_lrp}. Finally, Lipschitz continuity implies H\"older continuity with exponent $\beta=1$.

\subsubsection{Coordinate-wise convexity}
Since the second order partial derivatives are continuous and nonzero, they must have fixed sign. As in the preceding monotonicity proof, continuity and compactness give uniform bounds on the second-order partial derivatives, which coordinate-averaging preserves as their signs are fixed. Thus, $0<c\leq |\bar p_{A,d}''(t)|\leq C<\infty$. We may use a simplified version of the proof of
Proposition~\ref{prop:ck} for $k=2$, omitting the finite
subcover because the second derivative is bounded away
from zero uniformly, which gives
\begin{align*}
    \sigma_{A,d}^2 \gtrsim |I_d|^4.
\end{align*}
The coordinate-wise convexity thus prevents interactions from canceling out all of the variation when coordinate-averaging.

Next, we require a tighter argument for connecting $\sigma_A^2$  to $\sum_{d = 1}^D\sigma_{A,d}^2$ based on smoothness due to the above quartic lower bound, as the bound used in the previous monotonic case is too weak. We carry out a linear approximation of $p(x)$ to connect $p$ with $\bar{p}_{A,d}$, and use the boundedness of the Hessian to control the error. Let $z$ be the centroid of $A$, which will be useful under averaging shortly.
Taylor's theorem gives
\begin{align}\label{eq:expan}
p(x)=p(z)+\nabla p(z)^\top(x-z)+R(x),
\end{align}
where boundedness of the Hessian from $p \in C^2([0,1]^D)$ implies
$$
\sup_{x\in A}|R(x)|
\lesssim \sum_{d=1}^D|I_d|^2.
$$
Since $z$ is the center of $A$, for every coordinate $d$,
$$
\frac1{|I_d|}\int_{I_d}(x^{(d)}-z^{(d)})\,dx^{(d)}=0.
$$
Thus, averaging \eqref{eq:expan} over $x \in A$
removes the linear term, and averaging over
$x_{-d}\in A_{-d}$ removes all but the $d$-th linear term. Consequently, averaging over $A$ and $A_{-d}$ gives
\begin{align*}
\bar p_A
&=p(z)+\bar R_A, \qquad 
\bar p_{A,d}(t)
=p(z)+\frac{\partial p(z)}{\partial z^{(d)}}(t-z^{(d)})+R_d(t),
\end{align*}
where
$$
\bar R_A=\frac1{H_A}\int_A R(x)\,dx,
\qquad
R_d(t)=\frac1{H_{A_{-d}}}
\int_{A_{-d}}R(t,x_{-d})\,dx_{-d},
$$
and both $|\bar R_A|$ and $|R_d(t)|$ are bounded by
$\sup_{x\in A}|R(x)|$. The above then implies
\begin{align*}
p(x)-\bar p_A
&= \sum_{d=1}^D[\bar p_{A,d}(x^{(d)})-\bar p_A] + 
R(x)-\sum_{d=1}^D R_d(x^{(d)})+(D-1)\bar R_A\\
&= \sum_{d=1}^D[\bar p_{A,d}(x^{(d)})-\bar p_A] + 
\widetilde{R}_A(x)
\end{align*}
where the combined remainder satisfies
$\sup_{x\in A}|\widetilde R_A(x)| \lesssim \sum_{d=1}^D |I_d|^2$ by the triangle inequality. Squaring both sides, repeated application of Cauchy-Schwarz gives
\begin{align*}
[p(x)-\bar p_A]^2 \lesssim  \sum_{d=1}^D[\bar p_{A,d}(x^{(d)})-\bar p_A]^2 + \sum_{d = 1}^D |I_d|^4.
\end{align*}
Integrating the above over $x \in A$ and dividing by $H_A$ gives
\begin{align*}
\sigma_A^2 &\lesssim  \sum_{d=1}^D\sigma_{A,d}^2 + \sum_{d=1}^D|I_d|^4.
\end{align*}
Combining this with
$\sigma_{A,d}^2\gtrsim |I_d|^4$ yields
$$
\sigma_A^2\lesssim\sum_{d=1}^D\sigma_{A,d}^2
\leq D\max_d\sigma_{A,d}^2,
$$
thus establishing Assumption~\ref{as:marg_vis}.

Continuing with the argument of Proposition \ref{prop:ck} gives that $\bar{p}_{A,d}$ satisfies LRP with a constant uniform over $A,d$, so we have Assumption \ref{as:coord_lrp}. Finally, since $p\in C^2([0,1]^D)$ has bounded gradient, it is Lipschitz and hence H\"older continuous with exponent $\beta=1$.

\subsubsection{Variable sparsity}
Suppose the marginal density $g(x^{(1:D_0)})$ satisfies Assumptions \ref{as:bounded density}, \ref{as:greedy identifiable} and \ref{as:decay} with some $\kappa$ and $\gamma$ (and some $m\geq 2$ and $\delta \in (0,1)$). Let $A  =B \times [0,1]^{D - D_0}$ where $B$ is a positive volume subrectangle in $[0,1]^{D_0}$. Here, quantities indexed by $B$ or its descendants are computed under $g$, and near-optimal trees for $p$ and $g$ use the same $m$ and $\delta$.

For $x \in A$, we have $p(x) = g(x^{(1:D_0)})$ and $p(x \mid A) = g(x^{(1:D_0)} \mid B)$, which then give
\begin{align*}
    P_A = P_B, \qquad S_A = S_B.
\end{align*}
 Next, we consider splitting $A$ at some $(d,q)$. If $d\leq D_0$, let $B_0$ and $B_1$ be the children obtained by splitting $B$ at the same $(d,q)$. The corresponding children of $A$ are then $A_i=B_i\times[0,1]^{D-D_0}$ for $i=0,1$.
 In this case, one can similarly show that $$P_{A_i} = P_{B_i},  \quad P_{A_i \mid A} = P_{B_i \mid B}, \quad H_{A_i \mid A} = H_{B_i \mid B},$$
which results in $s_A(d,q) = s_B(d,q)$.  If $d>D_0$, then conditional on $X\in A$, the independent noise $X^{(d)}$ remains uniform on $[0,1]$ under both $P$ and $H$. Hence $P_{A_i\mid A}=H_{A_i\mid A}$ giving $s_A(d,q)=0$. Summarizing, we have
\begin{align}\label{eq:sa_sb_ineq}
 s_A(d,q) = \begin{cases}
        s_B(d,q) &\quad \text{if } d \leq D_0,\\
        0 &\quad \text{otherwise.}
    \end{cases}
\end{align}
From the above, if additionally $B$ belongs to a near-optimal tree for $g$, then we have
\begin{align*}
    \max_{d,q} s_A(d,q) = \max_{d \leq D_0, q} s_B(d,q) \geq \kappa S_B = \kappa S_A.
\end{align*}
Furthermore, for $\gamma < \infty$, we have
\begin{align*}
    S_A = S_B \leq \left(C_1 P_{B}\right)^{\gamma} = \left(C_1 P_{A}\right)^{\gamma}. 
\end{align*}
The same argument works for $\gamma = \infty$. For a node $A$ of the aforementioned form, $\kappa$ and $\gamma$ are thus valid for the inequalities in Assumptions \ref{as:greedy identifiable} and \ref{as:decay} to hold for $p$.

We now proceed by induction to show that every node $A$ in a near-optimal tree for $p$ with $S_A>0$ has the form
$A=B\times[0,1]^{D-D_0}$, where
$B$ belongs to a near-optimal tree for $g$.
This holds at the root, with $B=[0,1]^{D_0}$. For the induction step, suppose the claim holds for a
node $A$ in a near-optimal tree of $p$ with $S_A > 0$, and let
$(\hat d,\hat q)$ be any near-optimal split of $A$.
The assumption on $B$ and (\ref{eq:sa_sb_ineq}) give
$$
s_A(\hat d,\hat q)
\geq (1-\delta)\max_{d,q}s_A(d,q)
\geq (1-\delta)\kappa S_A>0.
$$
Note that we must have $\hat{d} \leq D_0$, as $s_A(\hat{d}, \hat{q}) > 0$, so the children also take the form
$A_i=B_i\times[0,1]^{D-D_0}$.
Since (\ref{eq:sa_sb_ineq}) implies the split of $B$ is also near-optimal for $g$, we have that $B_i$ must belong to a near-optimal tree for $g$. Finally, by Lemma~\ref{lem:split_gain}, every node with $S_A > 0$ has
only ancestors with $S_{A'} > 0$, so the induction covers all
nodes relevant to Assumptions~\ref{as:greedy identifiable}
and~\ref{as:decay}. Assumptions~\ref{as:greedy identifiable} and~\ref{as:decay}
thus hold for $p$ with the same constants as for $g$, and Assumption \ref{as:bounded density} is obvious.

\subsubsection{Quadratic examples}\label{app:copula}
In this section, we discuss an illustrative quadratic example for $D = 2$. Let $p$ be a general quadratic polynomial in 2 dimensions, where $p > 0$ on $[0,1]^2$ and $p$ integrates to 1. From Example \ref{ex:convex}, if the coefficients for the terms $(x^{(1)})^2$ and $(x^{(2)})^2$ are non-zero, we then have greedy identifiability for sufficiently large $m$. This example shows that quadratic densities can include interactions while still satisfying marginal visibility.

A quadratic counterexample for marginal visibility is 
\begin{align*}
    p(x) = 1 + \theta\left(1-2x^{(1)}\right)\left(1-2x^{(2)}\right)
\end{align*}
for $\theta \in (-1,1) \setminus \{0\}$ which is the FGM bivariate copula density.  The above quadratic has coefficients zero for $(x^{(1)})^2$ and $(x^{(2)})^2$, so does not satisfy the coordinate-wise convexity condition of  Example \ref{ex:convex}. In fact, as discussed in the main paper, the above has uniform marginals despite not being uniform on $[0,1]^2$, so Assumption \ref{as:greedy identifiable} fails. 

Fortunately, in this example, we can apply a transform coordinate-wise to repair this issue. Consider $y^{(j)} = [\exp(\lambda x^{(j)}) - 1]/[\exp(\lambda) - 1]$ for $\lambda > 0$, which maps each coordinate from  $[0,1] \to [0,1]$. Let $p_Y$ be the transformed density which is no longer a copula, and suppose $H$ is chosen to still be uniform on the $y$ scale. One can then verify with the chain rule that ${\partial^2 {p}_Y(y)}/{\partial (y^{(1)})^2}$ is a positive term times
\begin{align*}
  \lambda + \theta(1-2x^{(2)})[\lambda(1-2x^{(1)}) + 3]
\end{align*}
where $x^{(j)}$ is the suitable inverse transform of $y^{(j)}$. The above term is lower bounded by $\lambda - |\theta| (3+\lambda)$, and is positive for all $y$ if $\lambda > 3|\theta|/(1-|\theta|)$.
Symmetry gives the same bound for $y^{(2)}$, which is then sufficient for the coordinate-wise convexity condition of Example \ref{ex:convex}.

\section{Technical results}\label{app: technical results}
This section develops the complexity and probability bounds,
together with the log marginal likelihood expansion, used in
the proofs of the main theorems.  Throughout this section, we assume that Assumptions~\ref{as:bounded density}, \ref{as:greedy identifiable}, and~\ref{as:decay} hold, with $m\geq2$, $\delta\in(0,1)$ and all constants in the assumptions fixed independently of $n$.

\subsection{Admissible family complexity}\label{sec: complexity admissible}
In this section, we analyze the complexity of the admissible family by considering the contributions from the identifiable and non-identifiable parts separately.  For the identifiable part, we account for the multiple possible near-optimal splitting paths. For the non-identifiable part, we use the gap between strong- and weak-signal nodes to choose a fixed threshold $k_{\max}$ for the number of excess splits. Combining these results with the boundedness of $k_{\max}$, we obtain a bound on the complexity of admissible partitions.

For the remainder of the proof, it is convenient to write the identifiability threshold as
\begin{align*}
\tau_n=\frac{C_0\log n}{n},
\end{align*}
where $C_0> 0$ is a sufficiently large constant. We will initially regard $C_0$ and the positive integer $k_{\max}$ as fixed constants and specify their specific choices shortly below.
\subsubsection{Oracle nodes}\label{sec:oracle_complexity}
We begin by defining the set of oracle nodes below which will be useful for later parts of the proof, before analyzing their complexity in Proposition~\ref{def:complexity_exponents}.
\begin{definition}[Oracle nodes]\label{df: oracle-tree}
For all admissible trees $T\in\mathcal T_n$, truncate each
branch at its first non-identifiable node, and let $\mathcal I_n^*$ and $\mathcal L_n^*$ be the collections of all internal nodes and leaves respectively of all resulting trees. We refer to $\mathcal{I}_n^*$ as the identifiable oracle nodes, and $\mathcal{L}_n^*$ as the non-identifiable oracle leaves.
\end{definition}
Figure~\ref{fig:oracle_tree} shows the oracle nodes contributed by one truncated admissible tree, as described in the first part of Definition~\ref{def: admissible trees}. We now have the complexity bound.
\begin{figure}[!h]
  \centering
  \vspace{5mm}
  \resizebox{0.8\linewidth}{!}{%
\begin{tikzpicture}[
  x=0.85cm, y=1cm,
  nodepoint/.style={circle,draw=black!65,line width=0.65pt,
    fill=white,minimum size=6mm,inner sep=0pt},
  nearoptimal/.style={nodepoint,fill=blue!22!white,draw=blue!45!black},
  branch/.style={draw=black!65,line width=0.7pt},
  every node/.style={font=\small}
]
  % Shaded nodes are identifiable and split near-optimally.
  \node[nearoptimal] (r)  at (0,0) {$\mathcal A$};
  \node[nearoptimal] (a)  at (-4.25,-1.1) {};
  \node[nearoptimal] (b)  at ( 4.25,-1.1) {};
  \node[nodepoint] (a0) at (-5.75,-2.2) {};
  \node[nearoptimal] (a1) at (-2.75,-2.2) {};
  \node[nearoptimal] (b0) at ( 2.75,-2.2) {};
  \node[nearoptimal] (b1) at ( 5.75,-2.2) {};
  % The first non-identifiable nodes are a0 and l3,...,l8.
  \node[nodepoint] (l3) at (-3.5,-3.7) {};
  \node[nodepoint] (l4) at (-2.0,-3.7) {};
  \node[nodepoint] (l5) at ( 2.0,-3.7) {};
  \node[nodepoint] (l6) at ( 3.5,-3.7) {};
  \node[nodepoint] (l7) at ( 5.0,-3.7) {};
  \node[nodepoint] (l8) at ( 6.5,-3.7) {};
  % The dotted line separates strong signals from weak signals.
  \draw[black!55,line width=0.8pt,densely dotted]
    (-7.7,-1.65) -- (-4.9,-1.65)
    .. controls (-4.1,-1.65) and (-4.6,-3.1) .. (-3.8,-3.1)
    -- (7.4,-3.1);
  \node[anchor=west,inner sep=0pt] (identifiability)
    at (7.65,-3.1) {Identifiability};
  \node[yshift=2cm]
    at (identifiability.center) {$\mathcal I_n^*$};
  \node[anchor=north,yshift=-7pt]
    at (identifiability.south) {$\mathcal L_n^*$};
  \node[text=black!65] at (0,-2.45) {$S_A\geq \frac{C_0\log n}{n}$};
  \node[text=black!65] at (0,-3.7) {$S_A<\frac{C_0\log n}{n}$};
  \foreach \parent/\child in {
    r/a,r/b,a/a0,a/a1,b/b0,b/b1,
    a1/l3,a1/l4,b0/l5,b0/l6,b1/l7,b1/l8}
    \draw[branch] (\parent) -- (\child);
  % Legend.
  \node[nearoptimal,minimum size=5mm] at (-5.9,-4.65) {};
  \node[anchor=west] at (-5.45,-4.65) {Near-optimal split};
  \node[nodepoint,minimum size=5mm] at (0.6,-4.65) {};
  \node[anchor=west] at (1.05,-4.65) {Non-identifiable leaf};
\end{tikzpicture}
  }
  \centering
  \caption{Oracle nodes contributed by a truncated admissible tree. Shaded internal nodes belong to $\mathcal I_n^*$, and unshaded leaves belong to $\mathcal L_n^*$.}
  \label{fig:oracle_tree}
\end{figure}
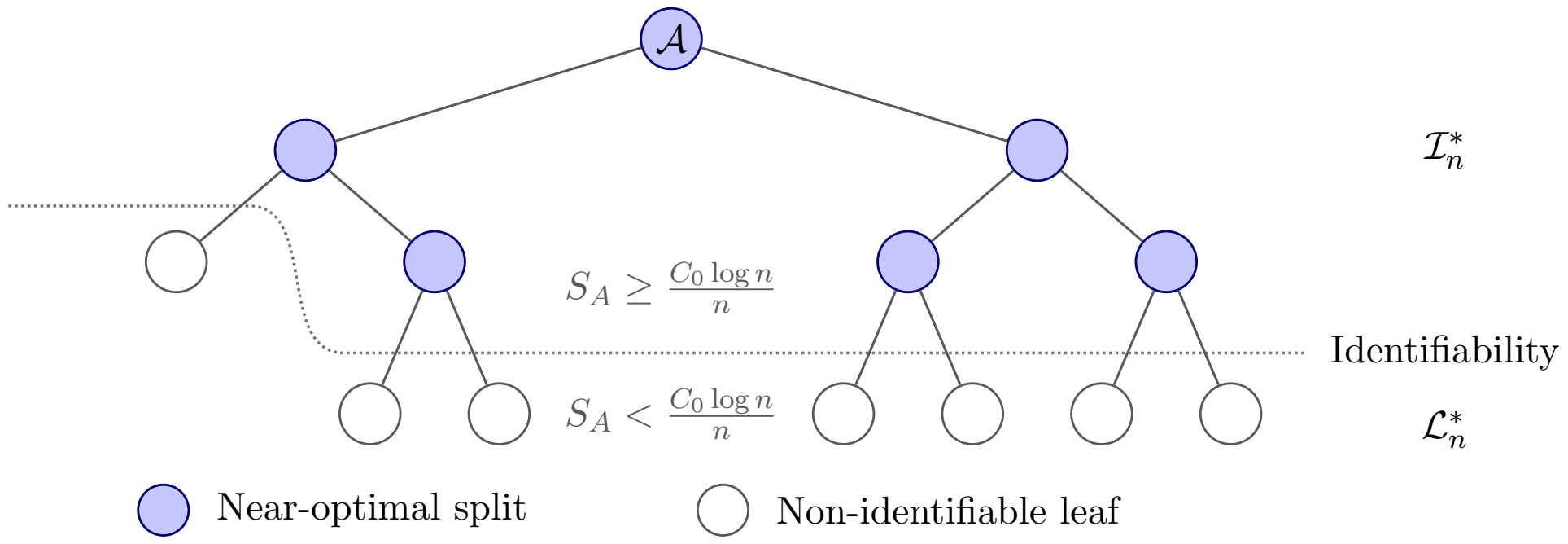
    \begin{proposition}[Complexity exponents]\label{def:complexity_exponents}
     For $n \geq 2$, the oracle nodes satisfy the inequality
    \begin{align*}
        |\mathcal{I}_n^*| + |\mathcal{L}_n^*| &\lesssim \tau_n^{-t_0},
    \end{align*}
    where the exponent is
    \begin{align}\label{eq:s0} 
        t_0=\frac{\log[2D(m-1)]}{\log[m/(m-1)]}.
    \end{align}
\end{proposition}
\begin{proof}
Every grid split reduces the volume by a factor at most
$(m-1)/m$, so a node $A$ at depth $\ell$ satisfies
$H_A\leq [(m-1)/m]^\ell$.
Assumption~\ref{as:bounded density} gives
$S_A\leq P_A\log(C_u/C_l)\lesssim H_A$.
Hence, if $A$ is identifiable, we have
$\tau_n\leq S_A\lesssim [(m-1)/m]^\ell$. Taking logarithms yields
$$
\ell\leq
\frac{\log \tau_n^{-1}}{\log[m/(m-1)]}+C,
$$
where $C>0$ is independent of $n$. Every non-identifiable oracle leaf $A \in \mathcal{L}_n^*$ that is not the root has an identifiable parent in $\mathcal{I}_n^*$,
so the same bound holds for oracle leaves after
increasing $C$ by one. The bound also holds for the root, since $\tau_n\le C_0/e\lesssim1$ for $n\ge2$.
Thus, the maximum depth $\ell_n$ attained by an oracle
node is finite and satisfies
$$
\ell_n\leq
\frac{\log \tau_n^{-1}}{\log[m/(m-1)]}+C. 
$$
Across all candidate trees, there are at most
$[2D(m-1)]^\ell$ distinct nodes at depth $\ell$.
Summing over depths and applying the geometric series
formula therefore gives
\begin{align*}
|\mathcal I_n^*|+|\mathcal L_n^*|
&\leq\sum_{\ell=0}^{\ell_n}[2D(m-1)]^\ell\lesssim[2D(m-1)]^{\ell_n}.
\end{align*}
Applying the upper bound on $\ell_n$ then gives the result. 
\end{proof}
While the above bound works for any $C_0$, we now specify a specific value for the remainder of the proofs. We will see that a sufficiently large choice is
\begin{align}\label{eq: tau_n C_0}
C_0 = 864 \frac{C_u}{C_l}\frac{t_0} {\kappa \delta^2}
\end{align}
where $(C_l,C_u)$ and $\kappa$ are from Assumptions \ref{as:bounded density} and \ref{as:greedy identifiable} respectively, and $\delta$ is from Definition~\ref{df: near-optimal}. Note that since $t_0 \geq 1$, our specific choice satisfies $C_0 > 1$.

\subsubsection{Excess splits}\label{app: splitting bound}
Next, we handle the additional complexity induced by the excess splitting on non-identifiable nodes in the admissible trees. We first introduce the notion of a \emph{stopping split}, which mirrors the definition of identifiable nodes in (\ref{eq:ident}), but is an upper threshold for weak split gains.
\begin{definition}[Stopping split]\label{df: stopping}
    A split $(d,q)$ on $A$ is a {stopping split} if
    $$s_A(d, q) \leq \frac{C_s \log n}{n},$$ where the constant takes the form
    \begin{align*}
        C_s = \frac{1}{40}\frac{C_l}{C_u}.
    \end{align*}
\end{definition}
Intuitively, a stopping split is one where the split gain is too weak relative to the penalty at a given $n$, so the penalty dominates. Again, we have chosen a specific value for convenience which will be used later. For now, we highlight that $C_s < 1 < C_0$, so the stopping signal threshold is below the identifiable one. 

We now leverage the limited amount of splitting that occurs between the identifiability and stopping thresholds to control the non-identifiable part of admissible trees.
\begin{lemma}[Splitting beyond identifiability]\label{lemma:splitting_bound}
    Let $k_{\max}$ be the integer defined as
    \begin{align}\label{eq: k_max}
 k_{\max}=\lceil C_0/C_s\rceil+\lceil15t_0\rceil
    \end{align}
    where $t_0$ is defined in (\ref{eq:s0}). Then any subtree rooted at a non-identifiable node $A$ with \(k_{\max}\) splits contains at least \(\lceil15t_0\rceil\) stopping splits.
\end{lemma}
\begin{proof}
    Since $A$ is non-identifiable, \eqref{eq:ident} upper bounds its signal with
    $$
        S_A <  \frac{C_0\log n}{n}.
    $$
    For any non-stopping split within a node $B \subseteq A$,
    Definition~\ref{df: stopping} guarantees that the split gain satisfies
    $$
        s_B(d, q) >  \frac{C_s\log n}{n}.
    $$
    By the signal decomposition of Lemma~\ref{lem:split_gain}, the sum of all split gains in the subtree is at most $S_A$. Hence the number of non-stopping splits is strictly less than $C_0/C_s$. The remaining splits must thus be stopping splits.
\end{proof}
The choice $\lceil15t_0\rceil$ allows the later probability bounds to offset the oracle node complexity $\tau_n^{-t_0}$. Finally, we bound the number of leaves of an admissible tree.
\begin{lemma}[Leaf count for admissible trees]
\label{lem:admissible_leaf_count}
Let $T\in\mathcal T_n$ be any admissible tree, and let
$\mathcal P$ be its leaf partition. For $n \geq 2$, we have
\begin{align}\label{lemma:leaf_count}
    |\mathcal P|\lesssim\tau_n^{-1/\gamma},
\end{align}
where the implicit constant is independent of $T$ and $n$.
\end{lemma}
\begin{proof}
Consider truncating each branch of $T$ at its first non-identifiable node, as in Definition \ref{df: oracle-tree}. Let $L$ be the leaf partition of the
resulting tree. If the root is identifiable, every $B\in L$ has an
identifiable parent $A$. The grid restriction and
Assumption~\ref{as:bounded density} give $H_B\geq H_A/m\gtrsim P_A$. For $\gamma<\infty$, identifiability and
Assumption~\ref{as:decay} imply $\tau_n\leq S_A\leq(C_1P_A)^\gamma$,
so $P_A\gtrsim\tau_n^{1/\gamma}$. For $\gamma=\infty$,
the same conclusion follows from $P_A\geq C_1^{-1}$.
Thus every $B\in L$ satisfies
$$
H_B\gtrsim\tau_n^{1/\gamma}.
$$
If the root is non-identifiable, then
$ L=\{\mathcal A\}$, and this bound again holds trivially
for $n \geq 2$. Since $ L$ partitions $\mathcal A$, we have
\[
1=\sum_{B\in L}H_B
\gtrsim| L|\tau_n^{1/\gamma},
\]
and hence $| L|\lesssim\tau_n^{-1/\gamma}$.
By admissibility, each $B\in L$ undergoes fewer than
$k_{\max}$ additional splits, producing at most
$k_{\max}$ leaves. Since $k_{\max}$ is fixed, we have
\[
 |\mathcal P|
 \leq k_{\max}|L|
 \lesssim\tau_n^{-1/\gamma}. \qedhere
\]
\end{proof}

\subsubsection{Partition complexity}
Lemma~\ref{lem:admissible_leaf_count} controls the leaf count
uniformly over all admissible trees. We now control the
complexity of the possible partitions across these trees.
\begin{lemma}[Partition complexity]\label{lemma:complexity}
Let $\mathscr B_n$ be the collection of all sets that can
be expressed as a union of leaves of a single admissible
tree $T\in\mathcal T_n$. For $n\geq 2$, we have
\begin{align}\label{lemma:B_P_n}
    \log|\mathscr B_n|
    \lesssim\tau_n^{-1/\gamma}\log n.
\end{align}
\end{lemma}
\begin{proof}
Starting from a node $A$, consider an ordered sequence of
$k$ grid splits, each acting on a current leaf. Before the
$j$-th split, there are exactly $j$ leaves, each admitting
$D(m-1)$ candidate splits. Hence the number of such
sequences is at most
\begin{align}\label{eq: subtree complexity}
\prod_{j=1}^k jD(m-1)=k![D(m-1)]^k.
\end{align} 
This also holds for $k=0$, when there is one empty sequence.

For the global bound, let $k_n$ be the maximum number
of leaves of an admissible tree, where Lemma~\ref{lem:admissible_leaf_count} gives $k_n\lesssim\tau_n^{-1/\gamma}$ if $n \geq 2$.
Each admissible tree has at most $k_n-1$ internal nodes.
For each $k=0,\ldots,k_n-1$, there are at most
$k![D(m-1)]^k$ ordered split sequences, each producing
$k+1$ leaves and $2^{k+1}$ cell unions. Therefore,
\begin{align*}
|\mathscr B_n|
&\leq\sum_{k=0}^{k_n-1}k![D(m-1)]^k2^{k+1} \leq k_n![D(m-1)]^{k_n}2^{k_n},
\end{align*}
where the last inequality follows by bounding each of
the $k_n$ summands by the largest term and using
$D(m-1)\geq1$.
Taking logarithms gives
$$
\log|\mathscr B_n|
\leq k_n\log\{2D(m-1)k_n\}
\lesssim\tau_n^{-1/\gamma}\log n,
$$
where the first inequality used $\log(k_n!) \leq k_n \log k_n$. 
\end{proof}

\subsection{Nodewise decision conditions}\label{app: sufficient conditions}
This section contains a technical study of the decomposition \eqref{eq:splitBF_decomp}, where we formally control the size of the error. We then summarize the sufficient conditions for near-optimal splitting and stopping based on the decomposition of the split Bayes factor. 
\subsubsection{Key events}\label{app: E_{A, n}}
To control the error term in \eqref{eq:splitBF_decomp}, we will need to define useful events on the counts within a node $A$.  To that end, for a node $A$ with split $(d,q)$ and left child $A_0$, we write the empirical estimates of the node probabilities for $N_A > 0$ as
\begin{align*}
    \widehat{P}_A = \frac{N_A}{n}, \qquad \widehat{P}_{A_0 \mid A} = \frac{N_{A_0}}{N_A},
\end{align*}
where $N_A  = \sum_{i = 1}^n\mathbbm{1}(X_i \in A)$ with $X_i \overset{\mathrm{i.i.d.}}{\sim} P$. For convenience, let $\widehat{P}_A = \widehat{P}_{A_0 \mid A} = 0$ if $N_A = 0$. For a given $A$ and $(d,q)$, in the events below, $N_A$, $N_{A_0}$, $\widehat{P}_A$ and $\widehat{P}_{A_0 \mid A}$ are random variables.
\begin{definition}[Node count event]\label{def:E_{A, n}}
    For a node $A$ with $P_A > 0$, let
    \begin{align*}
        \mathcal E_{A, n} := \left\{ \left|\frac{\widehat{P}_A}{P_A} - 1 \right| \leq \frac{1}{\log n} \right\}.
    \end{align*}
\end{definition}
\newpage
\begin{definition}[Split count event]\label{def:D_events}
Let $A$ be a node with $P_A > 0$, and $(d,q)$ be a split of $A$ resulting in left child $A_0$. For a finite constant $t > 0$, let
    \begin{align*}
        \mathcal{D}_{A, n}(t, d, q) &:= \left\{N_A > 0, \quad   \frac{\left(\widehat{P}_{A_0 \mid A} - P_{A_0 \mid A}\right)^2}{P_{A_0 \mid A}\,(1-P_{A_0 \mid A})} < 6t \frac{\log n}{N_A} \right\}, \quad \widetilde{\mathcal{D}}_{A, n}(t) := \bigcap_{d,q} \mathcal{D}_{A, n}(t, d, q),
    \end{align*}
where the intersection ranges over all candidate $m$-grid splits of $A$.
\end{definition}
The first event controls the accuracy of the empirical node
probability, while the split count events control the
empirical conditional child probabilities. Note that  $\widetilde{\mathcal{D}}_{A, n}(t)$ corresponds to the inequality holding uniformly over all candidate splits. The choice of $t$ will aid with technical proofs later, with splitting using $t = t_0$ and stopping using the smaller $t = 1/15$. Section \ref{app: nodewise split and stop prob} studies the probability of the above events. 

We now state a lemma summarizing some regularity properties under the above events, which is useful for later proofs.
\begin{lemma}\label{lem:reg}
Let $t>0$ be a fixed constant. There exists a constant $b\in(0,1/2)$
such that, for sufficiently large $n$, the following
properties hold uniformly over nodes $A$ of admissible
trees and candidate splits $(d,q)$ of $A$.

The node probabilities satisfy the deterministic bounds
\begin{align}\label{lemma:admissible_mass_lower}
    P_A \gtrsim \frac{(\log n)^5}{n}, \qquad 
    P_{A_0\mid A}
    &\in[b,1-b].
\end{align}
On the event $\mathcal{E}_{A,n}$, we then have
\begin{align}\label{cor:N_A_lower}
    \widehat P_A \gtrsim \frac{(\log n)^5}{n}.
\end{align}
Finally, on $\mathcal{E}_{A,n}\cap\mathcal D_{A,n}(t,d,q)$, we have
\begin{align}
    \left|\widehat P_{A_0\mid A}-P_{A_0\mid A}\right|
    &\lesssim (\log n)^{-2}, \qquad \widehat P_{A_0\mid A}\in[b,1-b].
  \label{lemma:delta_bound_under_Dc}\end{align}
\end{lemma}
\begin{proof}
We begin with the deterministic results.
Let $B$ be the last identifiable node on the path from the
root to $A$, including $A$ itself. If there is no such node,
take $B$ to be the root. In either case,
$P_B\gtrsim\tau_n^{1/\gamma}$, as for an identifiable $B$ this
follows from $S_B\geq\tau_n$ and
Assumption~\ref{as:decay}, while for the root $P_B=1$.

By Definition~\ref{def: admissible trees}, the path from
$B$ to $A$ contains at most $k_{\max}$ splits. Each split
retains at least $1/m$ of the parent volume, so
Assumption~\ref{as:bounded density} gives
$$
P_A
\geq \frac{C_l}{C_u}m^{-k_{\max}}P_B
\gtrsim \tau_n^{1/\gamma}.
$$
Since $\tau_n=C_0\log n/n$ and $1/\gamma<1$, we have
$n\tau_n^{1/\gamma}/(\log n)^5\to\infty$. Hence, for sufficiently large $n$, we have $P_A\gtrsim {(\log n)^5}/{n}$ uniformly
over admissible nodes.  
Next, the $m$-grid restriction and
Assumption~\ref{as:bounded density} imply
$$
\frac{C_l}{mC_u}
\leq P_{A_0\mid A}
\leq 1-\frac{C_l}{mC_u}.
$$
We now proceed to the implications of the events. On $\mathcal{E}_{A,n}$, we have
$\widehat P_A/P_A\in[1-1/\log n,1+1/\log n]$,
so for $n \geq 3$, we have $\widehat P_A\asymp P_A$. Combining this with the preceding probability bound gives
\eqref{cor:N_A_lower} uniformly over $A$ for sufficiently large $n$. 

If $\mathcal D_{A,n}(t,d,q)$ also holds,  $P_{A_0\mid A}(1-P_{A_0\mid A})\leq1/4$ gives
$$
\left|\widehat P_{A_0\mid A}-P_{A_0\mid A}\right|
\leq \sqrt{\frac{3t\log n}{2N_A}}
\lesssim (\log n)^{-2}
$$
since $N_A=n\widehat P_A\gtrsim(\log n)^5$ on $\mathcal{E}_{A,n}$
by \eqref{cor:N_A_lower}, and $t>0$ is fixed. Finally, $\left|\widehat P_{A_0\mid A}-P_{A_0\mid A}\right|$ will eventually be sufficiently small uniformly over $A$ and $(d,q)$, so we can choose $b = C_l/(2mC_u)$.
\end{proof}

\subsubsection{Decomposition error control}\label{app: decomposition}
We now turn to the key technical lemma for controlling the error term in the split Bayes factor expansion from (\ref{eq:splitBF_decomp}).
\begin{lemma}\label{theorem:decompose-log-marginal-likelihood}
Fix a constant $0<t\leq t_0$. For sufficiently large $n$,
the following bound holds uniformly over nodes $A$ of an admissible tree $T\in\mathcal T_n$ and candidate splits $(d,q)$ whenever the event
$\mathcal E_{A,n}\cap\mathcal D_{A,n}(t,d,q)$ holds:
\[
 |\varepsilon_{A, n}(d,q)|
 \leq
 2\sqrt3\left(\frac{C_u}{C_l}\right)^{1/2}
 \sqrt{s_A(d,q)}\sqrt{\frac{tP_A\log n}{N_A}}
 +3\frac{tP_A\log n}{N_A}
 +\frac{C}{n},
\]
where $C_l,C_u$ are from Assumption~\ref{as:bounded density}, and $C$ is independent of $A$, $(d,q)$ and $n$.
\end{lemma}
\begin{proof}
Throughout the proof, we work on
$\mathcal E_{A,n}\cap\mathcal D_{A,n}(t,d,q)$. All thresholds on $n$ below can be chosen independently of $A$ and $(d,q)$, since the bounds and constants used are uniform.

To recover the split gain and penalty in \eqref{eq:splitBF_decomp},
we first approximate the two beta functions in \eqref{eq: node_bf}, which will also yield the expression for $\varepsilon_{A,n}(d,q)$. We then bound its components separately under the count events and apply the triangle inequality to give the target bound.

For the prior contribution term $-\log B(\alpha H_{A_0},\alpha H_{A_1})$,
we use properties of the gamma function. Writing $z=\alpha H_A$
and $r=H_{A_0\mid A}$, the identity
$\Gamma(1+z)=z\Gamma(z)$ gives
\begin{align*}
    \log\frac{\Gamma(z)} {\Gamma(zr)\, \Gamma(z(1-r))}
 =\log H_A+
 \log\left\{
 \frac{\alpha r(1-r)\,\Gamma(1+z)}
      {\Gamma(1+zr)\,\Gamma(1+z(1-r))}
 \right\}.
\end{align*}
The first term is our approximation, while the second term is
continuous on the compact set $[0,\alpha]\times[1/m,1-1/m]$ as a function of $(z,r)$, so its absolute value is bounded by a constant depending only on $m$ and $\alpha$.

For the numerator beta function, Stirling's formula for the gamma function gives
\begin{align*}
    &\log B\left(\alpha H_{A_0} + N_{A_0}, \alpha H_{A_1} + N_{A_1}\right) \\
    &\quad = \left[\sum_{i=0}^1 N_{A_i}\log N_{A_i} - N_A \log N_A\right]\\
    &\quad + \left[ \sum_{i=0}^1 \left(\alpha H_{A_i}-\frac{1}{2}\right)\log N_{A_i} - \left(\alpha H_{A}-\frac{1}{2}\right)\log N_{A} \right] + R_{A,n}(d,q).
\end{align*}
Henceforth for convenience, the notation $R_{A,n}(d,q)$ is used to denote a remainder that is bounded uniformly in $A$, $(d,q)$, and $n$, where its meaning may change between lines. In the above, the remainder absorbs the constant $\frac12\log(2\pi)$,
the errors in Stirling's approximation, and the errors from
replacing $\log(N_B+\alpha H_B)$ with $\log N_B$
for $B\in\{A_0,A_1,A\}$. 

We now proceed to show the remainder is indeed uniformly bounded. First, \eqref{cor:N_A_lower} and
\eqref{lemma:delta_bound_under_Dc} ensure that
$N_B\gtrsim(\log n)^5$ uniformly for sufficiently large $n$.
The absolute error in Stirling's expansion of
$\log\Gamma(N_B+\alpha H_B)$ is bounded by
$$
\frac{1}{12(N_B+\alpha H_B)}
\lesssim(\log n)^{-5},
$$
and is therefore uniformly bounded. Additionally, we now show that replacing $\log(N_B+\alpha H_B)$ with $\log N_B$
incurs a uniformly bounded error. For sufficiently large $n$,
the uniform count lower bound ensures $N_B\geq\max\{1,\alpha\}$.
Since $H_B\leq1$, we have
$N_B+\alpha H_B-\frac12>0$ and $\alpha H_B\leq N_B$.
Using $\log(1+x)\leq x$ for $x\geq0$, we therefore obtain
$$
0\leq
\left(N_B+\alpha H_B-\frac12\right)
\log\left(1+\frac{\alpha H_B}{N_B}\right)
\leq (N_B+\alpha H_B)\frac{\alpha H_B}{N_B}
\leq2\alpha H_B\leq2\alpha.
$$

Next, we seek to combine the two terms in the square brackets. The first part equals a log-likelihood:
$$
 \sum_{i=0}^1N_{A_i}\log N_{A_i}-N_A\log N_A
 =\sum_{i=0}^1N_{A_i}\log\widehat P_{A_i\mid A}.
$$
In the second bracket, $\mathcal E_{A,n}$ gives
$nP_A/2\leq N_A\leq2nP_A$ for sufficiently large $n$,
while \eqref{lemma:delta_bound_under_Dc} gives
$bN_A\leq N_{A_i}\leq N_A$. Together with
$C_lH_A\leq P_A\leq C_uH_A$, these imply
\[
\frac{bC_l}{2}
\leq \frac{N_B}{nH_A}
\leq 2C_u,
\qquad B\in\{A_0,A_1,A\}.
\]
Thus $\log N_B-\log(nH_A)=\log\{N_B/(nH_A)\}$ is
uniformly bounded. Since the coefficients
$\alpha H_B-1/2$ are also uniformly bounded, replacing
each $\log N_B$ in the second bracket with $\log(nH_A)$
incurs only a uniformly bounded error. The resulting coefficient of $\log(nH_A)$ is then
$$
\sum_{i=0}^1\left(\alpha H_{A_i}-\frac12\right)
-\left(\alpha H_A-\frac12\right)=-\frac12.
$$
Combining the two beta function approximations with the
remaining term in \eqref{eq: node_bf} gives
\begin{align*}
\log\mathcal B_A(d,q)
&=\sum_{i=0}^1N_{A_i}
  \log\frac{\widehat P_{A_i\mid A}}{H_{A_i\mid A}}
  +\log H_A-\frac12\log(nH_A)+R_{A,n}(d,q)\\
&=N_AK(\widehat P_{A_0\mid A})
  -\frac12\log\frac n{H_A}+R_{A,n}(d,q),
\end{align*}
where $K(u)$ for $u \in (0,1)$ is the Bernoulli KL divergence defined as 
\begin{align*}
    K(u) &= u \log \frac{u}{H_{A_0 \mid A}} + (1 - u) \log \frac{1-u}{1-H_{A_0 \mid A}}.
\end{align*}
Finally, \eqref{lemma:admissible_mass_lower} and
Assumption~\ref{as:bounded density} give
$H_A \geq 1/n$ for sufficiently large $n$, which implies $\log(n/H_A)\leq2\log n$.
On $\mathcal E_{A,n}$, inserting the factor
$\widehat P_A/P_A$ into the penalty therefore changes it
by at most
$$
\frac12\left|\frac{\widehat P_A}{P_A}-1\right|
\log\frac n{H_A}\leq1.
$$
Absorbing this uniformly bounded error into $R_{A,n}(d,q)$ gives
\begin{equation}\label{eq: empirical KL decomposition}
\log\mathcal B_A(d,q)
=N_AK(\widehat P_{A_0\mid A})
-\frac{\widehat P_A}{2P_A}\log\frac n{H_A}
+R_{A,n}(d,q),
\end{equation}
which yields the desired penalty in \eqref{eq:splitBF_decomp}
after multiplication by $P_A/N_A$.

Armed with the above, we next expand $K(\widehat{P}_{A_0 \mid A})$ about $P_{A_0 \mid A}$. Let $\Delta_n=\widehat P_{A_0\mid A}-P_{A_0\mid A}$, suppressing its dependence on $A$ and $(d,q)$, and note that $K$ has the derivatives
\begin{align}\label{eq: K_u and derivatives}
    K'(u) &= \log \frac{u (1-H_{A_0 \mid A})}{(1 - u) H_{A_0 \mid A}}, \qquad K''(u) = \frac{1}{u (1 - u)}, \qquad K'''(u) = \frac{2u - 1}{u^2(1-u)^2}.
\end{align}
 Since $\widehat P_{A_0\mid A}=P_{A_0\mid A}+\Delta_n$ and
$K(P_{A_0\mid A})=s_A(d,q)/P_A$, we can write the quadratic expansion:
\begin{align*}
 K(\widehat P_{A_0 \mid A}) = \frac{s_A(d,q)}{P_A}
 +K'(P_{A_0 \mid A})\,\Delta_n + \frac{1}{2}K''(P_{A_0 \mid A})\,\Delta_n^2 + R_n.
\end{align*}
By \eqref{lemma:admissible_mass_lower} and \eqref{lemma:delta_bound_under_Dc}, the segment between
$P_{A_0\mid A}$ and $\widehat P_{A_0\mid A}$ is away from the boundaries, so $K'''(u)$ is uniformly bounded on this interval. Taylor's theorem then gives $|R_n|\lesssim|\Delta_n|^3$. Moreover, Definition~\ref{def:D_events} gives
$N_A\Delta_n^2\lesssim\log n$, while
\eqref{lemma:delta_bound_under_Dc} gives
$|\Delta_n|\lesssim(\log n)^{-2}$. Hence
\[
 N_A|R_n|
 \lesssim N_A\Delta_n^2|\Delta_n|
 \lesssim\frac1{\log n},
\]
and we can absorb $N_A R_n$ into $R_{A,n}(d,q)$ in \eqref{eq: empirical KL decomposition}.

Multiplying both sides of \eqref{eq: empirical KL decomposition}
by $P_A/N_A$ and substituting the Taylor expansion gives the
decomposition in \eqref{eq:splitBF_decomp}, with error
\begin{align*}
    \varepsilon_{A, n}(d, q) = P_A \left[ K'(P_{A_0 \mid A}) \Delta_n + \frac{1}{2}K''(P_{A_0 \mid A})\Delta_n^2 +\frac{R_{A,n}(d,q)}{N_A} \right].
\end{align*}
We proceed to bound the three distinct terms above. On $\mathcal E_{A, n}$ for sufficiently large $n$, the earlier bound $N_A\geq nP_A/2$ and the uniform boundedness of $R_{A,n}(d,q)$ give
\begin{align}
    \left|\frac{P_A R_{A,n}(d,q)}{N_A}\right| \lesssim \frac{1}{n}.
\end{align}
The quadratic noise is directly controlled by the definition of
$\mathcal D_{A,n}(t,d,q)$:
\begin{equation}\label{eq:nodewise_quadratic_bound}
 \frac{P_A}{2}K''(P_{A_0\mid A})\Delta_n^2
 <3\frac{tP_A\log n}{N_A}.
\end{equation}
We now control the linear term.
By Assumption~\ref{as:bounded density}, any $u$ between
$H_{A_0\mid A}$ and $P_{A_0\mid A}$ satisfies
\begin{align*}
u
\geq \min\{H_{A_0\mid A},P_{A_0\mid A}\}
\geq \frac{C_l}{C_u}P_{A_0\mid A}, 
\end{align*}
and similarly $1-u \geq ({C_l}/{C_u})(1-P_{A_0\mid A})$. Hence \eqref{eq: K_u and derivatives} gives
\[
K''(u)=\frac1u+\frac1{1-u}
\leq\frac{C_u}{C_l}
\left(\frac1{P_{A_0\mid A}}+\frac1{1-P_{A_0\mid A}}\right)
=\frac{C_u}{C_l}K''(P_{A_0\mid A}).
\]
Recall that $K(H_{A_0\mid A})=K'(H_{A_0\mid A})=0$.
If $P_{A_0\mid A}=H_{A_0\mid A}$, the linear term therefore
vanishes. Otherwise, Cauchy's mean value theorem applied
to $(K')^2$ and $K$ gives some $\xi$ strictly between
$P_{A_0\mid A}$ and $H_{A_0\mid A}$ such that
$$
\frac{|K'(P_{A_0\mid A})|^2 - 0}{K(P_{A_0\mid A})- 0}
=
\frac{2K'(\xi)K''(\xi)}{K'(\xi)}
=2K''(\xi)
\leq2\frac{C_u}{C_l}K''(P_{A_0\mid A}).
$$
Here $K'(\xi)\neq0$ because $K''>0$ and
$K'(H_{A_0\mid A})=0$. Using $K(P_{A_0\mid A})=s_A(d,q)/P_A$, rearranging the above gives
$$
|K'(P_{A_0\mid A})|
\leq\sqrt2\left(\frac{C_u}{C_l}\right)^{1/2}
\sqrt{\frac{s_A(d,q)}{P_A}}\,
\sqrt{K''(P_{A_0\mid A})}.
$$
Multiplying the above by $P_A|\Delta_n|$
and applying \eqref{eq:nodewise_quadratic_bound} gives
$$
|P_AK'(P_{A_0\mid A})\Delta_n|
\leq2\sqrt3\left(\frac{C_u}{C_l}\right)^{1/2}
\sqrt{s_A(d,q)}
\sqrt{\frac{tP_A\log n}{N_A}}.
$$
The triangle inequality completes the proof.
\end{proof}

Lemma \ref{theorem:decompose-log-marginal-likelihood} crucially controls the error in the split Bayes factor decomposition from (\ref{eq:splitBF_decomp}) under the events on the node and split counts. For fixed $t$, we see that if $s_A(d,q)$ is large relative to $\log n/n$, it will dominate this error. Conversely, when $s_A(d,q)=0$, the linear term vanishes and the remaining error is
$O(t\log n/n)$, allowing the stopping penalty to dominate
for sufficiently small $t$.

\subsubsection{Near-optimal splitting and stopping}\label{app: split and stop sufficient}
We can now leverage Lemma \ref{theorem:decompose-log-marginal-likelihood} to connect the events $\mathcal{E}_{A,n}$ and $\mathcal{D}_{A,n}$ to splitting and stopping of the GEB algorithm. Throughout this subsection, for any node $A$, let $(\hat{d}, \hat{q})$ maximize $\mathcal{B}_A(d,q)$ over all $D(m-1)$ candidate splits. The maximizer need not be unique, so we fix a deterministic tie-breaking rule (the choice is unimportant), such as choosing the smallest $d$ and then the smallest $q$ among maximizers. For fixed $A$, the selected pair $(\hat d,\hat q)$ is then a random discrete vector. We also highlight that $\mathcal{B}_A(d,q)$ is random and depends on $n$, but we omit the $n$ dependence to match the main paper.

We highlight that this definition of $(\hat{d},\hat{q})$ is slightly different to Algorithm \ref{alg:geb}, where the algorithm only splits eligible nodes $A$ at eligible splits. Instead, we handle eligibility separately using the event defined below,
$$\mathcal{Q}_{A,n} = \bigcap_{d,q}\left\{N_{A_0}\geq1,\;N_{A_1}\geq1\right\},$$
where the intersection is over all $D(m-1)$ candidate splits of $A$. It is clear that if $\mathcal Q_{A,n}$ holds and $\mathcal{B}_A(\hat{d}, \hat{q}) > 1$, then $(\hat{d},\hat{q})$ defined above is equivalent to the split chosen by the GEB algorithm at node $A$, as all splits are eligible. 
\begin{lemma}[Sufficient condition for near-optimal splitting]
\label{lemma:main_sufficient_condition}
For each identifiable oracle node $A\in\mathcal I_n^*$,
define the event  $\mathcal{G}_{A,n}$ as
\begin{align*}
 \mathcal G_{A,n}
 :=\left\{
 s_A(\widehat d,\widehat q)
 \geq(1-\delta)\max_{d,q}s_A(d,q)
 \right\}
 \cap\{\log\mathcal B_A(\widehat d,\widehat q)>0\}\cap \mathcal{Q}_{A,n},
\end{align*}
where the maximum is over all candidate $m$-grid splits at $A$. Then for sufficiently large $n$, the following holds uniformly over the identifiable oracle nodes $A\in\mathcal I_n^*$: 
$$
 \mathcal{E}_{A,n}\cap \widetilde{\mathcal {D}}_{A,n}(t_0)\subseteq\mathcal G_{A,n}.
$$
\end{lemma}
\begin{proof}
All thresholds on $n$ in this proof can be chosen independently
of $A$ and $(d,q)$, since the bounds and constants used below
are uniform. Fix $A\in\mathcal I_n^*$ and work on
$\mathcal E_{A,n}\cap\widetilde{\mathcal D}_{A,n}(t_0)$,
which in particular ensures $N_A>0$.

We begin with the first near-optimal condition. Let $(d^*,q^*)$ maximize the population split gain $s_A(d,q)$, with the same deterministic tie-breaking rule. 
The penalty in \eqref{eq:splitBF_decomp} is common to all candidate splits, so we have the following split gain gap for any split $(d,q)$:
\begin{align*}
    \frac{P_A}{N_A}
 \{\log\mathcal B_A(d^*,q^*)-\log\mathcal B_A(d,q)\}=
 s_A(d^*,q^*)-s_A(d,q)
 +\varepsilon_{A, n}(d^*,q^*)-\varepsilon_{A, n}(d,q).
\end{align*}
We now show that for sufficiently large $n$, and uniformly over $A\in \mathcal{I}_n^*$, any candidate split that is not near-optimal will not be a maximizer. Specifically, suppose $(d',q')$ is {not} near-optimal, so $s_A(d',q') < (1-\delta) s_A(d^*,q^*)$. Then we have
\begin{align*}
       \frac{P_A}{N_A}
 \{\log\mathcal B_A(d^*,q^*)-\log\mathcal B_A(d',q')\} > \delta s_A(d^*,q^*) - 2\sup_{d,q}|\varepsilon_{A,n}(d,q)|. 
\end{align*}
From the above, it is clear that if 
\begin{equation}\label{eq:identification_error_bound}
 \sup_{d,q}|\varepsilon_{A, n}(d,q)|
 \leq\frac{\delta}{2}s_A(d^*,q^*),
\end{equation}
then $\log\mathcal B_A(d^*,q^*)-\log\mathcal B_A(d',q') > 0$, so $(d',q')$ cannot be chosen. We now proceed to show that the above holds true for sufficiently large $n$ under $\mathcal{E}_{A,n}\cap \widetilde{\mathcal {D}}_{A,n}(t_0)$.

We will rely on Assumption~\ref{as:greedy identifiable} and identifiability which give
\begin{equation}\label{eq:identifiable_gain_lower}
 s_A(d^*,q^*)\geq\kappa S_A
 \geq\frac{\kappa C_0\log n}{n}.
\end{equation}
On $\mathcal{E}_{A,n}\cap\widetilde{\mathcal D}_{A,n}(t_0)$,
Lemma~\ref{theorem:decompose-log-marginal-likelihood} with $t=t_0$ upper bounds $|\varepsilon_{A,n}(d,q)|$, and we now allocate $\delta s_A(d^*,q^*)/6$ to each of the bound's three terms. For the third term, choosing $ n\geq \exp\!\left\{{6C}/({\delta\kappa C_0})\right\}$ together
 with \eqref{eq:identifiable_gain_lower} gives the following bound uniformly over identifiable $A$:
$$
 \frac{C}{n}\leq\frac{\delta}{6}s_A(d^*,q^*).
$$
For the second error term, under $\mathcal{E}_{A,n}$, we can choose $n$ sufficiently large so that $P_A/N_A\leq2/n$. 
The second term in the error then satisfies
$$
 3t_0\frac{P_A\log n}{N_A}
 \leq6t_0\frac{\log n}{n}
 \leq\frac{\delta}{6}s_A(d^*,q^*),
$$
where the last inequality follows from
\eqref{eq:identifiable_gain_lower} and
$\kappa C_0\geq36t_0/\delta$ from the choice of $C_0$ in (\ref{eq: tau_n C_0}). For the first term, we use $s_A(d,q)\leq s_A(d^*,q^*)$ and
$\log n/n\leq s_A(d^*,q^*)/(\kappa C_0)$ from
\eqref{eq:identifiable_gain_lower} which gives
\begin{align*}
 &2\sqrt3\left(\frac{C_u}{C_l}\right)^{1/2}
       \sqrt{s_A(d,q)}
       \sqrt{\frac{t_0P_A\log n}{N_A}}\leq
 2\sqrt6\left(\frac{C_u}{C_l}\right)^{1/2}
       \sqrt{\frac{t_0}{\kappa C_0}}\,s_A(d^*,q^*)
 =\frac{\delta}{6}s_A(d^*,q^*),
\end{align*}
again using $P_A/N_A \leq 2/n$ and the choice of $C_0$ in
\eqref{eq: tau_n C_0}. The above equality motivates our exact choice of $C_0$. The three bounds prove
\eqref{eq:identification_error_bound}, so every maximizer of $\log \mathcal{B}_A(d,q)$ must be 
near-optimal.

We now handle the event $\{\log \mathcal{B}_A(\hat{d}, \hat{q}) > 0\}$, where $n$ is sufficiently large so all the above bounds hold. Assumption~\ref{as:bounded density} and
\eqref{lemma:admissible_mass_lower} give
$H_A\gtrsim(\log n)^5/n$, so $H_A>1/n$ for sufficiently
large $n$. Using \eqref{eq:identifiable_gain_lower} and
$\kappa C_0>2/\delta$ from \eqref{eq: tau_n C_0}, we obtain
$$
\frac1{2n}\log\frac n{H_A}
<\frac{\log n}{n}
\leq\frac{s_A(d^*,q^*)}{\kappa C_0}
<\frac{\delta}{2}s_A(d^*,q^*).
$$
Therefore, the split Bayes factor expansion in \eqref{eq:splitBF_decomp} and \eqref{eq:identification_error_bound} give
\begin{align*}
 \frac{P_A}{N_A}\log\mathcal B_A(\widehat d,\widehat q)
 &\geq\frac{P_A}{N_A}\log\mathcal B_A(d^*,q^*)\\
 &\geq s_A(d^*,q^*)-\frac1{2n}\log\frac n{H_A}
       -\frac{\delta}{2}s_A(d^*,q^*)\\
 &>(1-\delta)s_A(d^*,q^*)>0.
\end{align*}
Finally, Lemma~\ref{lem:reg} gives
$\widehat P_{A_0\mid A}\in[b,1-b]$ for every candidate
split, ensuring that both children contain observations, so $\mathcal Q_{A,n}$ holds. 
\end{proof}

The next lemma gives a sufficient condition for ruling out a
candidate split under the GEB algorithm, either because it is
ineligible or because its Bayes factor is at most one. 
\begin{lemma}[Sufficient condition for stopping]\label{lemma:stopping}
For any node $A$ of an admissible tree $T\in\mathcal T_n$
and candidate split $(d,q)$, define the event
$$
\mathcal W_A(d,q)
:=\{\log\mathcal B_A(d,q)\leq0\} \cup \{N_{A_0}\geq 1, N_{A_1}\geq 1\}^c.
$$
Then, for sufficiently large $n$, the following holds
uniformly over such nodes and stopping splits $(d,q)$
 satisfying
$s_A(d,q)\leq C_s\log n/n$:
$$
\mathcal E_{A,n}\cap\mathcal D_{A,n}(1/15,d,q)
\subseteq\mathcal W_A(d,q).
$$
\end{lemma}
\vspace{-5mm}
\begin{proof}
As before, all thresholds on $n$ in this proof can be chosen independently of $A$ and $(d,q)$, and we work on
$\mathcal E_{A,n}\cap {\mathcal D}_{A,n}(1/15,d,q)$,
which in particular ensures $N_A>0$. Since $H_A\leq1$, the penalty in \eqref{eq:splitBF_decomp}
is at least $\log n/(2n)$ in absolute value, so
$$
 \frac{P_A}{N_A}\log\mathcal B_A(d,q)
 \leq-\frac{\log n}{2n}+s_A(d,q)+|\varepsilon_{A, n}(d,q)|.
$$
We thus seek to show that the above term is non-positive under our assumptions. We now carry out a similar exercise as in Lemma \ref{lemma:main_sufficient_condition} and study the split gain and the three error terms in Lemma \ref{theorem:decompose-log-marginal-likelihood}. For the third term, we can choose $n$ sufficiently large so that
\begin{align*}
    \frac{C}{n} \leq \frac{\log n}{40n}.
\end{align*}
For the second term, on $\mathcal{E}_{A,n}$, we can again choose $n$ sufficiently large so that $P_A/N_A \leq 5/(4n)$. This then gives the second term as
$$
 \frac{3}{15}\frac{P_A\log n}{N_A}
 \leq\frac14\frac{\log n}{n}.
$$
We then handle the first term, where the stopping condition in Definition~\ref{df: stopping} gives
$s_A(d,q)\leq C_s\log n/n$. Applying Lemma~\ref{theorem:decompose-log-marginal-likelihood} with $t=1/15$, the above bounds on $s_A(d,q)$ and $P_A/N_A$ control its first term with
\begin{align*}
 2\sqrt3\left(\frac{C_u}{C_l}\right)^{1/2}
       \sqrt{s_A(d,q)}\sqrt{\frac{P_A\log n}{15N_A}}
 &\leq\left(\frac{C_u}{C_l}\right)^{1/2}
       \sqrt{C_s}\,\frac{\log n}{n} =\frac1{\sqrt{40}}\frac{\log n}{n}
 \leq\frac15\frac{\log n}{n}
\end{align*}
where the final equality uses the definition of $C_s$. Consequently, the above error bound and the fact that $s_A(d,q) \leq \log n/ (40n)$ since $C_s \leq 1/40$ give
$$
 \frac{P_A}{N_A}\log\mathcal B_A(d,q)
 \leq-\left[\frac12-\frac1{40} - \left(\frac{1}{40}+\frac15+\frac14\right)\right]\frac{\log n}{n} \leq 0
$$
for sufficiently large $n$. Since $P_A/N_A>0$, this implies
$\log\mathcal B_A(d,q)\leq0$, so its enlargement $\mathcal{W}_A(d,q)$ also holds.
\end{proof}
\subsection{Nodewise probabilistic properties}\label{app: nodewise split and stop prob}
We have derived interpretable sufficient conditions for both splitting and stopping in the previous section, which ultimately depend on the two count events from Definitions~\ref{def:E_{A, n}} and \ref{def:D_events}. We now proceed to obtain probabilistic bounds on near-optimal splitting and stopping.

\subsubsection{Probabilistic bounds for count events}\label{sec: bounds for two events}
We begin with a technical lemma on a probabilistic bound for the count events in Definitions~\ref{def:E_{A, n}} and \ref{def:D_events}.
\begin{lemma}
\label{lem:main_ep}
Fix a constant $t>0$. For sufficiently large $n$, the following probabilistic bounds hold uniformly over nodes $A$ of admissible
trees and candidate splits $(d,q)$. First,
\[
\mathbb P(\mathcal E_{A,n}^c)\leq n^{-\log n}.
\]
Second, for every integer \(k\in\{1,\ldots,n\}\) with
\(\mathbb P(N_A=k)>0\) satisfying the condition of
\(\mathcal E_{A,n}\), namely
\[
\left|\frac{k}{nP_A}-1\right|\leq\frac1{\log n},
\]
we have
\begin{align*}
\mathbb P\left(
\mathcal D_{A,n}(t,d,q)^c\mid N_A=k
\right)
&\leq \frac{2}{n^{t}}\,, \qquad
\mathbb P\left(
\widetilde{\mathcal D}_{A,n}(t)^c\mid N_A=k
\right)
\leq \frac{2Dm}{n^{t}}.
\end{align*}
\end{lemma}
\begin{proof}
All thresholds on \(n\) below can be chosen independently of
\(A\) and \((d,q)\). For the first result, we apply the binomial Chernoff bound
to \(N_A\sim\mathrm{Binomial}(n,P_A)\). For \(0<\lambda<1\),
\[
\mathbb P\left(
\left|\frac{\widehat P_A}{P_A}-1\right|\geq\lambda
\right)
=
\mathbb P\left(|N_A-nP_A|\geq\lambda nP_A\right)
\leq
2\exp\left(-\frac{\lambda^2 nP_A}{3}\right).
\]
Take $\lambda=1/\log n$, which lies in $(0,1)$ for $n\geq3$.
By \eqref{lemma:admissible_mass_lower},
$nP_A/(\log n)^4\to\infty$, so for sufficiently large $n$, we have $nP_A\geq 6(\log n)^4$ for every such $A$. Substituting this lower bound into the Chernoff bound gives
$$
\mathbb P(\mathcal E_{A,n}^c)
\leq 2\exp\left\{-\frac{nP_A}{3(\log n)^2}\right\}
\leq 2e^{-2(\log n)^2}
\leq e^{-(\log n)^2}
= n^{-\log n}.
$$

We now turn to the split count events. The event
\(\mathcal D_{A,n}(t,d,q)\) is unchanged when $A_0$ and $A_1$ are
interchanged, so we may assume without loss of generality that
\(P_{A_0\mid A}\leq1/2\). Since $(N_{A_0},N_{A_1},n-N_A) \sim\mathrm{Multinomial}\!\left(n;\,P_{A_0},P_{A_1},1-P_A\right)$, conditioning gives \(N_{A_0}\mid N_A=k\sim\mathrm{Binomial}(k,P_{A_0\mid A})\). For the next conditional Chernoff bound, we set
\[
\eta
=
\sqrt{
\frac{6t\log n}{k}
\frac{(1-P_{A_0\mid A})}{P_{A_0\mid A}}
}.
\]
With this choice, on $\{N_A=k\}$, the event
$\mathcal D_{A,n}(t,d,q)^c$ is equivalent to
\[
 |N_{A_0}-kP_{A_0\mid A}|
 \geq\eta kP_{A_0\mid A}.
\]
The binomial Chernoff bound therefore applies once we verify
$0<\eta<1$. By the condition on $k$ and \eqref{lemma:admissible_mass_lower},
\[
 k\geq\left(1-\frac1{\log n}\right)nP_A
 \gtrsim(\log n)^5.
\]
As \eqref{lemma:admissible_mass_lower} also directly gives $P_{A_0\mid A}\in[b,1-b]$, we have for sufficiently large $n$,
\[
 0<\eta^2
 =\frac{6t\log n}{k}
   \frac{(1-P_{A_0\mid A})}{P_{A_0\mid A}}
 \lesssim(\log n)^{-4}<1,
\]
 uniformly over admissible $A$
and $(d, q)$. Thus
the Chernoff bound applies and gives
\begin{align*}
\mathbb P\left(
\mathcal D_{A,n}(t,d,q)^c\mid N_A=k
\right)
&\leq
2\exp\left(-\frac{\eta^2 kP_{A_0\mid A}}{3}\right)\\
&=
2\exp[-2t(1-P_{A_0\mid A})\log n]\\
&\leq 2n^{-t},
\end{align*}
where the last inequality uses \(1-P_{A_0\mid A}\geq1/2\).

For the simultaneous event \(\widetilde{\mathcal D}_{A,n}(t)\),
its complement is the union of the individual complements $\mathcal D_{A,n}(t,d,q)^c$ over
all candidate splits. There are \(D(m-1)\) such splits, so the
union bound simply gives
\begin{align*}
\mathbb P\left(
\widetilde{\mathcal D}_{A,n}(t)^c\mid N_A=k
\right)
&\leq
\sum_{d,q}
\mathbb P\left(
\mathcal D_{A,n}(t,d,q)^c\mid N_A=k
\right)\\
&\leq 2D(m-1)n^{-t}
\leq 2Dm\,n^{-t}.
\end{align*}
\end{proof}

\subsubsection{Splitting and stopping probabilities}
Lemmas~\ref{lemma:main_sufficient_condition} and \ref{lemma:stopping}
express the splitting/stopping decisions in terms of the count events. We now use this connection and the just shown probabilistic bound to obtain
probabilistic bounds for the two requirements of admissibility in
Definition~\ref{def: admissible trees}, namely near-optimal splitting at
identifiable oracle nodes and bounded splitting below non-identifiable
oracle leaves. We first control the probability of a near-optimal split
at an identifiable oracle node.
\begin{lemma}[Near-optimal splitting probability]
\label{lemma: nodewise splitting probability}
Let $\mathcal G_{A,n}$ denote the near-optimal splitting event in
Lemma~\ref{lemma:main_sufficient_condition}. For sufficiently
large $n$, uniformly over identifiable oracle nodes
$A\in\mathcal I_n^*$, we have
\[
 \mathbb P(\mathcal G_{A,n}^c)
 \leq \frac{2Dm}{n^{t_0}}+\frac1{n^{\log n}},
\]
where $t_0$ is defined in (\ref{eq:s0}).
\end{lemma}

\begin{proof}
By Lemma~\ref{lemma:main_sufficient_condition},
$\mathcal E_{A,n}\cap\widetilde{\mathcal D}_{A,n}(t_0)
\subseteq\mathcal G_{A,n}$. Thus, a failure of near-optimal
splitting requires either the node count event to fail, or the
node count event to hold while the simultaneous split count
event fails:
\[
 \mathcal G_{A,n}^c
 \subseteq\mathcal E_{A,n}^c
 \cup\left\{
 \mathcal E_{A,n}\cap\widetilde{\mathcal D}_{A,n}(t_0)^c
 \right\}.
\]
Lemma~\ref{lem:main_ep} bounds the first probability by
$n^{-\log n}$. On $\mathcal E_{A,n}$, the observed count $N_A$ satisfies
the condition for the conditional bound in Lemma~\ref{lem:main_ep}.
Since $\mathcal E_{A,n}$ is determined by $N_A$, conditioning
on $N_A$ and applying the tower property gives
\begin{align*}
 \mathbb P\left(
 \mathcal E_{A,n}\cap\widetilde{\mathcal D}_{A,n}(t_0)^c
 \right)
 &=\mathbb E\left[
 \mathbbm{1}_{\mathcal E_{A,n}}
 \mathbb P\left(
 \widetilde{\mathcal D}_{A,n}(t_0)^c\mid N_A
 \right)\right] \leq\frac{2Dm}{n^{t_0}}.
\end{align*}
Combining the above with the union bound proves the claim.
\end{proof}

We next control splitting below a non-identifiable oracle leaf.
Admissibility allows fewer than $k_{\max}$
refinement splits, so the required guarantee concerns the total
number of splits rather than the decision at a single node.
By Lemma~\ref{lemma:splitting_bound}, any sequence reaching
$k_{\max}$ splits must contain several stopping splits.
Lemma~\ref{lemma:stopping} then relates continued splitting to
failures of the count events. The following result bounds the
probability of these repeated failures.

\begin{lemma}[Excess splitting probability]
\label{lemma:excess_splitting_leaf}
For each non-identifiable oracle leaf $A\in\mathcal L_n^*$, consider applying the GEB splitting and stopping rules starting at $A$, and let $\mathcal H_{A,n}$ denote the event that this recursion makes at least $k_{\max}$ splits.

For sufficiently large $n$, uniformly over non-identifiable
oracle leaves $A\in\mathcal L_n^*$, we have
\begin{align*}
 &\mathbb P(\mathcal{H}_{A,n})\leq
 k_{\max}!\{D(m-1)\}^{k_{\max}}
 \left\{
 (2n^{-1/15})^{\lceil15t_0\rceil}
 +\lceil15t_0\rceil n^{-\log n}
 \right\},
\end{align*}
where $t_0$ is defined in \eqref{eq:s0} and $k_{\max}$ is
given in \eqref{eq: k_max}.
\end{lemma}

\begin{proof}
Throughout this proof, we consider a fixed node $A\in\mathcal L_n^*$ and take $A$ as the root of the GEB recursion. We will now proceed to upper bound  $\mathbb P(\mathcal{H}_{A,n})$ using a conditional independence argument.

Fix a deterministic sequence of $k_{\max}$ splits rooted at $A$, where at least $k_s:=\lceil15t_0\rceil$ of these are stopping splits by Lemma~\ref{lemma:splitting_bound}. For \(i=1,\ldots,k_s\), let \((d_i,q_i)\) denote the \(i\)-th stopping split in the sequence and \(B_i\) the node at which it occurs. Since $A$ is a leaf of a truncated admissible tree,
applying the first $k_{\max}-1$ splits at $A$ produces
an admissible tree by Definition~\ref{def: admissible trees}.
Every node at which one of the $k_{\max}$ splits occurs
already belongs to this tree, including the one for the last split. Thus all $B_i$ belong to a common
admissible tree.

Next, we write the count events for the nodes $B_i$ with stopping splits as 
\[
 \mathcal E_i=\mathcal E_{B_i,n},
 \qquad
 \mathcal D_i=\mathcal D_{B_i,n}(1/15,d_i,q_i).
\]
Suppose the GEB recursion begins with the fixed sequence of $k_{\max}$ splits, where it may continue to split further. It must thus select $(d_i,q_i)$ at $B_i$ for every $i=1,\ldots,k_s$.  If all the node count events
$\mathcal E_i$ hold, Lemma~\ref{lemma:stopping} implies that every
$\mathcal D_i$ must fail, otherwise GEB could not have selected
the corresponding split. The only other possibility is that at least one $\mathcal E_i$ fails. For $j = 1,\ldots, k_s$, define the event
\[
 F_j=\bigcap_{i=1}^j(\mathcal D_i^c\cap\mathcal E_i),
 \qquad F_0=\Omega.
\]
Thus, the event that the GEB recursion begins with the fixed sequence of $k_{\max}$ splits is contained in the event 
\begin{align}\label{eq:inclusion}
   F_{k_s}\cup\bigcup_{i=1}^{k_s}\mathcal E_i^c. 
\end{align}
The above form will allow us to leverage our probabilistic bounds. The node count failures $\mathcal{E}_i^c$ have small probabilities guaranteed by Lemma~\ref{lem:main_ep}, which are easy to handle. The harder task is to control $F_{k_s}$ probabilistically, which will require conditional independence along the split sequence, which we show now.

For a fixed node $B$ with $P_B>0$ and children $B_0,B_1$,
and any finite collection of fixed measurable sets
$C_1,\ldots,C_r\subseteq B^c$, we have
\begin{align}
 (N_{B_0},N_{B_1})
 &\perp\!\!\!\perp(N_{C_1},\ldots,N_{C_r})\mid N_B,
 \label{lemma:multinomial_conditional_independence}\\
   N_{B_0}\mid N_B
 &\sim\mathrm{Binomial}(N_B,P_{B_0\mid B}).\notag
\end{align}
To see this, first suppose that $0<P_B<1$ and that
$C_1,\ldots,C_r$ form a partition of $B^c$.
The conclusion then follows directly from the properties of the multinomial distribution, e.g. \cite[Section 2.1.5]{agresti2013categorical}. For arbitrary \(C_1,\ldots,C_r\subseteq B^c\), choose a finite partition of \(B^c\) such that each \(C_i\) is a union of cells. Since each \(N_{C_i}\) is a sum of cell counts, the same conditional independence holds for \((N_{C_1},\ldots,N_{C_r})\).
If $P_B=1$, all external counts are zero a.s., and
the conclusion is immediate.

We now leverage the above for a fixed stopping split $j$. For each \(i<j\), the events \(\mathcal E_i\) and \(\mathcal D_i\) depend only on the counts in \(B_i\) and \(B_{i0}\), where $B_{i0}$ is the left child of $B_i$. Since no earlier split cuts through \(B_j\), each such region \(C\) either contains \(B_j\) or is disjoint from it. Hence
$$
N_C=N_{C\setminus B_j}+\mathbbm{1}\{B_j\subseteq C\}\,N_{B_j}.
$$
The indicator is deterministic because the split sequence is fixed. Thus, given \(N_{B_j}\), counts on fixed subsets of \(B_j^c\) determine \(F_{j-1}\), while the child count \(N_{B_{j0}}\) determines \(\mathcal D_j\). By \eqref{lemma:multinomial_conditional_independence},
\begin{align}\label{eq:condit_indep}
\mathcal D_j\perp\!\!\!\perp F_{j-1}\mid N_{B_j}.
\end{align}
We will now bound \(\mathbb P(F_j)\) recursively, using \(F_j=F_{j-1}\cap\mathcal E_j\cap\mathcal D_j^c\). Since \(N_{B_j}\) determines \(\mathcal E_j\), conditioning on \(N_{B_j}\) and \(\mathbbm{1}_{F_{j-1}}\) gives
\begin{align*}
\mathbb P(F_j)
&=\mathbb E\!\left[
  \mathbbm{1}_{F_{j-1}}\mathbbm{1}_{\mathcal E_j}
  \mathbb P\!\left(\mathcal D_j^c\mid N_{B_j},\mathbbm{1}_{F_{j-1}}\right)
\right]\\
&=\mathbb E\!\left[
  \mathbbm{1}_{F_{j-1}}\mathbbm{1}_{\mathcal E_j}
  \mathbb P\!\left(\mathcal D_j^c\mid N_{B_j}\right)
\right],
\end{align*}
where the first equality follows from the tower property, and the second equality follows from the conditional independence in \eqref{eq:condit_indep}.

On \(\mathcal E_j\), the observed count \(k=N_{B_j}\) satisfies \(\lvert k/(nP_{B_j})-1\rvert\leq1/\log n\). Lemma~\ref{lem:main_ep} with \(t=1/15\) therefore bounds $\mathbb P\!\left(\mathcal D_j^c\mid N_{B_j}\right)$ by \(2n^{-1/15}\) on this event. Hence, for sufficiently large \(n\), uniformly over \(A\), the candidate sequences, and \(j\),
\[
\mathbb P(F_j)
\leq 2n^{-1/15}\mathbb P(F_{j-1}\cap\mathcal E_j)
\leq 2n^{-1/15}\mathbb P(F_{j-1}).
\]
For $j=1$, the same calculation holds with $F_0=\Omega$.
Iteration yields
\[
 \mathbb P(F_{k_s})\leq(2n^{-1/15})^{k_s}.
\]
Lemma~\ref{lem:main_ep} also gives
$\mathbb P(\mathcal E_i^c)\leq n^{-\log n}$ for every $i$.
By \eqref{eq:inclusion}, the probability that the GEB recursion begins with the fixed sequence of $k_{\max}$ splits is at most
\[
 (2n^{-1/15})^{k_s}+k_sn^{-\log n}.
\]
By \eqref{eq: subtree complexity}, there are at most \(k_{\max}!\{D(m-1)\}^{k_{\max}}\) candidate sequences of length \(k_{\max}\). 
The event \(\mathcal H_{A,n}\) occurs exactly when the recursion begins with one of these sequences, so taking a union bound over all possible $k_{\max}$ split sequences completes the proof.
\end{proof}

The two probabilistic bounds in this section correspond to the two requirements
of admissible trees, which we now leverage to prove 
Theorem~\ref{theorem:geb_tree_bound} in the next section. 
 
\section{Proof of Theorem~\ref{theorem:geb_tree_bound}}
\label{app: admissible}

By Definition~\ref{def: admissible trees}, an admissible tree
is obtained by splitting identifiable nodes near-optimally until
the leaves are non-identifiable, followed by fewer than $k_{\max}$
additional splits below each leaf. We show that GEB satisfies
both requirements with probability tending to one.

Suppose the event
$\bigcap_{A\in\mathcal I_n^*}\mathcal G_{A,n}$ holds, with
$\mathcal G_{A,n}$ defined in
Lemma~\ref{lemma:main_sufficient_condition}.
For each $A\in\mathcal I_n^*$, the event $\mathcal G_{A,n}$
ensures that all candidate splits are eligible and that
the selected split is near-optimal with a positive log split
Bayes factor. If the root is identifiable, it belongs to
$\mathcal I_n^*$, so GEB splits it near-optimally.
After a near-optimal split on a node $A \in \mathcal{I}_n^*$, each identifiable child also belongs
to $\mathcal I_n^*$. Induction therefore shows that under $\bigcap_{A\in\mathcal I_n^*}\mathcal G_{A,n}$, GEB
starting at the root splits every identifiable node it reaches near-optimally
without stopping. 

By Lemma~\ref{lemma: nodewise splitting probability} and Proposition~\ref{def:complexity_exponents}, a union bound over
$\mathcal I_n^*$ gives
\begin{align*}
 \mathbb P\left(
 \bigcup_{A\in\mathcal I_n^*}\mathcal G_{A,n}^c
 \right)
 &\lesssim \left(\frac{n}{C_0\log n}\right)^{t_0}
       \left(\frac{2Dm}{n^{t_0}}+\frac1{n^{\log n}}\right)\\
 &\lesssim(\log n)^{-t_0}
       \left(1+n^{t_0-\log n}\right)
 \to0,
\end{align*}
since $t_0$ is a positive constant. Thus, with probability tending to one, GEB splits every identifiable node it reaches near-optimally, establishing the first requirement of admissibility.

For the second requirement, suppose the event
$\bigcap_{A\in\mathcal L_n^*}\mathcal H_{A,n}^c$ holds, with $\mathcal{H}_{A,n}$ defined in Lemma~\ref{lemma:excess_splitting_leaf}.
On this event, applying the GEB splitting and stopping
rules starting at any $A\in\mathcal L_n^*$ produces
a subtree with fewer than $k_{\max}$ splits. When the first requirement of admissibility holds, the first non-identifiable node \(A\) on each branch of \(T_n\) belongs to \(\mathcal L_n^*\).  For each such \(A\), the event \(\mathcal H_{A,n}^c\) holds, so the subtree of \(T_n\) rooted at \(A\) contains fewer than \(k_{\max}\) splits in total, establishing the second requirement of admissibility.

By Lemma~\ref{lemma:excess_splitting_leaf} and
Proposition~\ref{def:complexity_exponents}, a union bound gives
\begin{align*}
 \mathbb P\left(
 \bigcup_{A\in\mathcal L_n^*}\mathcal H_{A,n}
 \right)
 &\lesssim
 \left(\frac{n}{C_0\log n}\right)^{t_0}k_{\max}!\{D(m-1)\}^{k_{\max}}
 \left\{
 (2n^{-1/15})^{\lceil15t_0\rceil}
 +\lceil15t_0\rceil n^{-\log n}
 \right\}\\
&\lesssim (\log n)^{-t_0}
\left\{
n^{t_0-\lceil15t_0\rceil/15}
+n^{t_0-\log n}
\right\}
\to 0,
\end{align*}
since $k_{\max}$ is a constant and $\lceil15t_0\rceil/15\geq t_0$.

Both requirements of an admissible tree from Definition~\ref{def: admissible trees} thus hold on the event
\begin{align}\label{eq:joint_event}
\left(\bigcap_{A\in\mathcal I_n^*}\mathcal G_{A,n}\right)
\cap
\left(\bigcap_{A\in\mathcal L_n^*}\mathcal H_{A,n}^c\right).
\end{align}
Additionally, if the root $\mathcal A$ is non-identifiable, the first requirement of admissibility holds automatically. By the two preceding probabilistic bounds and a union bound argument, the event \eqref{eq:joint_event} has probability tending to one, so
\[
\mathbb P(T_n\in\mathcal T_n)\to1.
\]
\section{Posterior contraction}
\label{app: contraction}

By Theorem~\ref{theorem:geb_tree_bound}, the GEB tree is
admissible with probability tending to one. Admissibility
controls both the remaining signal in its leaves and the
number of leaf probabilities to be estimated. We first use these properties
to establish convergence of the posterior mean.

Let $\mathcal P_n$ be the leaf partition of the GEB tree $T_n$.
Recall from \eqref{eq: posterior predictive} that the posterior
mean density on this partition is
\[
 \bar f_n(x)
 =\sum_{A\in\mathcal P_n}\bar G_A h(x\mid A),
 \qquad
 \bar G_A=\frac{\alpha H_A+N_A}{n+\alpha}.
\]
To separate the prior contribution from the error due to
partitioning and estimating the cell probabilities, define
\[
 \hat{f}_n(x)
 =\sum_{A\in\mathcal P_n}\widehat P_A h(x\mid A),
 \qquad
 \widehat P_A=\frac{N_A}{n},
\]
and its population counterpart
\[
 f_n(x)
 =\sum_{A\in\mathcal P_n}P_A h(x\mid A).
\]
Since $H$ is uniform, both are constant on each leaf.
The triangle inequality gives 
\begin{equation}\label{eq: bias-variance}
 \| p - \bar f_n\|_{L_1}
 \leq \| \hat{f}_n - \bar f_n\|_{L_1}
     +\|p-f_n\|_{L_1}
     +\|f_n-\hat{f}_n\|_{L_1}.
\end{equation}
The first term measures the effect of the prior on the cell
probabilities. The second is the approximation bias of the partition,
and the third is the sampling error in estimating its cell
probabilities. With the prior-bias-variance decomposition, we are able to show the posterior mean convergence based on the admissibility of the GEB tree.
\vspace{-2mm}
\begin{lemma}[Posterior mean convergence]
\label{lemma:posterior_mean_convergence}
Fix $m\geq2$ and $\alpha>0$, and suppose
Assumptions~\ref{as:bounded density},
\ref{as:greedy identifiable}, and~\ref{as:decay} hold.
There exists a constant $C>0$, independent of $n$, such that
\[
 \mathbb P\left\{
 \|p-\bar f_n\|_{L_1}
 \leq C\tau_n^{(1-1/\gamma)/2}
 \right\}\to1.
\]
\end{lemma}
\vspace{-2mm}
\begin{proof}
We bound the three terms in \eqref{eq: bias-variance}
in order.
For the prior contribution, observe that
\[
 \bar G_A-\widehat P_A
 =\frac{\alpha}{n+\alpha}(H_A-\widehat P_A).
\]
The cells are disjoint and each conditional base density
integrates to one, so
\begin{equation}\label{eq:approx_error}
 \|\hat{f}_n-\bar f_n\|_{L_1}
 =\sum_{A\in\mathcal P_n}|\bar G_A-\widehat P_A|
 \leq\frac{\alpha}{n+\alpha}
       \sum_{A\in\mathcal P_n}(H_A+\widehat P_A)
 =\frac{2\alpha}{n+\alpha}.
\end{equation}
This bound holds uniformly over any $\mathcal{P}_n$.

We next work on the admissibility event
$\{T_n\in\mathcal T_n\}$ to control the bias and variance.
By Definition~\ref{def: admissible trees}, each leaf is
non-identifiable or is a descendant of a non-identifiable
node. Since Lemma~\ref{lem:split_gain} gives
$S_{A_0}+S_{A_1}\leq S_A$, signal cannot increase along a
path, so $S_A<\tau_n$ for every $A\in\mathcal P_n$.
Applying the signal identity \eqref{eq:KL_signal} and
the admissible leaf count \eqref{lemma:leaf_count} gives
\[
 K(p;f_n)
 =\sum_{A\in\mathcal P_n}S_A
 \leq|\mathcal P_n|\tau_n
 \lesssim\tau_n^{1-1/\gamma}.
\]
Pinsker's inequality therefore yields
\begin{equation}\label{eq:admissible_bias_bound}
 \|p-f_n\|_{L_1}
 \leq\sqrt{2K(p;f_n)}
 \lesssim\tau_n^{(1-1/\gamma)/2}.
\end{equation}

For the variance term, following
the proof of \citet[Theorem~1]{data-adaptive-consistency},
we first express the sampling error through cell probability
differences:
\[
 \|f_n- \hat{f}_n\|_{L_1}
 =\sum_{A\in\mathcal P_n}|\widehat P_A-P_A|.
\]
Since the empirical and population cell probabilities both sum to one, the total excess probability on cells with \(\widehat P_A>P_A\) equals the total missing probability on cells with \(\widehat P_A<P_A\).
As in the proof of \citet[Lemma~1]{data-adaptive-consistency}, this gives
\[
 \sum_{A\in\mathcal P_n}|\widehat P_A-P_A|
 =2\sum_{\substack{A\in\mathcal P_n\\
                   \widehat P_A > P_A}}
       (\widehat P_A-P_A).
\]
On $\{T_n\in\mathcal T_n\}$, the union of the cells in the
last sum belongs to the deterministic collection
$\mathscr B_n$ from Lemma~\ref{lemma:complexity}, so
\begin{equation}\label{eq:variance_measure_deviation}
 \|f_n- \hat{f}_n\|_{L_1}
 \leq2\sup_{B\in\mathscr B_n}|\widehat P_B-P_B|,
\end{equation}
where $\widehat P_B=n^{-1}\sum_{i=1}^n\mathbbm{1}(X_i\in B)$.
Thus it suffices to control empirical probability deviations
uniformly over $\mathscr B_n$. 

We now bound the above supremum under the unconditional sampling law. For each fixed $B\in\mathscr B_n$, Hoeffding's inequality
gives, for any fixed $L>0$,
\begin{align*}
 \mathbb P\left(
 |\widehat P_B-P_B|>L\tau_n^{(1-1/\gamma)/2}
 \right)
 &\leq2\exp\left(-2nL^2\tau_n^{1-1/\gamma}\right) =2\exp\left(-2L^2C_0\tau_n^{-1/\gamma}\log n\right).
\end{align*}
By the complexity bound \eqref{lemma:B_P_n}, there exists
a constant $C_{\mathscr B}>0$, independent of $n$, such that
\[
 \log|\mathscr B_n|
 \leq C_{\mathscr B}\tau_n^{-1/\gamma}\log n.
\]
Choose $L$ sufficiently large that
$2L^2C_0>C_{\mathscr B}$. A union bound then gives
\begin{align*}
 &\mathbb P\left(
 \sup_{B\in\mathscr B_n}|\widehat P_B-P_B|
 >L\tau_n^{(1-1/\gamma)/2}
 \right)\leq
 2\exp\left\{
 -(2L^2C_0-C_{\mathscr B})
 \tau_n^{-1/\gamma}\log n
 \right\}
 \to0.
\end{align*}
By the above bound and Theorem~\ref{theorem:geb_tree_bound}, we have that the following joint event holds with probability tending to $1$:
\[
 \left\{
 T_n\in\mathcal T_n,\;
 \sup_{B\in\mathscr B_n}|\widehat P_B-P_B|
 \leq L\tau_n^{(1-1/\gamma)/2}
 \right\}
\]
On this event, \eqref{eq:variance_measure_deviation} gives
\[
 \|f_n-\hat f_n\|_{L_1}
 \leq2L\tau_n^{(1-1/\gamma)/2}.
\]
Moreover, \eqref{eq:approx_error} and
\eqref{eq:admissible_bias_bound} show that the prior
contribution is of smaller order and the approximation
bias is of order at most $\tau_n^{(1-1/\gamma)/2}$.
Combining the three bounds in \eqref{eq: bias-variance} gives, on this event and for sufficiently large $n$,\vspace{-2mm}
$$
\|p-\bar f_n\|_{L_1}
\lesssim \tau_n^{(1-1/\gamma)/2}.
$$\vspace{-2mm}
Since this event has probability tending to one, the result follows.
\end{proof}

\subsection{Proof of Theorem~\ref{th:posterior_contraction}}
Let \(\varepsilon_n=\tau_n^{(1-1/\gamma)/2}\), let \(C\) be the constant in Lemma~\ref{lemma:posterior_mean_convergence}, and fix \(M>C\). 

Conditional on the data, $\mathcal{P}_n$ is fixed,
and the posterior cell probabilities satisfy
\[
 (G_A:A\in\mathcal P_n)\mid x_{1:n}
 \sim\operatorname{Dirichlet}
       (\alpha H_A+N_A:A\in\mathcal P_n).
\]
Their conditional means and variances are
\[
 \mathbb E_\Pi[G_A\mid x_{1:n}]=\bar G_A,
 \qquad
 \operatorname{Var}_\Pi(G_A\mid x_{1:n})
 =\frac{\bar G_A(1-\bar G_A)}{n+\alpha+1}.
\]
Since the cells are disjoint and $h( \cdot \mid A)$ integrates to 1 over $A$, Cauchy--Schwarz gives
\[
 \|f-\bar f_n\|_{L_1}^2
 =\left(\sum_{A\in\mathcal P_n}|G_A-\bar G_A|\right)^2
 \leq|\mathcal P_n|
       \sum_{A\in\mathcal P_n}(G_A-\bar G_A)^2.
\]
Taking posterior expectations therefore yields
\begin{align}
 \mathbb E_\Pi[\|f-\bar f_n\|_{L_1}^2\mid x_{1:n}]
 &\leq|\mathcal P_n|
       \sum_{A\in\mathcal P_n}
       \operatorname{Var}_\Pi(G_A\mid x_{1:n})
 \notag\\
 &=\frac{|\mathcal P_n|}{n+\alpha+1}
       \left(1-\sum_{A\in\mathcal P_n}\bar G_A^2\right)
 \notag\\
 &\leq\frac{|\mathcal P_n|}{n+\alpha+1}.
 \label{eq:posterior_second_moment}
\end{align}
For a singleton partition, its posterior cell probability is one almost surely, and the same bound holds. Thus \eqref{eq:posterior_second_moment} holds for every observed sample.

Next, fix an observed sample $x_{1:n}$. If $\|p-\bar f_n\|_{L_1}\leq C\varepsilon_n$, then by the triangle inequality, every posterior draw of $f$ satisfying $\|p-f\|_{L_1}\geq M\varepsilon_n$ also satisfies
\[
\|f-\bar f_n\|_{L_1}
\geq \|p-f\|_{L_1}-\|p-\bar f_n\|_{L_1}
\geq (M-C)\varepsilon_n.
\]
 If instead $\|p-\bar f_n\|_{L_1}>C\varepsilon_n$, we bound the posterior tail probability by one. Combining the two cases, the below holds for $\mathbb P$-almost all $X_{1:n}$:
\begin{align*}
\Pi\!\left\{\|p-f\|_{L_1}\geq M\varepsilon_n\mid X_{1:n}\right\}
&\leq
\mathbbm{1}_{\{\|p-\bar f_n\|_{L_1}>C\varepsilon_n\}}
+\Pi\!\left\{\|f-\bar f_n\|_{L_1}\geq(M-C)\varepsilon_n
             \mid X_{1:n}\right\}\\
&\leq
\mathbbm{1}_{\{\|p-\bar f_n\|_{L_1}>C\varepsilon_n\}}
+\frac{\mathbb E_\Pi[\|f-\bar f_n\|_{L_1}^2\mid X_{1:n}]}
      {(M-C)^2\varepsilon_n^2}\\
&\leq
\mathbbm{1}_{\{\|p-\bar f_n\|_{L_1}>C\varepsilon_n\}}
+\frac{|\mathcal P_n|}
      {(n+\alpha+1)(M-C)^2\varepsilon_n^2}.
\end{align*}
The second inequality applies Markov's inequality to
$\|f-\bar f_n\|_{L_1}^2$ under the posterior, and the third
uses \eqref{eq:posterior_second_moment}.

By Lemma~\ref{lemma:posterior_mean_convergence}, the indicator converges to zero in probability. By Theorem~\ref{theorem:geb_tree_bound} and Lemma~\ref{lem:admissible_leaf_count}, \(|\mathcal P_n|\lesssim\tau_n^{-1/\gamma}\) with probability tending to one, so the second term is of order $O_{\mathbb P}\{(\log n)^{-1}\}$ which also converges to zero in \(\mathbb P\)-probability. Substituting \(\tau_n=C_0\log n/n\) and absorbing $C_0^{(1-1/\gamma)/2}$ into \(M\) gives the stated rate.

\section{Experimental details}
\subsection{Model hyperparameters}
\label{app:model_hyperparameters}

For completeness, we list the hyperparameter settings of all methods used in the experiments. 
\begin{itemize}
    \item \textbf{GEB:} We set the shrinkage parameter to \(\alpha=1\) and divide each coordinate
into \(m=32\) grid subintervals. For the density-estimation experiments, we
use the PPB posterior mean given the GEB selected partition. For the
posterior-uncertainty experiment, we generate \(B=1{,}000\) posterior draws
of \(\theta\) using Algorithm~\ref{alg: ppb}.

\item \textbf{HMPT:} We use maximum resolution \(15\), partition-location penalty \(\eta=0.1\),
\(32\) grid subintervals, and \(1{,}000\) SMC particles as suggested by \cite{LM_Hidden}. The code is available at
\url{https://github.com/MaStatLab/HMPT_experiments}.

\item \textbf{DPMM:} In the density-estimation experiments, we use the variational inference implementation of \texttt{sklearn}, with truncation level \(50\), concentration parameter
\(1\), at most \(1{,}000\) iterations, and five initializations. Note that the DP concentration of $1$ matches our GEB choice of $\alpha = 1$. For the
posterior uncertainty experiment, we use the MCMC sampler implemented in
\texttt{BNPmix::PYdensity}. We again set the concentration parameter to \(1\) and run each chain for \(2{,}000\)
iterations, discarding the first \(1{,}000\) as burn-in and retaining
\(1{,}000\) posterior draws.

\item \textbf{PT:} We use a conventional P\'olya tree with midpoint splits, maximum depth of \(15\),
and Dirichlet parameter \(0.1k^2\) for each child at level \(k\) as in \cite{LM_Hidden}.
\end{itemize}

\newpage
\subsection{Real-data preprocessing}
\label{app:preprocessing}
We outline dataset-specific preprocessing below.
\begin{itemize}
    \item \textbf{POWER:}
    We carry out the same preprocessing steps as the MAF benchmark
    \cite{NIPS2017_6c1da886}
    (\url{https://github.com/gpapamak/maf}). In each replication, \(100{,}000\) observations are sampled without replacement from the \(2{,}049{,}280\) observations, with \(50{,}000\) assigned to training and \(50{,}000\) to testing.
    \item \textbf{Dry Bean:}
    No additional feature preprocessing is applied, and all \(16\) original numerical variables are used.

    \item \textbf{ISOLET:} 
    In each replication, the \(7{,}797\) observations are randomly
    divided into \(3{,}898\) training and \(3{,}899\) test
    observations. To stay within the $D$ range reported for HMPT in \cite{LM_Hidden}, PCA is then fitted to the
    standardized training sample, and the first $100$ principal components of the $617$ original dimensions are used for analysis.
\end{itemize}

\subsection{Hyperparameter sensitivity}
\label{app:geb-sensitivity}

\begin{figure}[htb!]
    \centering
    \includegraphics[width=\linewidth]{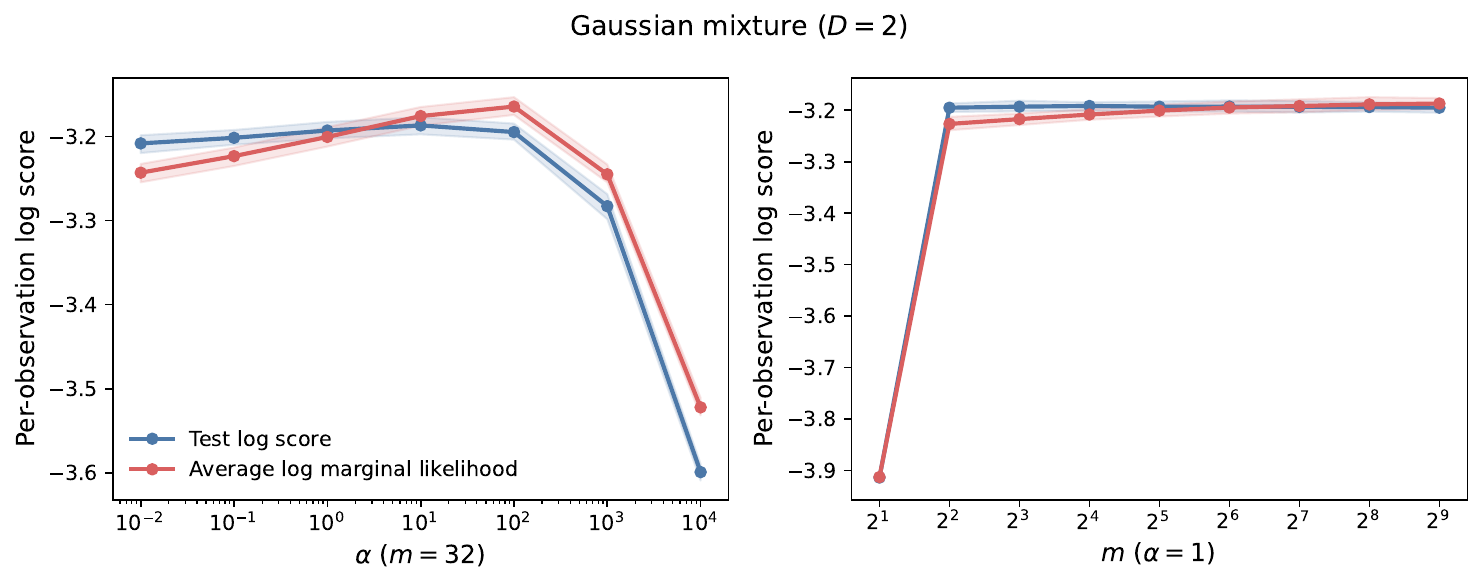}\\
    \includegraphics[width=\linewidth]{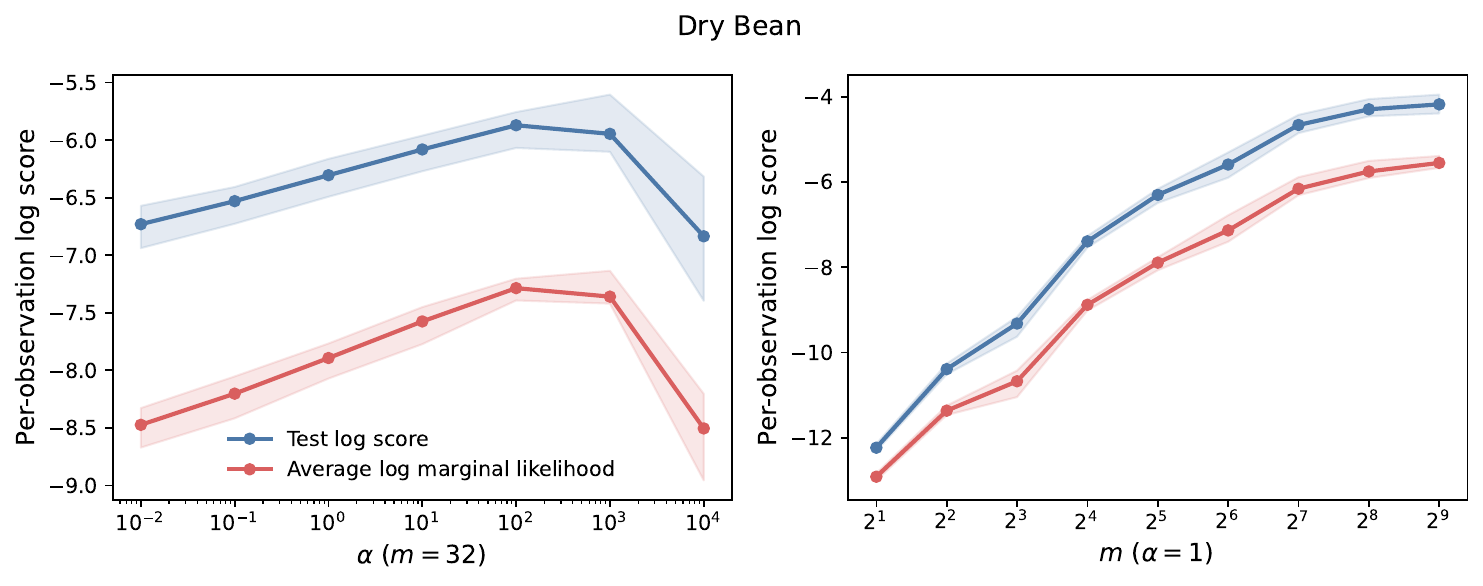}
    \caption{Sensitivity of GEB to $\alpha$ and $m$ for the
    Gaussian mixture simulation with $D=2$ (top) and Dry Bean
    (bottom). The left panels vary $\alpha$ from $10^{-2}$ to
    $10^4$ with $m=32$; the right panels vary $m$ from $2$ to
    $512$ with $\alpha=1$. Lines show the mean test log score
    and log marginal likelihood per training observation across
    $20$ train--test splits, and shaded regions show the
    corresponding interquartile ranges.}
    \label{fig:parameter-sensitivity}
\end{figure}

We examine the sensitivity of GEB to the shrinkage parameter
$\alpha$ and grid size $m$ using the Gaussian mixture simulation
with $D=2$ and the Dry Bean data. For each dataset, we use the
same $20$ train--test splits as in the main experiments and apply
the Gaussian marginal transformation fitted to the training sample.
We evaluate model fit by the log marginal likelihood per training observation and predictive performance by the average test log score:
\[
 \frac{1}{n_{\mathrm{train}}}\log\mathcal M(\mathcal P_n), \qquad
 \frac{1}{n_{\mathrm{test}}}
 \sum_{i=1}^{n_{\mathrm{test}}}\log\widehat p(X_i^{\mathrm{test}}).
\]
We vary $\alpha = 10^{\ell}$ for integer $\ell$ from $-2$ to $4$ with $m=32$ fixed, and $m = 2^{j}$ for integer $j$ from $1$ to $9$ with $\alpha=1$ fixed. Figure~\ref{fig:parameter-sensitivity} shows
the means and interquartile ranges of both criteria across
the $20$ splits.

At $m=32$, the test log score and log marginal likelihood both
initially increase with $\alpha$ and then decline. This pattern
is consistent with the competing effects of partition resolution
and posterior shrinkage discussed in the main text. We retain
$\alpha=1$ as a default for robustness, while the similar trends
of the two criteria suggest that additional marginal-likelihood tuning may
improve predictive performance.

At $\alpha=1$, both criteria generally increase with $m$, with diminishing
gains beyond a moderate grid size. The resolution needed depends
on the density structure, as smooth densities may require
fewer candidate splits, whereas sharp local features
can benefit from a denser grid. Increasing $m$ also raises the
candidate-search cost (but see Theorem \ref{theorem: maximum data point}). Following the grid choice in
\citet{LM_Hidden}, we use $m=32$ in the main paper to balance resolution and computation. Across both datasets, the agreement between
the marginal likelihood and test log score supports the use of
empirical Bayes tuning for $\alpha$ and $m$ as well, with more noticeable improvements for Dry Bean.

\end{appendices}

\end{document}